\documentclass[11pt]{article}
\usepackage[utf8]{inputenc}
\usepackage{amsmath,amsfonts,amssymb}
\usepackage{psfrag}
\usepackage{enumerate}
\usepackage{mathrsfs}
\usepackage{graphicx}
\usepackage{wrapfig}
\usepackage{xcolor}
\usepackage{caption}
\usepackage{amsthm}
\usepackage{subfig}
\usepackage{xcolor}
\usepackage{esint}

\usepackage{comment}

\usepackage{jheppub}
\allowdisplaybreaks

 \newcommand{\be}{\begin{equation}}
\newcommand{\beq}{\begin{equation}}
 \newcommand{\ee}{\end{equation}}
 \newcommand{\bea}{\begin{align}}
 
 \newcommand{\eea}{\end{align}}

\def\nn{\nonumber}

\global\long\def\mJ{\mathcal{J}}%

\global\long\def\mS{\mathcal{S}}%

\global\long\def\mX{\mathcal{X}}%

\global\long\def\ra{\rightarrow}%

\global\long\def\avg#1{\left\langle #1\right\rangle }%

\global\long\def\f#1#2{\frac{#1}{#2}}%
 
\global\long\def\del{\partial}%
 
\global\long\def\t{\theta}%
 
\global\long\def\a{\alpha}%
 
\global\long\def\b{\beta}%
 
\global\long\def\g{\gamma}%
 
\global\long\def\G{\Gamma}%
 
\global\long\def\s{\sigma}%
 
\global\long\def\r{\rho}%
 
\global\long\def\d{\delta}%
 
\global\long\def\Tr{\text{Tr}}%

\global\long\def\ket#1{\left\langle #1\right|}%
 
\global\long\def\bra#1{\left|#1\right\rangle }%

\global\long\def\I{\mathbb{I}}%
 
\global\long\def\Z{\mathbb{Z}}%

\global\long\def\R{\mathbb{R}}%

\global\long\def\app{\approx}%

\preprint{MIT-CTP/5511}

\title{Commuting SYK: a pseudo-holographic model}

\author{Ping Gao}%

\affiliation{Center for Theoretical Physics,\\ Massachusetts Institute of Technology,
Cambridge, MA 02139, USA}

\emailAdd{pgao@mit.edu}

\abstract{In this work, we study a type of commuting SYK model in which all terms in the Hamiltonian are commutative to each other. Because of the commutativity, this model has a large number of conserved charges and is integrable. After the ensemble average of random couplings, we can solve this model exactly in any $N$. Though this integral model is not holographic, we do find that it has some holography-like features, especially the near-perfect size winding in high temperatures. Therefore, we would like to call it pseudo-holographic. We also find that the size winding of this model has a narrowly peaked size distribution, which is different from the ordinary SYK model. We apply the traversable wormhole teleportation protocol in the commuting SYK model and find that the teleportation has a few features similar to the semiclassical traversable wormhole but in different parameter regimes. We show that the underlying physics is not entirely determined by the size-winding mechanism but involves the peaked-size mechanism and thermalization. Lastly, we comment on the recent simulation of the dynamics of traversable wormholes on Google's quantum processor.  }

\begin{document}

\maketitle

\section{Introduction}

Sachdev-Ye-Kitaev (SYK) model \cite{sachdev1993gapless,kitaev2015simple} has been studied extensively in recent years as a candidate for low dimensional holography \cite{kitaev2015simple, Maldacena:2016hyu}. This model consists of $N$ Majorana fermions $\psi_i$ with a $q$-local all-to-all random Hamiltonian 
\be  
H=\sum_{\{i_k\}}\mJ_{i_1\cdots i_q} \psi_{i_1}\cdots \psi_{i_q}
\ee 
where $\mJ_{i_1\cdots i_q}$ obeys a Gaussian distribution. This 0+1 dimensional quantum mechanical model in large $N$ limit is conjectured to be dual to 2-dimensional quantum gravity. Since this model does not involve spatial direction, it is the simplest (potentially) holographic model, which might be realized in an experiment in the foreseeable future. Several experimental proposals have been made, e.g. \cite{Danshita:2016xbo, Garcia-Alvarez:2016wem, Pikulin:2017mhj, Chen:2018wvb,Brzezinska:2022mhj,Uhrich:2023ddx}. Along this direction, it has been proposed by Susskind and his collaborators as an exciting project called ``quantum gravity in the lab" \cite{Susskind:2017ney, Brown:2019hmk,Nezami:2021yaq}, which aims to utilize a near-future quantum computer or specially designed condensed matter system to simulate the dynamics of quantum gravity in asymptotic AdS background through holographic duality. 

One of the very first nontrivial tasks for ``quantum gravity in the lab" is to verify ER=EPR \cite{Maldacena:2013xja} in terms of the traversable wormhole teleportation protocol \cite{Gao:2019nyj}. Euclidean path integral formalism of quantum gravity suggests that a pair of identical entangled black holes (labeled as $l$ and $r$) in the thermofield double state is dual to a non-traversable wormhole (Einstein-Rosen bridge) connecting the two black holes behind their horizons. However, ER bridge does not grant a causal connection between two black holes behind the horizons, which makes the existence of ER bridge hard to verify. Nonetheless, turning on a generic coupling $\d H=\mu O_l O_r$ with a specific sign of $\mu$ between these two black holes will backreact on the geometry such that the ER bridge, if it exists, changes to a traversable wormhole \cite{Gao:2016bin}. Here $O_{l,r}$ are two generic identical operators respectively in the two black holes. Through this traversable wormhole, one can send a qubit into one black hole and receive it from the horizon of the other black hole. One can directly observe this causal connection between the two sides that is based on the ER bridge. This bulk process is dual to quantum teleportation in many-body systems on the boundary \cite{Gao:2016bin}, which was realized by a concrete protocol in the SYK model \cite{Gao:2019nyj}. Therefore, if we can implement this quantum teleportation protocol on a holographic model (e.g. SYK model) in an experimental setting, we will be able to simulate the dynamics of a traversable wormhole in quantum gravity.

Even though the SYK model is the best candidate for an experimentally realizable holographic model, it is still too complicated for state-of-art technology because it involves a massive number of random coupling terms in the Hamiltonian that scales as $N^q$ with $N$. If we implement the unitary evolution with this Hamiltonian on a quantum computer, the complexity of the circuit is much beyond current fault tolerance. Therefore, reducing the complexity of the Hamiltonian while keeping its essential holographic properties is necessary for the simulation of quantum gravity in the lab. In \cite{Garcia-Garcia:2020cdo,Xu:2020shn}, a sparse version of the SYK model has been studied, in which only $k N$ randomly selected terms in the full Hamiltonian exist. For large enough but still of order unity $k$, it has been shown in \cite{Garcia-Garcia:2020cdo} that the global spectral density of the sparse SYK model around its ground state energy $E_0$ has a form of $\sinh(\sqrt{\g (E-E_0)})$, which indicates a gravitational dual that is effectively described by a Schwarzian derivative. This was also discussed in \cite{Xu:2020shn} with a different method to suggest that a maximally chaotic gravitational sector exists in the sparse SYK model in low temperatures. Even though both methods rely on some approximations and the duality is not rigorously proven, minimizing the number of terms in the Hamiltonian while preserving holography to some extent becomes possible.

Following this idea of the sparse SYK model, the recent paper \cite{Jafferis:2022crx} shows that one can learn a $N=7$ sparse SYK Hamiltonian, which only contains 5 terms (see \eqref{eq:77}), to simulate the dynamics of a traversable wormhole on Google's quantum processor Sycamore. With this 5-term learned Hamiltonian, we can construct the thermofield double state of two identical SYK models, and then measure how much causal relation is built between them after turning on a two-sided coupling $e^{-i\mu V}$ that in the dual gravity will generate a traversable wormhole for $\mu>0$. A strong causal relation built in this way will guarantee high fidelity of the teleportation \cite{Gao:2019nyj}. However, the five terms in this learned Hamiltonian are all commutative to each other, which causes some debates \cite{Kobrin:2023rzr,Jafferis:2023moh} on to what extent this model is holographic. It is sort of peculiar and also surprising that this small $N$, five-term commuting SYK model exhibits some holographic features that (at least) qualitatively match with the dynamics of the traversable wormhole because, after all, it is an integrable model in essence.

Putting aside the debates \cite{Kobrin:2023rzr,Jafferis:2023moh}, it is indeed very interesting to understand why a commuting SYK model could exhibit some holographic features. Would this be a small $N$ behavior or it extends to large $N$? Does this only exist for specific learned Hamiltonians or for a generic ensemble of random commuting Hamiltonians? How different are these features from the authentic holography of the full SYK model with a non-commuting Hamiltonian? In this paper, we will try to study these questions by focusing on a type of commuting SYK model with random couplings for any even number $N$. This model is equivalent to the $q$ generalization of the well-known Sherrington-Kirkpatrick (SK) model \cite{sherrington1975solvable, thouless1977solution, parisi1979infinite} though we define it in terms of the fundamental Majorana fermions. In large $N$ limit, this model has a critical temperature $T_c$, below which there is a spin glass phase. We will only focus on the non-spin glass phase in this paper. Interestingly, we find that the commuting SYK model, on the one hand, is integrable, but on the other hand, after the ensemble average and in high temperature, it shows the size-winding property \cite{Brown:2019hmk,Nezami:2021yaq}, which is a very special feature only observed in holographic models before and is shown to be the mechanism for quantum teleportation through a traversable wormhole. Roughly speaking, size-winding means the coefficients of a (scrambled) growing operator expanded in terms of the basis of fixed size all have a phase proportional to the size (see more details in Section \ref{subsec:Size-winding}). Because of this, we would like to call the commuting SYK model {\it psuedo-holographic}. As we drop the temperature but still above $T_c$, the size-winding of the commuting SYK model is damped because the phases of the coefficients of the same size are not well aligned though their averaged phase is still proportional to their size.

After looking into this model in more detail, we find that this size winding in the large $N$ limit is quite different from the size winding in an ordinary SYK model. It has a narrowly peaked operator size distribution, which reflects that this model does not scramble as fast as a holographic model. For a holographic model, the scrambling speed measured by the out-of-time-ordered-correlator (OTOC) is exponential, while for the commuting SYK, the scrambling speed is just quadratic due to integrability. This leads to the conclusion that in the large $N$ limit and near the scrambling time, the quantum teleportation protocol works based on the peaked-size mechanism \cite{Schuster:2021uvg} rather than the size-winding mechanism though the latter property {\it is} present. 

The peaked size distribution persists only in the large $N$ limit and near the scrambling time regime. For a time much earlier than the scrambling time, the peaked-size mechanism fails to work for quantum teleportation and we find that an effective characteristic frequency due to the thermalization of this integrable model plays a crucial role in the sign difference effect of $\mu$, which means positive (negative) $\mu$ leads to a better (worse) 
 teleportation. In this regime, we also find the signal ordering is preserved in the teleportation protocol, which is surprisingly compatible with a semiclassical traversable picture, though in a much shorter time scale.
 
 For small $N$ systems, the size distribution is never narrowly peaked and the size-winding property still emerges after a short evolution of time in high temperatures. In this regime, the size winding starts to affect the fidelity of teleportation though thermalization is still equally important because the optimal parameter for teleportation has an order one deviation from the value required by the size-winding property. This is indeed the regime that \cite{Jafferis:2022crx} probes, and we suggest that the mechanism behind the simulation in \cite{Jafferis:2022crx} is an interplay between thermalization and size winding (see more analysis in Section \ref{sec:Conclusion-and-discussion}).

The paper is organized as follows. In Section \ref{sec2}, we define a type of commuting SYK model and show that it is not holographic by checking its spectrum, two-point function, and four-point functions. In Section \ref{sec:Commuting-SYK-has}, we set up the two-sided version of the commuting SYK and show that it has near-perfect size-winding in large $N$ limit and high temperature. We point out that this size-winding feature also comes with peaked size distribution. In Section \ref{sec4}, we apply the traversable wormhole teleportation protocol in the commuting SYK model in both large $N$ limit and small $N$ cases. We focus on the sign difference effect of $\mu$ and the preservation/inversion of the signal ordering. In Section  \ref{sec:Conclusion-and-discussion}, we summarize the conclusion and discuss a few questions. Appendix \ref{sec:Computation-of} includes the technical details of computing the correlation function in the traversable wormhole teleportation protocol.

\section{Commuting SYK is not holographic} \label{sec2}

\subsection{The model}\label{sec2.1}

The commuting SYK model consists of $N$ Majorana fermions obeying 
\begin{equation}
\{\psi_{i},\psi_{j}\}=\d_{ij}
\end{equation}
Define
\begin{equation}
X_{i}=\psi_{2i-1}\psi_{2i}\implies(X_{i})^{2}=-1/4,\quad i=1,\cdots,N/2
\end{equation}
and consider the Hamiltonian
\begin{equation}
H=\sum_{i_{k}}\mJ_{i_{1}\cdots i_{q/2}}X_{i_{1}}\cdots X_{i_{q/2}}\equiv\sum_{I}\mJ_{I}\mX_{I},\quad\,\mX_{I}\equiv X_{i_{1}}\cdots X_{i_{q/2}}\label{eq:3}
\end{equation}
where $I$ is the collective indices for all $q$ fermions in the
string $X_{i_{1}}\cdots X_{i_{q/2}}$. Note that $X_i$ takes eigenvalues of $\pm i/2$, the Hamiltonian \eqref{eq:3} indeed defines a $q$-local generalization \cite{derrida1980random, gardner1985spin, kirkpatrick1987dynamics} of the well-known Sherrington-Kirkpatrick (SK) model ($q=4$) \cite{sherrington1975solvable, thouless1977solution, parisi1979infinite}. Unlike these classic papers of the SK model focusing on the thermodynamics and the spin glass phase in low temperatures in $N\ra\infty$ 
 thermodynamic limit, we will instead mainly study the scrambling and teleportation features of this model in the context of comparison with the ordinary SYK model.\footnote{The comparison of free energy in low temperatures between bosonic SYK-like models and the ordinary fermionic SYK model are studied in \cite{Baldwin:2019dki}.}  In particular, for large $N$ case, we will limit our discussion {\it above} the critical temperature $T_c$ of spin glass phase. For finite and small $N$ cases, since no sharp phase transition and definite critical temperature exist, we will discuss the scenarios for all temperatures as long as it is not too low. Here we have a little abuse
of notation because for $\mJ_{I}$ the $I$ has $q/2$ length long
but for string $\mX_{I}$ the $I$ has $q$ length long recording
all fermionic indices. Similar to the SYK model, we take the random couplings
$\mJ_{I}$ as symmetric tensors whose ensemble average is 
\begin{equation}
\avg{\left(\mJ_{i_{1}\cdots i_{q/2}}\right)^{2}}=\s^{2}=\f{(q/2-1)!(N/2-q/2)!2^{q-1}}{(N/2-1)!}\mJ^{2}\label{eq:4}
\end{equation}
This mode is integrable because each $X_{i}$ is an independent conserved
charge. For simplicity, we assume $q\in4\Z$ throughout this paper. 

Note that \eqref{eq:3} is not the unique way to define a commuting SYK-like Hamiltonian. Here we construct a commuting SYK Hamiltonian using a set
of bi-fermion bosonic operators $X_{i}$ for $i=1,\cdots,N/2$, which only works for even $N$. To define more general
commuting SYK Hamiltonian, we can first set a few groups labeled by
$\a$ and for each $\a$ we define a set of bi-fermion bosonic
operators $X_{i}^{\a}$. For each group, we can define a Hamiltonian
$H^{\a}=\sum_{\{i_{k}\}}J_{i_{1}\cdots i_{q/2}}^{\a}X_{i_{1}}^{\a}\cdots X_{i_{q/2}}^{\a}$
and the total Hamiltonian is $H=\sum_{\a}H^{\a}$. Note that $X_{i}^{\a}$
defined for different $\a$ could have overlapped fermions but we
need to make sure that each term in $H^{\a}$ has even numbers of
overlapped fermions with all terms in other $H^{\a'}$. Besides, another type of integrable commuting SYK-like model has been studied in \cite{Balasubramanian:2021mxo,Craps:2022ese}, in which the complexity growth was mainly discussed. For simplicity, we will only focus on the commuting SYK model defined in \eqref{eq:3} and leave other constructions as a future direction of research.

Since each term in the Hamiltonian is commutative to each other, we
can exponentiate it exactly before the ensemble average. Using the the
fact
\begin{equation}
(\mX_{I})^{2}=(-1/4)^{q/2}=1/2^{q}\label{eq:5-1}
\end{equation}
we have
\begin{equation}
e^{\tau H}=\prod_{I}e^{\tau\mJ_{I}\mX_{I}}=\prod_{I}\left(\cosh\f{\tau\mJ_{I}}{2^{q/2}}+2^{q/2}\mX_{I}\sinh\f{\tau\mJ_{I}}{2^{q/2}}\right)\label{eq:8}
\end{equation}
For $e^{\tau H}$ moving across $\psi_{i}$, we have 
\begin{align}
e^{\tau H}\psi_{i} & =\prod_{I}\left(\cosh\f{\tau\mJ_{I}}{2^{q/2}}+2^{q/2}\mX_{I}\sinh\f{\tau\mJ_{I}}{2^{q/2}}\right)\psi_{i}\nonumber \\
 & =\psi_{i}\prod_{i\notin I}\left(\cosh\f{\tau\mJ_{I}}{2^{q/2}}+2^{q/2}\mX_{I}\sinh\f{\tau\mJ_{I}}{2^{q/2}}\right)\prod_{i\in I}\left(\cosh\f{\tau\mJ_{I}}{2^{q/2}}-2^{q/2}\mX_{I}\sinh\f{\tau\mJ_{I}}{2^{q/2}}\right)\label{eq:9}
\end{align}
where $\in$ means the index $i$ is in the string $I$ of indices
of fermions and we used the fact that $[\psi_{i},\mX_{I}]=0$ if $i\notin I$
and $\{\psi_{i},\mX_{I}\}=0$ if $i\in I$. We define a new notation
\begin{equation}
F_{\text{condition}}(\tau)\equiv\prod_{\text{condition}}\left(\cosh\f{\tau\mJ_{I}}{2^{q/2}}+2^{q/2}\mX_{I}\sinh\f{\tau\mJ_{I}}{2^{q/2}}\right)
\end{equation}
where the condition could be $i\in I$, $i,j\notin I$, etc. Each
term in the product commutes with others. $F$ has normalization $F_{c}(0)=\I$.
The product of two $F$'s with the same condition has an addition rule
\begin{equation}
F_{c}(\tau_{1})F_{c}(\tau_{2})=F_{c}(\tau_{1}+\tau_{2})\label{eq:9-1}
\end{equation}
The product of two $F$ with the same argument but different conditions
has a union rule
\begin{equation}
F_{c_{1}}(\tau)F_{c_{2}}(\tau)=F_{c_{1}\cup c_{2}}(\tau)\label{eq:10}
\end{equation}
For the product of all $I$, we just simply write $F(\tau)$. Then
(\ref{eq:9}) can be written as
\begin{equation}
F(\tau)\psi_{i}=\psi_{i}F_{i\notin I}(\tau)F_{i\in I}(-\tau)\label{eq:11}
\end{equation}
As we will see as follows, all computations of this model boil down
to repetitively using the rules (\ref{eq:9-1}), (\ref{eq:10}), (\ref{eq:11})
and their variants.

\subsection{Partition function and spectrum}

The partition function is
\begin{equation}
Z=\Tr e^{-\b H}=F(-\b)=\Tr\prod_{I}\left(\cosh\f{\b\mJ_{I}}{2^{q/2}}-2^{q/2}\mX_{I}\sinh\f{\b\mJ_{I}}{2^{q/2}}\right)
\end{equation}
Taking the ensemble average for the above expression and using
\begin{equation}
\f 1{\sqrt{2\pi\s^{2}}}\int d\mJ_{I}e^{-\mJ_{I}^{^{2}}/(2\s^{2})}\sinh a\mJ_{I}=0,\quad\f 1{\sqrt{2\pi\s^{2}}}\int d\mJ_{I}e^{-\mJ_{I}^{^{2}}/(2\s^{2})}\cosh a\mJ_{I}=e^{\f 12a^{2}\s^{2}}\label{eq:14}
\end{equation}
we have
\begin{equation}
\overline{Z}=\exp\left(\f c{2^{q+1}}\b^{2}\s^{2}\right)
\end{equation}
where $c=C_{N/2}^{q/2}=\f{(N/2)!}{(N/2-q/2)!(q/2)!}$ is the total
number of index choices of $I$. By (\ref{eq:4}), we have
\begin{equation}
\overline{Z}=\exp(N\b^{2}\mJ^{2}/(4q))\label{eq:14-1}
\end{equation}
Using the inverse Laplace transformation, we can derive the averaged
spectrum
\begin{equation}
\overline{Z}=\int dEe^{-\b E}\r(E)\implies\r(E)=\f 1{\mJ}\sqrt{\f q{\pi N}}\exp(-qE^{2}/(N\mJ^{2}))
\end{equation}
which is a Gaussian distribution \cite{derrida1980random}. This spectrum has no $\sqrt{E}$
edge behavior that is from the ordinary SYK model \cite{Garcia-Garcia:2016mno, Maldacena:2016hyu}. Since it does not have
a dense spectrum near the edge, we should expect that this model is
non-holographic. The easiest way to understand the Gaussian spectrum
is by noting that each term in the Hamiltonian is commuting and has
eigenvalue $\pm1/2^{q/2}\mJ_{I}$ because of (\ref{eq:5-1}). As we
assume that $\mJ_{I}$ obeys Gaussian distribution, the energy spectrum
should simply follow.

\subsection{Two-point function}

Consider the two-point function in finite temperature
\begin{equation}
G_{i}(\tau)=\f 1Z\Tr\left(e^{-\b H}\psi_{i}(\tau)\psi_{i}(0)\right),\quad\psi_{i}(\tau)=e^{\tau H}\psi_{i}e^{-\tau H}
\end{equation}
Using (\ref{eq:11}), it follows that
\begin{equation}
G_{i}(\tau)=\f 1{2Z}\Tr\left[F_{i\notin I}(-\b)F_{i\in I}(2\tau-\b)\right]
\end{equation}
For ensemble average, in this paper, we will take an approximation
by averaging the numerator and denominator independently (self-averaging), which gives
an error scales as $N^{1-q/2}$ by the highly narrow Gaussian distribution
(\ref{eq:4}). As we mentioned in Section \ref{sec2.1}, this model has a spin glass phase when $N\ra\infty$ and $T<T_c$, in which the ensemble average between replicas becomes crucial. Therefore, for large $N$ case the self-average approximation only holds for $T>T_c$. For finite and small $N$, there is no phase transition and we expect our self-average approximation should hold (at least qualitatively) as long as the temperature is not too low. Using (\ref{eq:14}) we have
\begin{align}
\overline{G_{i}(\tau)}\app &\f 1{2\overline{Z}}\overline{\Tr\left[\prod_{i\notin I}\left(\cosh\f{\b\mJ_{I}}{2^{q/2}}\right)\prod_{i\in I}\left(\cosh\f{(2\tau-\b)\mJ_{I}}{2^{q/2}}\right)\right]}\nn\\
=&\f 1{2\overline{Z}}\exp\left(\f{(c_{i\notin I}\b^{2}+c_{i\in I}(2\tau-\b)^{2})\s^{2}}{2^{q+1}}\right)\label{eq:16}
\end{align}
where $c_{i\in I}=C_{N/2-1}^{q/2-1}=\f{(N/2-1)!}{(q/2-1)!(N/2-q/2)!}$
is the number of $I$ that obeys $i\in I$. By (\ref{eq:14-1}), we
have 
\begin{equation}
\overline{G_{i}(\tau)}=\f 12\exp\left(-\f{c_{i\in I}\s^{2}\tau(\b-\tau)}{2^{q-1}}\right)\label{eq:17}
\end{equation}
which has correct normalization being $1/2$ at $\tau=0,\b$ and decays
away from $\tau=0,\b$ as expected. By (\ref{eq:4}), we have 
\begin{equation}
\overline{G_{i}(\tau)}=\f 12\exp\left(-\mJ^{2}\tau(\b-\tau)\right)\label{eq:21}
\end{equation}
Analytic continuation to Lorentzian time $\tau\ra it$, we have Gaussian
decay two-point function
\begin{equation}
\overline{G_{i}(t)}=\f 12\exp\left(-\mJ^{2}t^{2}-i\mJ^2\b t\right) \label{eq:thG}
\end{equation}

Clearly, this two-point function implies that this model is non-holographic
because it decays as a Gaussian tail with oscillation rather than
an exponential $e^{-\a t}$ with rate proportional to temperature
$1/\b$. The decay of the two-point function indicates the ``thermalization" process in which the excitation $\psi_i$ mixes with other degrees of freedom. Here we quote the term ``thermalization" to indicate that it is different from the ordinary thermalization in the sense of Eigenvalue Thermalization Hypothesis (ETH) (for a review see e.g. \cite{deutsch2018eigenstate}). Instead, throughout this paper we simply refer the ``thermalizaiton" as the decay of the two-point function in the commuting SYK model due to degrees of freedom mixing. Notably, the thermalization has an additional feature of oscillation, which reflects the integrability of the underlying model, in which the excitation $\psi_i$ has an effective (ensemble-averaged) characteristic frequency $\mJ^2\b$. This frequency can be understood as the energy change due to a $\psi_i$ excitation on the thermal state in leading order of $\b$, namely $\overline{\d E_\psi}\app\overline{\Tr (e^{-\b H}\psi_i H \psi_i)/Z}\app \b \mJ^2+O(\b^2)$. In a holographic model (e.g. full SYK model), there is no oscillation in two-point functions.

\subsection{Four-point function}

The next to consider is four-point function. Let us take (for $i\neq j$)
\begin{equation}
W_{ij}(\tau_{1},\tau_{2},\tau_{3},\tau_{4})=\f 1Z\Tr\left(e^{-\b H}\psi_{i}(\tau_{1})\psi_{j}(\tau_{2})\psi_{i}(\tau_{3})\psi_{j}(\tau_{4})\right),\quad\b>\tau_{1}>\tau_{2}>\tau_{3}>\tau_{4}>0
\end{equation}
Using (\ref{eq:11}), we can move two $\psi_{i}$'s and $\psi_{j}$'s
next to each other and annihilate them, which leads to 
\begin{align}
 W_{ij}= & \f 1Z\Tr\left(F(\tau_{14}-\b)\psi_{i}F(\tau_{21})\psi_{j}F(\tau_{32})\psi_{i}F(\tau_{43})\psi_{j}\right)\nonumber \\
= & -\f 1Z\Tr\left(F(\tau_{14}-\b)\psi_{i}\psi_{i}F_{i\notin I}(\tau_{21})F_{i\in I}(\tau_{12})\psi_{j}\psi_{j}F_{i\notin I,j\notin I}(\tau_{32})F_{i\notin I,j\in I}(\tau_{23})\right.\nonumber \\
 & \left.F_{i\in I,j\notin I}(\tau_{23})F_{i\in I,j\in I}(\tau_{32})F_{j\notin I}(\tau_{43})F_{j\in I}(\tau_{34})\right)\nonumber \\
= & -\f 1{4Z}\Tr\left(F_{i\notin I,j\notin I}(-\b)F_{i\notin I,j\in I}(2\tau_{24}-\b)F_{i\in I,j\notin I}(2\tau_{13}-\b)F_{i\in I,j\in I}(2(\tau_{12}+\tau_{34})-\b)\right)\label{eq:24}
\end{align}
where in the last step we used (\ref{eq:9-1}) and (\ref{eq:10})
to organize all $F$'s into four groups. For (\ref{eq:24}), we can
take the ensemble average independently, which basically replaces
each $F(\tau)$ with Gaussian function
\begin{equation}
\overline{W_{ij}}=-\f 1{4\overline{Z}}\exp\left(\f{(c_{i\notin I,j\in I}(2\tau_{24}-\b)^{2}+c_{i\in I,j\notin I}(2\tau_{13}-\b)^{2}+c_{i\in I,j\in I}(2(\tau_{12}+\tau_{34})-\b)^{2})\s^{2}}{2^{q+1}}\right)
\end{equation}
where $c_{\text{condition}}$ means the number of $I$ that obeys
the condition. 

There are two cases. For $\psi_{i},\psi_{j}$ not in one $X_{k}$
\begin{equation}
c_{i\notin I,j\in I}=c_{i\notin I,j\in I}=C_{N/2-2}^{q/2-1}\app\f{(N/2-2)^{q/2-1}}{(q/2-1)!},\quad c_{i\in I,j\in I}=C_{N/2-2}^{q/2-2}\app\f{(N/2-2)^{q/2-2}}{(q/2-2)!}
\end{equation}
and for $\psi_{i},\psi_{j}$ in one $X_{k}$ (e.g. $i=2k-1$ and $j=2k$)
\begin{equation}
c_{i\notin I,j\in I}=c_{i\notin I,j\in I}=0,\quad c_{i\in I,j\in I}=C_{N/2-1}^{q/2-1}\app\f{(N/2-1)^{q/2-1}}{(q/2-1)!}
\end{equation}
Here the approximation is under the large $N$ limit. It follows that
\begin{align}
\overline{W_{ij}} & =-\f 14\exp\left(-\s^{2}\f{c_{i\notin I,j\in I}\left(\tau_{24}(\b-\tau_{24})+\tau_{13}(\b-\tau_{13})\right)}{2^{q-1}}-\s^{2}\f{c_{i\in I,j\in I}(\tau_{12}+\tau_{34})(\b-\tau_{12}-\tau_{34})}{2^{q-1}}\right)\nonumber \\
 & \app-\f 14\begin{cases}
e^{-\mJ^{2}\left(\tau_{24}(\b-\tau_{24})+\tau_{13}(\b-\tau_{13})\right)-\f{(q-2)\mJ^{2}}N(\tau_{12}+\tau_{34})(\b-\tau_{12}-\tau_{34})} & \psi_{i}\in X_{k},\psi_{j}\in X_{k'}\\
e^{-\mJ^{2}(\tau_{12}+\tau_{34})(\b-\tau_{12}-\tau_{34})} & \psi_{i},\psi_{j}\in X_{k}
\end{cases} \label{eq:2.28}
\end{align}
where the first term in the first line is just the factorized two
point function in large $N$ limit and the second term is the $1/N$
correction for the non-factorized part. It is interesting that for the second case, there is no factorized piece but the non-factorized piece is enhanced by $N$. If we take $i=j$, following the same computation, the result is the same as the second line of \eqref{eq:2.28} with an additional minus sign.

Quantum chaos can be diagnosed by out-of-time-ordered correlators
(OTOC) near the scrambling time \cite{Shenker:2013pqa, kitaev2014hidden, Roberts:2014ifa}. For a quantum system with large $N$ degrees of freedom, a typical OTOC of interest
is 
\begin{equation}
W_{OTOC}(t)=\f 1Z\Tr\left(e^{-\b H}O_{1}(t)O_{2}(0)O_{1}(t)O_{2}(0)\right)\label{eq:29}
\end{equation}
whose leading behavior is
\begin{equation}
W_{OTOC}(t)\sim d_{1}-d_{2}f(t)/N\label{eq:30}
\end{equation}
with $f(t)=e^{\lambda t}$ for fast-scrambling systems \cite{Shenker:2013pqa, kitaev2014hidden, Roberts:2014ifa,Shenker:2013yza,Shenker:2014cwa,Maldacena:2016hyu} and
non-exponential $f(t)$ for other slow-scrambling systems.
Holographic systems should have fast-scrambling and in particular
in the strong coupling regime with semiclassical gravity dual should saturate
the chaos bound $\lambda=2\pi/\b$ \cite{Maldacena:2015waa} because it is related
to the boost symmetry near horizon \cite{Shenker:2013pqa}. SYK model is a well-known
example that has $\lambda=2\pi/\b$ in low temperature limit and $\lambda<2\pi/\b$
in finite temperature \cite{Maldacena:2016hyu}. For the commuting SYK model, we would
like to check the scrambling feature by OTOC. Taking $\tau_{1}=\tau_{3}=it$
and $\tau_{2}=\tau_{4}=0$, we have
\begin{equation}
W_{OTOC}(t)=-\f 14\begin{cases}
\exp\left(-\f {(q-2)}N\mJ^{2}(4t^{2}+2i\b t)\right)\app1-4(q-2)\mJ^{2}t^{2}/N & i\in X_{k},j\in X_{k'}\\
\exp\left(-\mJ^{2}(4t^{2}+2i\b t)\right)\app1-4\mJ^{2}t^{2} & i,j\in X_{k}
\end{cases}\label{eq:31}
\end{equation}
where we expand the exponent in leading order of $1/N$ and assume $t\gg \b$. Comparing with the general form (\ref{eq:30}), we find that this
model is slow-scrambling because $f(t)\sim t^{2}$ is quadratic. This slow scrambling behavior for $q=4$  was studied before, for example, in \cite{Swingle:2016jdj}. Due
to the quadratic growth, we can also identify the scrambling time
as $O(\sqrt{N})$ from (\ref{eq:31}). On the other hand, the imaginary part of the exponent in \eqref{eq:31} again has effective frequency of oscillation proportional to $\b\mJ^2$.

This difference to the ordinary SYK model can be qualitatively understood
as follows. The OTOC (\ref{eq:29}) essentially uses $O_{2}(0)$ to
probe how fast the operator $O_{1}(t)$ is scrambled along time. The
fast scrambling of ordinary SYK model is analogous to the pandemic
model, in which the all-to-all random non-integrable Hamiltonian couples
each fermion with any other fermions such that the operator size growth
rate of $\psi_{i}(t)$ is proportional to the size itself \cite{Roberts:2018mnp, Qi:2018bje}.
This leads to the exponential growth of OTOC because the probability
being probed by another fermion $\psi_{i}$ should be proportional
to the size of $\psi_{i}(t)$. On the other hand, though the commuting
SYK also has all-to-all random coupling in (\ref{eq:3}), the coupling
is much more sparse and in a form with integrability. Existence of
large number of conserved charges $X_{i}$ prevents many quantities
from scrambling. For example, $e^{iHt}X_{i}e^{-iHt}=X_{i}$ implies
that any product of $X_{i}$ does not scramble. As we will see in
Section \ref{subsec:Average-size-and}, the operator size growth in
commuting SYK model is quadratic and thus the growth of OTOC follows
the same rule. Another thing to mention is that the second case of
(\ref{eq:31}) has order $N$ enhancement to the first case because
when the source $\psi_{i}$ and probe $\psi_{j}$ are in the same
$X_{k}$, there are $C_{N/2-1}^{q/2-1}$ terms in the Hamiltonian
scrambling the source and is also detectable by the probe, but when the
source and probe are in two different $X_{k}$, there are only $C_{N/2-2}^{q/2-2}$
terms do the same job, which is $1/N$ smaller than the former case.
By above analysis of OTOC and scrambling features in the commuting
SYK model, we see again that this model is not holographic. 

\section{Commuting SYK has some holography-like features\label{sec:Commuting-SYK-has}}

In the last section, we have shown that the commuting SYK model is
not holographic by checking its spectrum, two-point, and four-point
functions. Nevertheless, this model is not as trivial as it looks
so far. In this section, we will show that the commuting SYK
model has some holography-like features, especially near-perfect size-winding
in high temperatures, which was thought as a significant property of
holographic systems \cite{Brown:2019hmk, Nezami:2021yaq}. 

By the essence of the ensemble average (\ref{eq:14}), we know that
the $\mJ$ dependence is only through $\b\mJ$ or $\tau\mJ$, where
$\tau$ is an Euclidean time variable. For a simpler notation, we
will rescale $\b$ and $\tau$ with a factor of $1/\mJ$, which is
equivalent to setting $\mJ=1$, in the rest of the paper. To recover
the $\mJ$ dependence, we just need to replace $\b\ra\b\mJ$ and $\tau\ra\tau\mJ$
in the following equations.

In order to check size distribution and size winding \cite{Brown:2019hmk, Nezami:2021yaq}, we
can consider the two-sided system. Define two commuting SYK models
labeled as left $l$ and right $r$
\begin{equation}
\{\psi_{i}^{a},\psi_{j}^{b}\}=\d^{ab}\d_{ij},\quad H_{a}=\sum_{I}\mJ_{I}\mX_{I}^{a},\quad a,b=l,r
\end{equation}
The size operator is 
\begin{equation}
S=N/2+\sum_{j}i\psi_{j}^{l}\psi_{j}^{r}=N/2+V
\end{equation}
which measures the size of a right size-basis of operators $\G_{I}^{r}=2^{s/2}i^{|I|(|I|-1)/2}\psi_{i_{1}}^{r}\cdots\psi_{i_{s}}^{r}=\G_{I}^{r\dagger}$
as $|I|=s$ by the expectation value \cite{Qi:2018bje}
\begin{equation}
\avg{0|\G_{I}^{r\dagger}S^{k}\G_{J}^{r}|0}=\avg{0|\G_{I}^{r}S^{k}\G_{J}^{r}|0}=s^{k}\d_{IJ}\label{eq:34-1}
\end{equation}
where the size-basis is normalized $(\G_{I}^{r})^{2}=1$ and $\bra 0$
is an EPR state defined by
\begin{equation}
\psi_{j}^{l}+i\psi_{j}^{r}\bra 0=0,\quad\avg{0|0}=1
\end{equation}
By this definition, the size of operator $\G_{I}^{r}$ means the length
$|I|$ of the string of indices $I$. By $q=4\Z$ and we have symmetry
\begin{equation}
H_{l}\bra 0=H_{r}\bra 0,\quad\psi_{j}^{l}(t)\bra 0=-i\psi_{j}^{r}(-t)\bra 0
\end{equation}

\subsection{Size distribution\label{subsec:Size-distribution}}

Introducing the left (auxiliary) system helps compute the size-related
quantities in the right system. We are interested in the size distribution
of $\psi_{j}^{r}(t)\r_{r}^{1/2}$ where $\r_{r}=\f 1Ze^{-\b H_{r}}$.
To define the size distribution, we first expand this right operator
in terms of the right size-basis
\begin{equation}
\psi_{j}^{r}(t)\r^{1/2}=\f 1{\sqrt{2}}\sum_{I}c_{I}(t)\G_{I}^{r}\label{eq:32}
\end{equation}
Note that this expansion is always possible because $\G_{I}^{r}$
are $2^{N}$ independent (and also orthogonal in the sense of (\ref{eq:34-1})
for $k=0$) operators spanning the full space of operators in the
right system. We define the size distribution of $\psi_{j}^{r}(t)\r_{r}^{1/2}$
as 
\begin{equation}
P_{n}(t)=\sum_{|I|=n}|c_{I}(t)|^{2}
\end{equation}
The distribution is unity normalized 
\begin{equation}
\sum_{n=0}^{N}P_{n}(t)=2\avg{0|\r_{r}^{1/2}\psi_{j}^{r}(t)\psi_{j}^{r}(t)\r_{r}^{1/2}|0}=1
\end{equation}
To compute the distribution, we can instead compute the generating
function \cite{Qi:2018bje}
\begin{equation}
K_{\mu}(t)=\avg{0|\r_{r}^{1/2}\psi_{j}^{r}(t)e^{-\mu S}\psi_{j}^{r}(t)\r_{r}^{1/2}|0}=\f 12\sum_{I}|c_{I}(t)|^{2}e^{-\mu|I|}\label{eq:33-3}
\end{equation}
The normalization of size distribution leads to $K_{0}(t)=1/2$. 

To compute this, we start with the Euclidean time correlation function
\begin{align}
k_{\mu}(\tau) & =\avg{0|\r_{r}^{1/2}\psi_{j}^{r}(\tau)e^{-\mu V}\psi_{j}^{r}(\tau)\r_{r}^{1/2}|0}\label{eq:33-2}
\end{align}
and analytically continue $\tau\ra it$ in the end. Let us first expand
$e^{-\mu V}$ as 
\begin{align}
e^{-\mu V} & =\prod_{j}e^{-\mu i\psi_{j}^{l}\psi_{j}^{r}}=\prod_{j}(\cosh\f{\mu}2-2i\psi_{j}^{l}\psi_{j}^{r}\sinh\f{\mu}2)\nonumber \\
 & =\sum_{I}(\cosh\f{\mu}2)^{N-|I|}(-i\sinh\f{\mu}2)^{|I|}\G_{I}^{l}\G_{I}^{r}\label{eq:35}
\end{align}
Taking this back to (\ref{eq:33-2}), we have
\begin{align}
k_{\mu}(\tau)= & \sum_{I}(\cosh\f{\mu}2)^{N-|I|}(-i\sinh\f{\mu}2)^{|I|}\avg{0|\r_{r}^{1/2}\psi_{j}^{r}(\tau)\G_{I}^{l}\G_{I}^{r}\psi_{j}^{r}(\tau)\r_{r}^{1/2}|0}\nonumber \\
= & \f 1Z\sum_{I}(\cosh\f{\mu}2)^{N-|I|}(-\sinh\f{\mu}2)^{|I|}\Tr_{r}\left(\G_{I}^{r}F(\tau-\b/2)\psi_{j}^{r}F(-\tau)\G_{I}^{r}F(\tau)\psi_{j}^{r}F(-\tau-\b/2)\right)\nonumber \\
= & \f 1{2Z}\sum_{I}(\cosh\f{\mu}2)^{N-|I|}(\sinh\f{\mu}2)^{|I|}(-)^{|j\cap I|}\nonumber \\
 & \times\Tr_{r}\left(\G_{I}^{r}F_{j\notin J}(-\b/2)F_{j\in J}(2\tau-\b/2)\G_{I}^{r}F_{j\notin J}(-\b/2)F_{j\in J}(-2\tau-\b/2)\right)\label{eq:43-2}
\end{align}
where in the second line we move $\G_{I}^{l}$ to the left to act
on $\ket 0$ and rewrite the expectation value as a trace in the right
system, and in the last line we move one $\psi_{j}^{r}$ across many
terms to annihilate the other $\psi_{j}^{r}$. The next step is to
move $\G_{I}^{r}$ across the two $F$'s and annihilate the other
$\G_{I}^{r}$. Use the notation $\Z_{\pm}$ referring to even/odd
integers respectively. We have the following property generalized
from (\ref{eq:11})
\begin{equation}
F_{c}(\tau)\G_{I}^{r}=\G_{I}^{r}F_{c,|I\cap J|\in\Z_{+}}(\tau)F_{c,|I\cap J|\in\Z_{-}}(-\tau),\quad c=j\in J\text{ or }j\notin J\label{eq:37}
\end{equation}
where $|I\cap J|$ means the number of overlapping fermionic indices
between $I$ and $J$. Moving $\G_{I}^{r}$ across two $F$'s in (\ref{eq:43-2})
leads to
\begin{equation}
k_{\mu}(\tau)=\f 1{2Z}\sum_{I}(\cosh\f{\mu}2)^{N-|I|}(\sinh\f{\mu}2)^{|I|}(-)^{|j\cap I|}\Tr\left(F_{|I\cap J|\in\Z_{+}}(-\b)F_{j\in J,|I\cap J|\in\Z_{-}}(-4\tau)\right)
\end{equation}
where we used (\ref{eq:9-1}) and (\ref{eq:10}) to combine a few
$F$'s together. After the ensemble average, we have 
\begin{align}
&\overline{k_{\mu}(\tau)}  =\f 1{2\overline{Z}}\sum_{I}(\cosh\f{\mu}2)^{N-|I|}(\sinh\f{\mu}2)^{|I|}(-)^{|j\cap I|}\exp\left(\f{c_{|I\cap J|\in\Z_{+}}\b^{2}+c_{j\in J,|I\cap J|\in\Z_{-}}(4\tau)^{2}}{2^{q+1}\s^{-2}}\right)\nonumber \\
 =& \f 12\sum_{I}(\cosh\f{\mu}2)^{N-|I|}(\sinh\f{\mu}2)^{|I|}(-)^{|j\cap I|}\exp\left(\f{-c_{|I\cap J|\in\Z_{-}}\b^{2}+c_{j\in J,|I\cap J|\in\Z_{-}}(4\tau)^{2}}{2^{q+1}\s^{-2}}\right)\label{eq:40-1-1}
\end{align}
where we used (\ref{eq:14-1}) and
\begin{equation}
c_{\text{all}}=c_{|I\cap J|\in\Z_{+}}+c_{|I\cap J|\in\Z_{-}}=C_{N/2}^{q/2}
\end{equation}

For simplicity, in the rest of this paper, we only consider $q=4$. This is the case for the original Sherrington-Kirkpatrick (SK) model. With our notation $\mJ=1$, the critical temperature is $T_c=1/\b_c=1$ in large $N$ limit \cite{sherrington1975solvable}.
Note that each $X_{i}=\psi_{2i-1}\psi_{2i}$ contains two successive
indices for fermions. Given an $I$, let us split it into two categories:
only one index of a $X_{i}$ is overlapped with $I$, or both indices
of a $X_{i}$ is overlapped with $I$. We count the number of the
former indices in $I$ as $i_{1}$ and others as $i_{2}.$ For example,
if $I=(1,2,5,9)$, then $(1,2)$ overlap with $X_{1}$, $(5)$ overlaps
with $X_{3}$ and $(9)$ overlaps with $X_{5}$, from which we count
$i_{1}=2$ and $i_{2}=1$. It is clear that $|I|=i_{1}+2i_{2}$. For
$q=4$, $\mX_{J}=X_{j_{1}}X_{j_{2}}$, and we only need to check two
$X$ in the counting of $c$ for all possible $J$. It does not matter
which $j$ we choose in (\ref{eq:40-1-1}), so we will take $j=1$.
Moreover, $|I\cap J|$ can only take values $0,1,2,3,4$. The counting
for $c$'s in (\ref{eq:40-1-1}) are as follows.
\begin{enumerate}
\item Compute $c_{j\in J,|I\cap J|\in\Z_{-}}$. For $1\in J$, say $j_{1}=1$.
There are two cases: a) 1 or $2\in I$, b) $1,2\notin I$ or $1,2\in I$.
For a) $|I\cap J|$ counts from 1, and $j_{2}$ needs to be chosen
to avoid $i_{1}-1$ choices from total $N/2-1$ choices, which gives
$|I\cap J|=1,3$ and $c_{j\in J,|I\cap J|\in\Z_{-}}=N/2-i_{1}$; For
b) $|I\cap J|$ counts from 0 or 2, and $j_{2}$ are those covering
$i_{1}$ choices, which gives $|I\cap J|=1,3$ and $c_{j\in J,|I\cap J|\in\Z_{-}}=i_{1}$. 
\item Compute $c_{j\notin J,|I\cap J|\in\Z_{-}}$. There are again the two
cases: a) 1 or $2\in I$, b) $1,2\notin I$, or $1,2\in I$. For a)
$j_{1}$ needs to be chosen from $i_{1}-1$ choices and$j_{2}$ can
be any other $N/2-i_{1}$ choices that lead to $|I\cap J|=1,3$ and
$c_{j\notin J,|I\cap J|\in\Z_{-}}=(i_{1}-1)(N/2-i_{1})$; For b) $j_{1}$
needs to be chosen from $i_{1}$ choices and $j_{2}$ can be any other
$N/2-1-i_{1}$ choices that lead to $|I\cap J|=1,3$ and $c_{j\notin J,|I\cap J|\in\Z_{-}}=i_{1}(N/2-i_{1}-1)$.
\end{enumerate}
The number $c_{|I\cap J|\in\Z_{-}}$ in (\ref{eq:40-1-1}) is the
sum over the above two numbers. 

There is a $(-)^{|j\cap I|}$ factor that one needs to be careful
with. Given $i_{1}$ and $i_{2}$, depending on $|j\cap I|=0,1$,
the total numbers of possible $I$ are different but they have opposite
sign. The counting is as follows
\begin{enumerate}
\item $|j\cap I|=0$, namely $1\notin I$. For case a), $2\in I$, the total
number of $I$ is $2^{i_{1}-1}C_{N/2-1}^{i_{1}-1}C_{N/2-i_{1}}^{i_{2}}$;
For case b), $2\notin I$, the total number of $I$ is $2^{i_{1}}C_{N/2-1}^{i_{1}}C_{N/2-i_{1}-1}^{i_{2}}$.
\item $|j\cap I|=1$, namely $1\in I$. For case a), $2\notin I$, the total
number of $I$ is $2^{i_{1}-1}C_{N/2-1}^{i_{1}-1}C_{N/2-i_{1}}^{i_{2}}$;
For case b), $2\in I$, the total number of $I$ is $2^{i_{1}}C_{N/2-1}^{i_{1}}C_{N/2-i_{1}-1}^{i_{2}-1}$.
\end{enumerate}
It is noteworthy in (\ref{eq:40-1-1}) that for case a) the exponential
piece is the same and the total number of $I$ in both $|j\cap I|=0$
and $|j\cap I|=1$ are also the same. The factor $(-)^{|j\cap I|}$
leads to complete cancellation between these two parts. For case b),
the cancellation is partial because 
\begin{equation}
2^{i_{1}}C_{N/2-1}^{i_{1}}C_{N/2-i_{1}-1}^{i_{2}}-2^{i_{1}}C_{N/2-1}^{i_{1}}C_{N/2-i_{1}-1}^{i_{2}-1}=\f{2^{i_{1}}(N/2-1)!(N/2-i_{1}-2i_{2})}{i_{1}!i_{2}!(N/2-i_{1}-i_{2})!}
\end{equation}

Taking the above analysis together, it follows that
\begin{align}
 & \f{\overline{k_{\mu}(\tau)}}{(N/2-1)!}=  \f 12\sum_{i_{1},i_{2}}(\cosh\f{\mu}2)^{N-i_{1}-2i_{2}}(\sinh\f{\mu}2)^{i_{1}+2i_{2}}\f{2^{i_{1}}(N/2-i_{1}-2i_{2})}{i_{1}!i_{2}!(N/2-i_{1}-i_{2})!}e^{\f{-i_{1}(N/2-i_{1})\b^{2}+i_{1}(4\tau)^{2}}{2(N-2)}}\nonumber \\
= & \f 12\sum_{i_{1}=0}^{N/2}\f{2^{i_{1}}(\cosh\f{\mu}2)^{N-i_{1}}(\sinh\f{\mu}2)^{i_{1}}}{i_{1}!}e^{\f{-i_{1}(N/2-i_{1})\b^{2}+i_{1}(4\tau)^{2}}{2(N-2)}}\sum_{i_{2}=0}^{N/2-i_{1}}(\tanh\f{\mu}2)^{2i_{2}}\f{N/2-i_{1}-2i_{2}}{i_{2}!(N/2-i_{1}-i_{2})!}\nonumber \\
= & \f 12\sum_{i_{1}=0}^{N/2-1}\f{(\cosh\mu)^{N/2-i_{1}-1}(\sinh\mu)^{i_{1}}}{i_{1}!(N/2-1-i_{1})!}e^{\f{-i_{1}(N/2-i_{1})\b^{2}+i_{1}(4\tau)^{2}}{2(N-2)}}\label{eq:43-1}
\end{align}
where we have used $\s^{-2}=(N/2-1)/(2^{3})$ for $q=4$ and $\mJ=1$.
From (\ref{eq:43-1}), we can confirm the normalization of the size distribution
\begin{equation}
K_{0}=k_{0}=\f 12
\end{equation}
where only $i_{1}=0$ term survives. Continuing $\tau\ra it$, we
have the generating function $K_{\mu}(t)$ as
\begin{align}
 & K_{\mu}(t)=e^{-\mu N/2}\overline{k_{\mu}(it)}\nonumber \\
= & \sum_{n=0}^{N/2-1}e^{-\mu(2n+1)}\sum_{i_{1}=0}^{N/2-1}\sum_{k=\max\{0,n+i_{1}+1-N/2\}}^{\min\{n,i_{1}\}}\f{(N/2-1)!C_{N/2-i_{1}-1}^{n-k}C_{i_{1}}^{k}(-)^{k}}{2^{N/2}i_{1}!(N/2-1-i_{1})!}e^{\f{-i_{1}(N/2-i_{1})\b^{2}-i_{1}(4t)^{2}}{2(N-2)}}
\end{align}
Comparing with (\ref{eq:33-3}) yields the size distribution
\begin{equation}
\overline{P_{2n+1}(t)}=\f 1{2^{N/2-1}}\sum_{i_{1}=0}^{N/2-1}\sum_{k=\max\{0,n+i_{1}+1-N/2\}}^{\min\{n,i_{1}\}}\f{(N/2-1)!(-)^{k}\exp\left(\f{-i_{1}(N/2-i_{1})\b^{2}-i_{1}(4t)^{2}}{2(N-2)}\right)}{(i_{1}-k)!k!(N/2-1-i_{1}-n+k)!(n-k)!}\label{eq:45-1}
\end{equation}

\subsection{Size-winding\label{subsec:Size-winding}}

While size distribution computes the magnitude of the coefficients
in the expansion (\ref{eq:32}), we are also interested in their phases.
For holographic systems, it has been argued in \cite{Brown:2019hmk, Nezami:2021yaq} that the
phase of $c_{I}(t)$ is linear in their size, namely
\begin{equation}
c_{I}(t)\equiv r_{I}(t)e^{i\phi_{I}(t)},\quad\phi_{I}(t)=a_{1}+a_{2}|I|,\quad r_{I}(t),\phi_{I}(t)\in\R
\end{equation}
This nontrivial feature is called size-winding and it is the microscopic mechanism for the traversable wormhole teleportation protocol. 

To check the size-winding, we need to compute a different generating
function
\begin{align}
G_{\mu}(t) & =-i\avg{0|\r_{l}^{1/2}\psi_{j}^{l}(-t)e^{-\mu S}\psi_{j}^{r}(t)\r_{r}^{1/2}|0}=\avg{0|\psi_{j}^{r}(t)\r_{r}^{1/2}e^{-\mu S}\psi_{j}^{r}(t)\r_{r}^{1/2}|0}\nonumber \\
 & =\f 12\sum_{I}c_{I}(t)^{2}\avg{0|\G_{I}^{r}e^{-\mu S}\G_{I}^{r}|0}=\f 12\sum_{I}c_{I}(t)^{2}e^{-\mu|I|}\label{eq:33-1}
\end{align}
and define
\begin{equation}
Q_{n}(t)=\sum_{|I|=n}c_{I}(t)^{2}\label{eq:55}
\end{equation}
By definition $|Q_{n}(t)|\leq P_{n}(t)$. Note that this generating function can only probe the averaged phase of coefficients with same
size rather than the phase of each individual coefficient. Nevertheless,
it is still a good measure for size winding, which can be characterized
by the following two properties of $P_{n}$ and $Q_{n}$ \cite{Jafferis:2022crx, Kobrin:2023rzr}
\begin{enumerate}
\item The phase of $Q_{n}(t)$ is a linear function of $n$
\item The ratio between $r_{n}=|Q_{n}(t)|/P_{n}(t)$ should be close to
one.
\end{enumerate}
The first property states that the averaged phase for the basis with
the same size is proportional to the size, and the second property states
that the phases of these bases with the same size are also aligned (otherwise
their cancellation leads to $r_{n}$ less than one).

The argument for the size-winding of holographic systems by \cite{Brown:2019hmk, Nezami:2021yaq}
is briefly reviewed as follows. For holographic systems, there is
a bulk computation for $G_{\mu}(t)$ by regarding it as a bulk scattering
process between the null shockwaves generated by $\psi_{j}^{l}(-t)\psi_{j}^{r}(t)$
and $V=i\sum_{j}\psi_{j}^{l}\psi_{j}^{r}$ at $t=0$ respectively
in the near-AdS$_{2}$ background \cite{Maldacena:2017axo}. It has been
shown that the ground state $\bra G$ of $\hat{E}=H_{l}+H_{r}+\mu V$
is dual to an eternal traversable wormhole \cite{Maldacena:2018lmt}. The ground state
$\bra G$ has order one overlap with the thermofield double state
$\bra{TFD}=\r_{r}^{1/2}\bra 0$ that has a boost symmetry $\hat{B}\bra{TFD}\equiv H_{r}-H_{l}\bra{TFD}=0$.
Therefore, we can regard $\hat{E}=H_{l}+H_{r}+\mu V-E_{0}$ as an
extra approximate symmetry of $\bra{TFD}$ with $\mu,E_{0}$ chosen
such that $\hat{E}\bra{TFD}\app0$. The bulk dual of thermofield double
state is near-AdS$_{2}$ that has an isometry of $SL(2)$. It was
shown in \cite{Lin:2019qwu} that $\hat{E}$ generates the global time translation
in near-AdS$_{2}$ spacetime, and $\hat{E}$, $\hat{B}$ and $[\hat{E},\hat{B}]$
indeed form the three generators of the $SL(2)$ isometry of near-AdS$_{2}$
spacetime. 

The translation generators along the two null directions of near-AdS$_{2}$
are linear combinations of $\hat{E}$ and $\hat{B}$: $\hat{P}_{\pm}=-\f 12(\hat{E}\pm\hat{B})$.
In particular, $-\hat{P}_{+}=H_{r}+\mu V/2+E_{0}/2$, where the $H_{r}$
can be replaced by its expectation value in thermofield double state
if we consider $t\gg\b$ \cite{Nezami:2021yaq}. Therefore, the size operator
$S$ is indeed approximately dual to the null momentum $\hat{P}_{+}$
up to a constant \cite{Maldacena:2017axo,Lin:2019qwu,Maldacena:2018lmt}. It follows that we can expand
$G_{\mu}(t)$ in the null momentum basis $\bra p$
\begin{equation}
G_{\mu}(t)\propto\int dpe^{-\mu p}\avg{TFD|\psi_{j}^{l}(-t)|p}\avg{p|\psi_{j}^{r}(t)|TFD}
\end{equation}
Comparing with (\ref{eq:33-1}), we can identify that 
\begin{equation}
Q_{p}(t)=\avg{TFD|\psi_{j}^{l}(-t)|p}\avg{p|\psi_{j}^{r}(t)|TFD}
\end{equation}
where the RHS is the wavefunction in AdS$_{2}$ and can be computed
directly using the $SL(2)$ isometry \cite{Maldacena:2016upp}. Similarly, there is
a bulk computation for $K_{\mu}(t)$ by understanding $\psi_{j}^{r}(t)$
as a left operator $\psi_{j}^{l}(i\b/2-t)$ acting on $\bra{TFD}$.
It turns out that the $SL(2)$ isometry leads to perfect size winding
\cite{Maldacena:2017axo,Lin:2019qwu,Nezami:2021yaq}
\begin{equation}
Q_{p}(t)=P_{p}(t)e^{ia_{1}+ia_{2}p},\quad a_2\propto e^{-2\pi t/\b}
\end{equation}

For the non-holographic commuting SYK model, it does not have a bulk
alternative to guarantee the size-winding. Therefore, it would more
appealing to understand how that could happen in some parameter regimes.
To compute $G_{\mu}(t)$, we again start with the Euclidean version
\begin{align}
g_{\mu}(\tau) & =-i\avg{0|\r_{l}^{1/2}\psi_{j}^{l}(-\tau)e^{-\mu V}\psi_{j}^{r}(\tau)\r_{r}^{1/2}|0}=\avg{0|\psi_{j}^{r}(\tau)\r_{r}^{1/2}e^{-\mu V}\psi_{j}^{r}(\tau)\r_{r}^{1/2}|0}\label{eq:33}
\end{align}
Taking (\ref{eq:35}) into (\ref{eq:33}) and following a similar
computation as (\ref{eq:43-2}) to annihilate two $\psi_{j}^{r}$
yields
\begin{align}
g_{\mu}(\tau)= & \f 1{2Z}\sum_{I}(\cosh\f{\mu}2)^{N-|I|}(\sinh\f{\mu}2)^{|I|}(-)^{|j\cap I|}\nonumber \\
 & \times\Tr_{r}\left(\G_{I}^{r}F_{j\notin J}(-\b/2)F_{j\in J}(2\tau+\b/2)\G_{I}^{r}F_{j\notin J}(-\b/2)F_{j\in J}(-2\tau-\b/2)\right)
\end{align}
Using (\ref{eq:37}), $(\G_{I}^{r})^{2}=1$ and $F_{c}(0)=1$, it
follows that
\begin{equation}
g_{\mu}(\tau)=\f 1{2Z}\sum_{I}(\cosh\f{\mu}2)^{N-|I|}(\sinh\f{\mu}2)^{|I|}(-)^{|j\cap I|}\Tr_{r}\left(F_{j\notin J,|I\cap J|\in\Z_{+}}(-\b)F_{j\in J,|I\cap J|\in\Z_{-}}(-4\tau-\b)\right)
\end{equation}
Let us take the ensemble average, which leads to
\begin{align}
\overline{g_{\mu}(\tau)} & =\f 1{2\overline{Z}}\sum_{I}(\cosh\f{\mu}2)^{N-|I|}(\sinh\f{\mu}2)^{|I|}(-)^{|j\cap I|}e^{\f{c_{j\notin J,|I\cap J|\in\Z_{+}}\b^{2}+c_{j\in J,|I\cap J|\in\Z_{-}}(4\tau+\b)^{2}}{2^{q+1}\s^{-2}}}\nonumber \\
 & =\f{e^{-\b^{2}/4}}2\sum_{I}(\cosh\f{\mu}2)^{N-|I|}(\sinh\f{\mu}2)^{|I|}(-)^{|j\cap I|}e^{\f{-c_{j\notin J,|I\cap J|\in\Z_{-}}\b^{2}+c_{j\in J,|I\cap J|\in\Z_{-}}(4\tau+\b)^{2}}{2^{q+1}\s^{-2}}}\label{eq:40-1}
\end{align}
where in the last step we used the fact
\begin{equation}
c_{j\notin J}=c_{j\notin J,|I\cap J|\in\Z_{+}}+c_{j\notin J,|I\cap J|\in\Z_{-}}=C_{N/2-1}^{q/2}
\end{equation}

Take $q=4$. The counting of coefficients are exactly the same as
Section \ref{subsec:Size-distribution}. By the cancellation due to
$(-)^{|j\cap I|}$ we have
\begin{align}
 & \f{\overline{g_{\mu}(\tau)}}{(N/2-1)!}\nonumber \\
= & \f{e^{-\b^{2}/4}}2\sum_{i_{1},i_{2}}(\cosh\f{\mu}2)^{N-i_{1}-2i_{2}}(\sinh\f{\mu}2)^{i_{1}+2i_{2}}\f{2^{i_{1}}(N/2-i_{1}-2i_{2})}{i_{1}!i_{2}!(N/2-i_{1}-i_{2})!}e^{\f{-i_{1}(N/2-i_{1}-1)\b^{2}+i_{1}(\b+4\tau)^{2}}{2(N-2)}}\nonumber \\
= & \f{e^{-\b^{2}/4}}2\sum_{i_{1}=0}^{N/2}\f{2^{i_{1}}(\cosh\f{\mu}2)^{N-i_{1}}(\sinh\f{\mu}2)^{i_{1}}}{i_{1}!}e^{\f{-i_{1}(N/2-i_{1}-1)\b^{2}+i_{1}(\b+4\tau)^{2}}{2(N-2)}}\sum_{i_{2}=0}^{N/2-i_{1}}(\tanh\f{\mu}2)^{2i_{2}}\f{N/2-i_{1}-2i_{2}}{i_{2}!(N/2-i_{1}-i_{2})!}\nonumber \\
= & \f{e^{-\b^{2}/4}}2\sum_{i_{1}=0}^{N/2-1}\f{(\cosh\mu)^{N/2-i_{1}-1}(\sinh\mu)^{i_{1}}}{i_{1}!(N/2-1-i_{1})!}e^{\f{-i_{1}(N/2-i_{1}-1)\b^{2}+i_{1}(\b+4\tau)^{2}}{2(N-2)}}\label{eq:43}
\end{align}
Multiplying $e^{-\mu N/2}$ and continuing $\tau\ra it$, we can expand
$G_{\mu}(t)$ in power series of $e^{-\mu}$
\begin{align}
 & G_{\mu}(t)=e^{-\mu N/2}\overline{g_{\mu}(it)}\nonumber \\
= & \sum_{n=0}^{N/2-1}e^{-\mu(2n+1)}\sum_{i_{1}=0}^{N/2-1}\sum_{k=\max\{0,n+i_{1}+1-N/2\}}^{\min\{n,i_{1}\}}\f{(N/2-1)!C_{N/2-i_{1}-1}^{n-k}C_{i_{1}}^{k}(-)^{k}}{e^{\b^{2}/4}2^{N/2}i_{1}!(N/2-1-i_{1})!}e^{\f{-i_{1}(N/2-i_{1}-1)\b^{2}+i_{1}(\b+4it)^{2}}{2(N-2)}}
\end{align}
Comparing with (\ref{eq:33-1}), we have
\begin{equation}
\overline{Q_{2n+1}(t)}=\f{e^{-\b^{2}/4}}{2^{N/2-1}}\sum_{i_{1}=0}^{N/2-1}\sum_{k=\max\{0,n+i_{1}+1-N/2\}}^{\min\{n,i_{1}\}}\f{(N/2-1)!(-)^{k}\exp\left(\f{-i_{1}(N/2-i_{1}-1)\b^{2}+i_{1}(\b+4it)^{2}}{2(N-2)}\right)}{(i_{1}-k)!k!(N/2-1-i_{1}-n+k)!(n-k)!}\label{eq:45}
\end{equation}

\subsection{Saddle approximation for size distribution and size winding\label{subsec:Saddle-approximation-for}}

Given the exact formula for $\overline{P_{2n+1}(t)}$ and $\overline{Q_{2n+1}(t)}$
by (\ref{eq:45-1}) and (\ref{eq:45}), it is not obvious to check
if the size winding is satisfied or not. However, in large $N$ limit,
we can do a saddle approximation and have better analytic control
of them. Let us first rewrite the sum in (\ref{eq:43-1}) and (\ref{eq:43})
in terms of an integral by the trick
\begin{equation}
e^{au^{2}+bu}=\f 1{\sqrt{a\pi}}\int dxe^{-\f 1ax^{2}+(2x+b)u}\label{eq:trick}
\end{equation}
It follows for the size distribution that 
\begin{align}
e^{-\mu N/2}\overline{k_{\mu}(\tau)} & =\f{e^{-\mu N/2}\G(N/2)}{2\sqrt{a'\pi}}\int dx\sum_{i_{1}=0}^{N/2-1}\f{(\cosh\mu)^{N/2-i_{1}-1}(\sinh\mu)^{i_{1}}}{i_{1}!(N/2-1-i_{1})!}e^{-\f 1{a'}x^{2}+(2x+b')i_{1}}\nonumber \\
 & =\f{e^{-\mu N/2}}{2\sqrt{a'\pi}}\int dxe^{-\f 1{a'}x^{2}}\left(\cosh\mu+\sinh\mu e^{2x+b'}\right)^{N/2-1}\nonumber \\
 & =\f{e^{-\mu}}{2^{N/2}\sqrt{a'\pi}}\int dxe^{-\f 1{a'}x^{2}}\left((1+e^{-2\mu})+(1-e^{-2\mu})e^{2x+b'}\right)^{N/2-1}\label{eq:109}
\end{align}
where 
\begin{equation}
a'=\f{\b^{2}}{4(N/2-1)},\quad b'=\f{16\tau^{2}-\b^{2}}{4(N/2-1)}-\b^{2}/4
\end{equation}
We can evaluate the integral in (\ref{eq:109}) using saddle approximation
as 
\begin{equation}
\f 1{\sqrt{2a'\pi}}\int dxe^{-\f 1{a'}F(x)}\app\f 1{\sqrt{F''(x_{0})}}e^{-\f 1{a'}F(x_{0})}\label{eq:70-1}
\end{equation}
where 
\begin{equation}
F(x)=x^{2}-\f{\b^{2}}4\log\left((1+e^{-2\mu})+(1-e^{-2\mu})e^{2x+b'}\right)\label{eq:68}
\end{equation}
The saddle equation is 
\begin{equation}
x=\f{\b^{2}}4\f{(1-e^{-2\mu})e^{2x+b'}}{(1+e^{-2\mu})+(1-e^{-2\mu})e^{2x+b'}}\label{eq:72}
\end{equation}
Since we are looking for a size distribution that could be as large
as $O(N)$, we need to set $\mu\sim O(1/N)$. For such a small $\mu$,
the solution of (\ref{eq:72}) is very close to zero and of order
$\sim(1-e^{-2\mu})$, which gives an approximate solution 
\begin{align}
x_{0} & \app\f{\b^{2}}4\f{(1-e^{-2\mu})e^{b'}}{(1+e^{-2\mu})+(1-e^{-2\mu})e^{b'}}\label{eq:x0}
\end{align}
Taking this back to (\ref{eq:68}), we have
\begin{align}
F(x_{0}) & =x_{0}^{2}-\f{\b^{2}}4\log\left((1+e^{-2\mu})+(1-e^{-2\mu})e^{2x_{0}}e^{b'}\right)\\
F''(x_{0}) & =2-\f{\b^{2}e^{2x_{0}+b'}(1-e^{-4\mu})}{(e^{2x_{0}}(1-e^{-2\mu})+e^{b'}(1+e^{-2\mu}))^{2}}
\end{align}
Taking these two equations into (\ref{eq:109}) and using (\ref{eq:70-1}),
we can expand it in powers series of $e^{-2\mu}$. This power series
need to be truncated at $N/2-1$ order because $N-1$ is the maximal
size of a fermionic operator. This expansion is complicated but in
high temperature $\b\ll1$ we will have a good simplification. In
this case $x_{0}$ is further suppressed by $\b^{2}e^{-\b^{2}/4}\ll1$
and we can approximate $x_{0}\app0$ in (\ref{eq:70-1}). It follows
that
\begin{align}
e^{-\mu N/2}\overline{k_{\mu}(\tau)} & \app\f{e^{-\mu}}{2^{N/2}}\left((1+e^{-2\mu})+(1-e^{-2\mu})e^{b'}\right)^{N/2-1}\nonumber \\
 & =\sum_{n=0}^{N/2-1}\f{e^{-(2n+1)\mu}}{2^{N/2}}C_{N/2-1}^{n}(1+e^{b'})^{N/2-1-n}(1-e^{b'})^{n}
\end{align}
which leads to
\begin{equation}
\overline{P_{2n+1}(t)}=\f 1{2^{N/2-1}}C_{N/2-1}^{n}(1+e^{-b_{p}(t)})^{N/2-1-n}(1-e^{-b_{p}(t)})^{n},\quad b_{p}(t)=\f{16t^{2}+\b^{2}}{4(N/2-1)}+\b^{2}/4\label{eq:117}
\end{equation}

\begin{figure}
\begin{centering}
\subfloat[$\protect\b=1/3,t=-1$\label{fig:1a}]{\begin{centering}
\includegraphics[height=2.3cm]{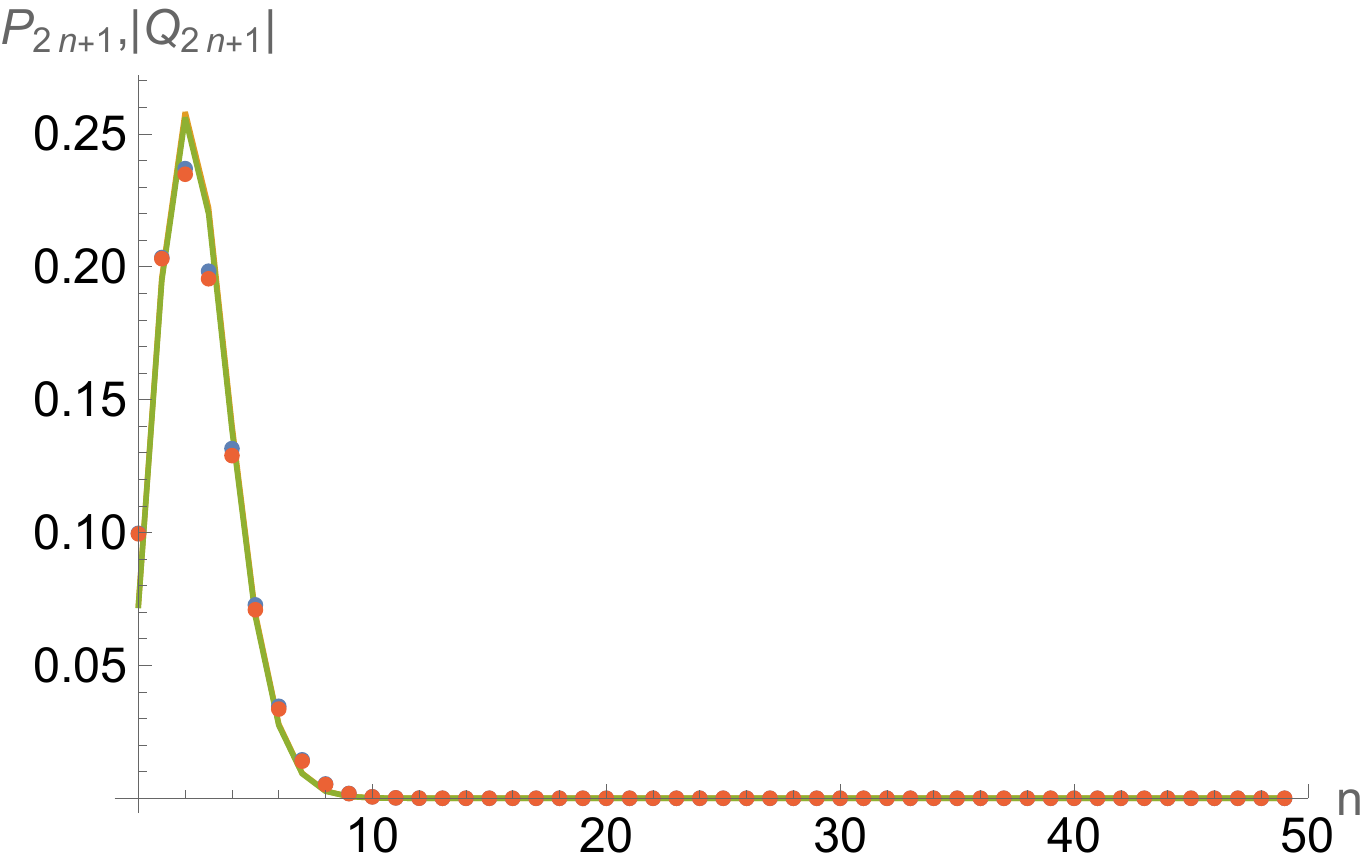}\includegraphics[height=2.3cm]{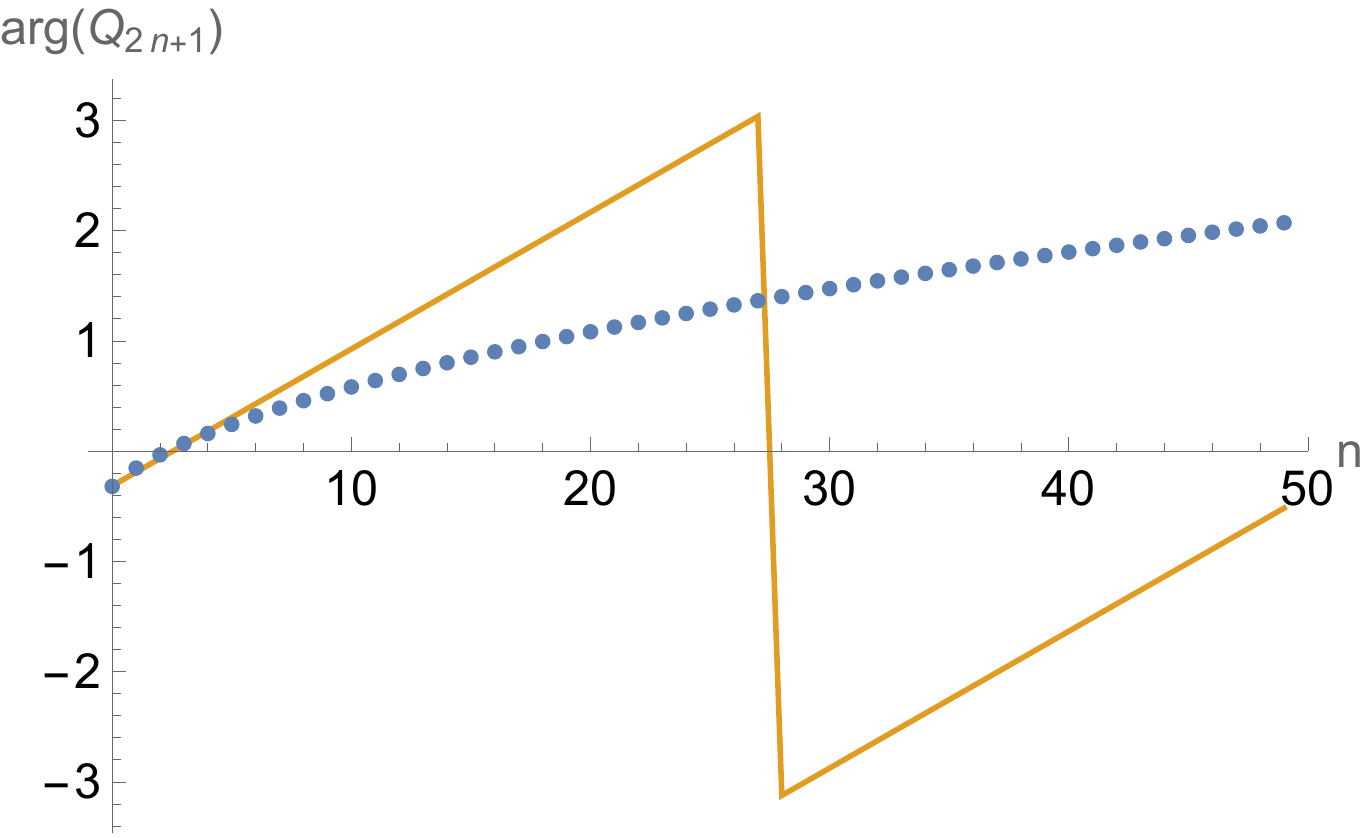}
\par\end{centering}
}\subfloat[$\protect\b=0.95,t=-1$]{\begin{centering}
\includegraphics[height=2.3cm]{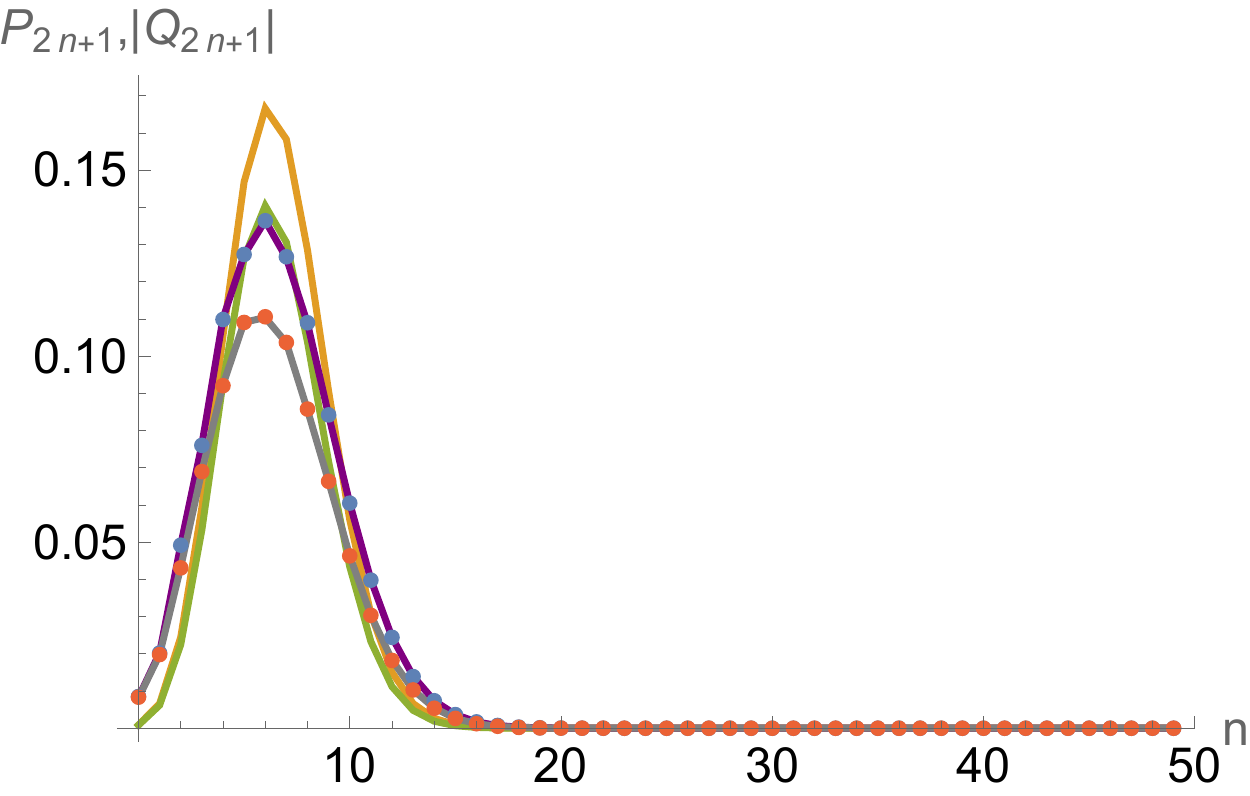}\includegraphics[height=2.3cm]{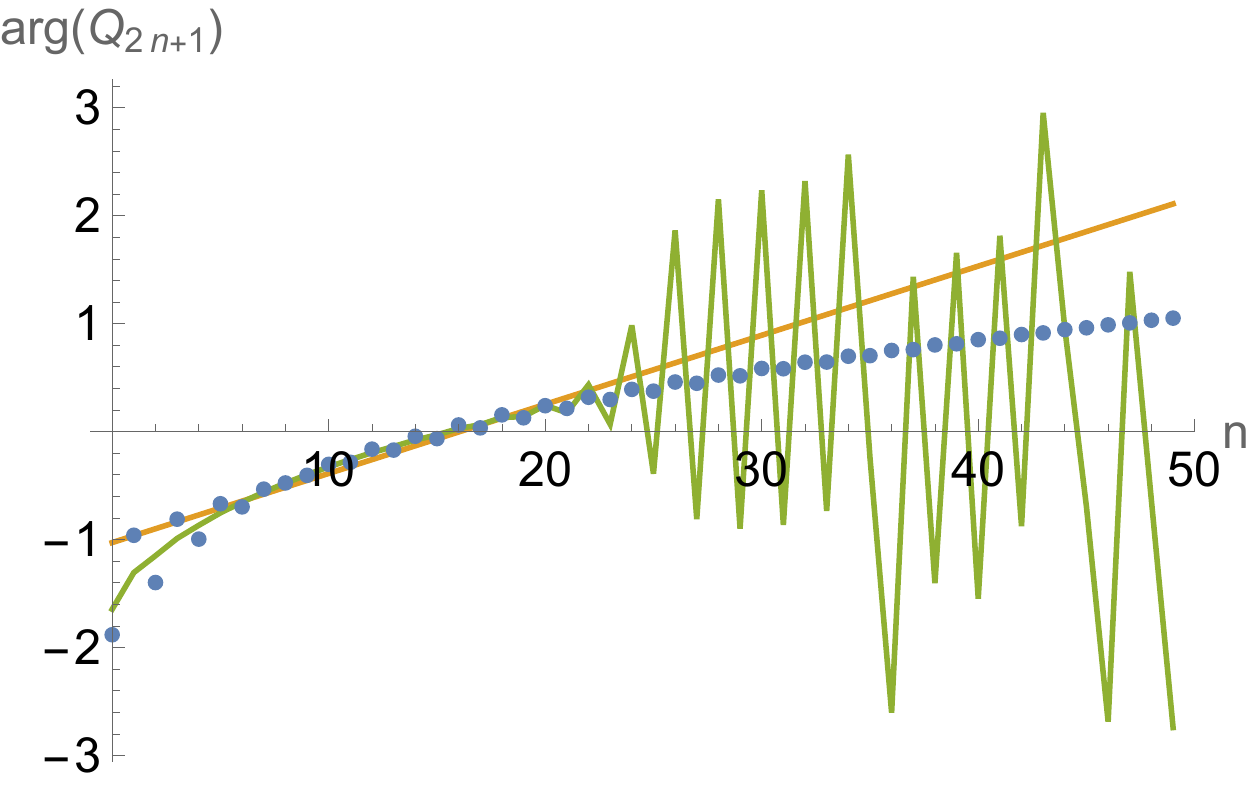}
\par\end{centering}
}\\
\subfloat[$\protect\b=1/3,t=-3$]{\begin{centering}
\includegraphics[height=2.3cm]{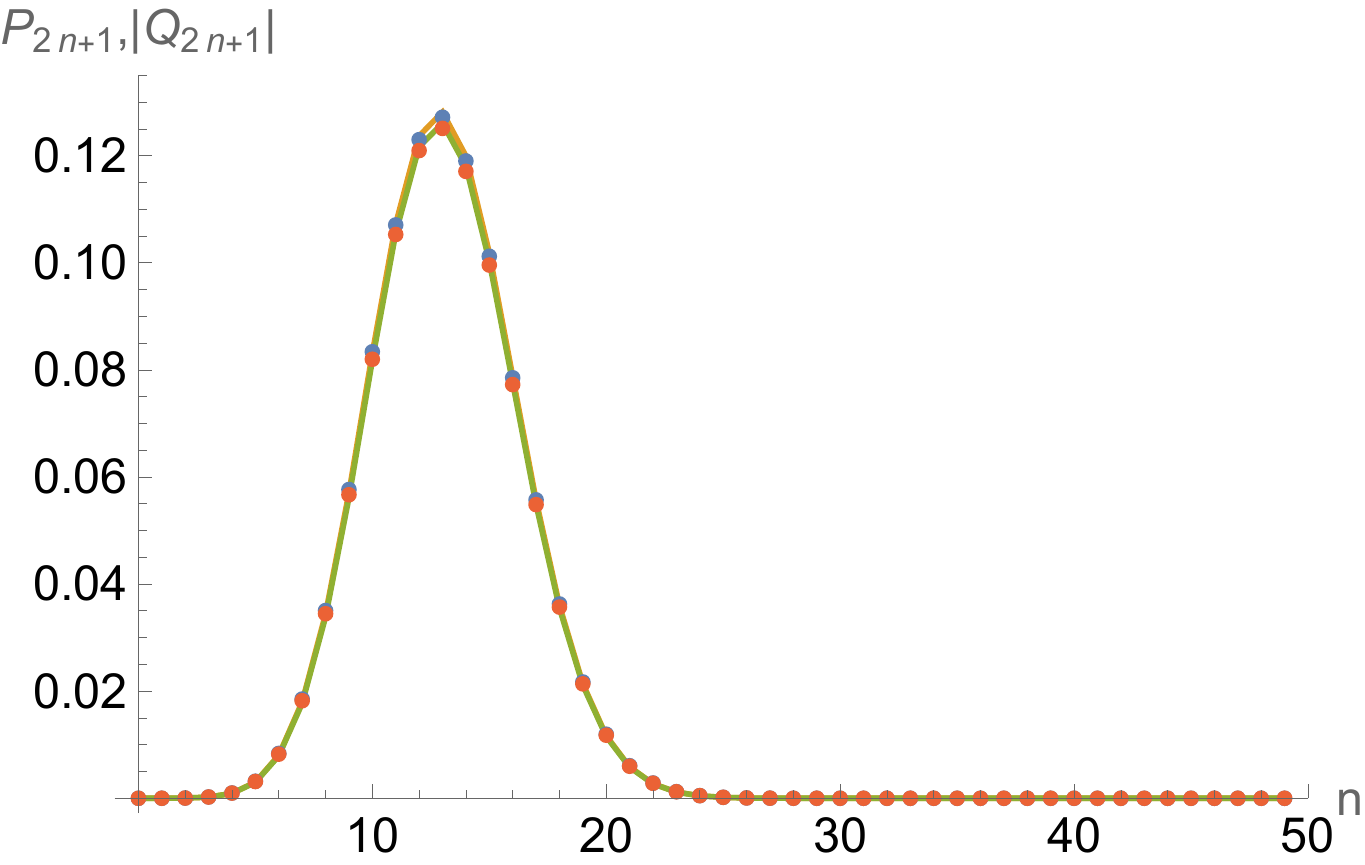}\includegraphics[height=2.3cm]{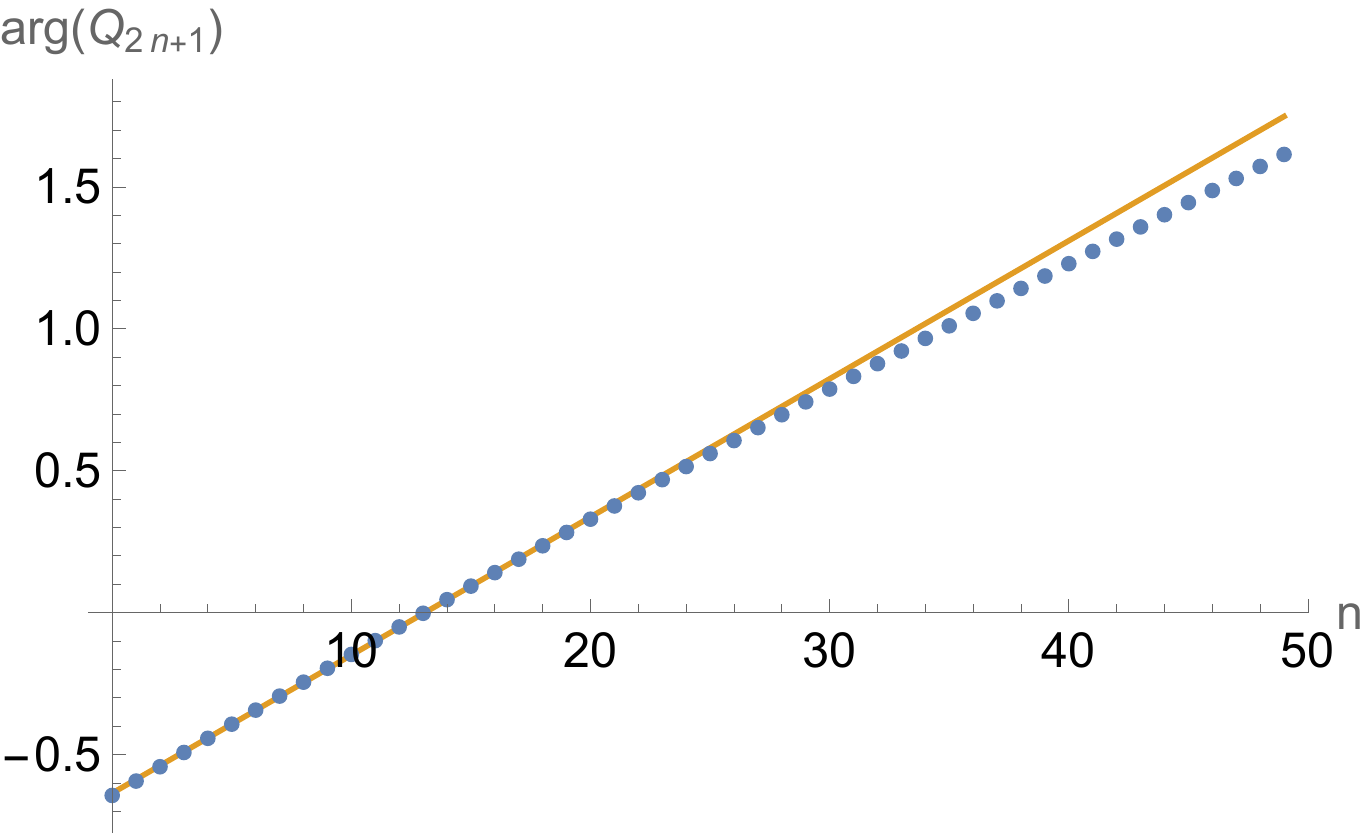}
\par\end{centering}
}\subfloat[$\protect\b=0.95,t=-3$]{\begin{centering}
\includegraphics[height=2.3cm]{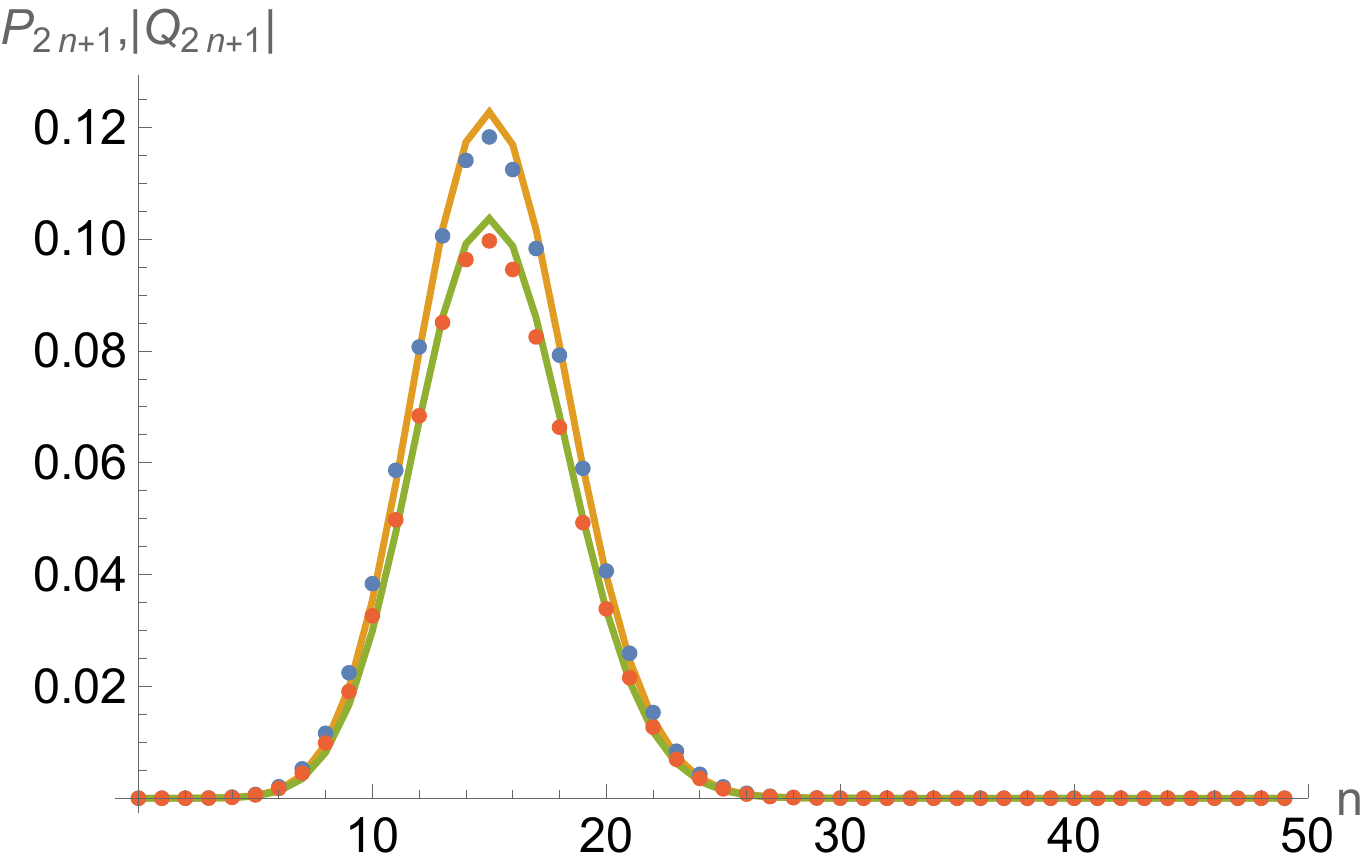}\includegraphics[height=2.3cm]{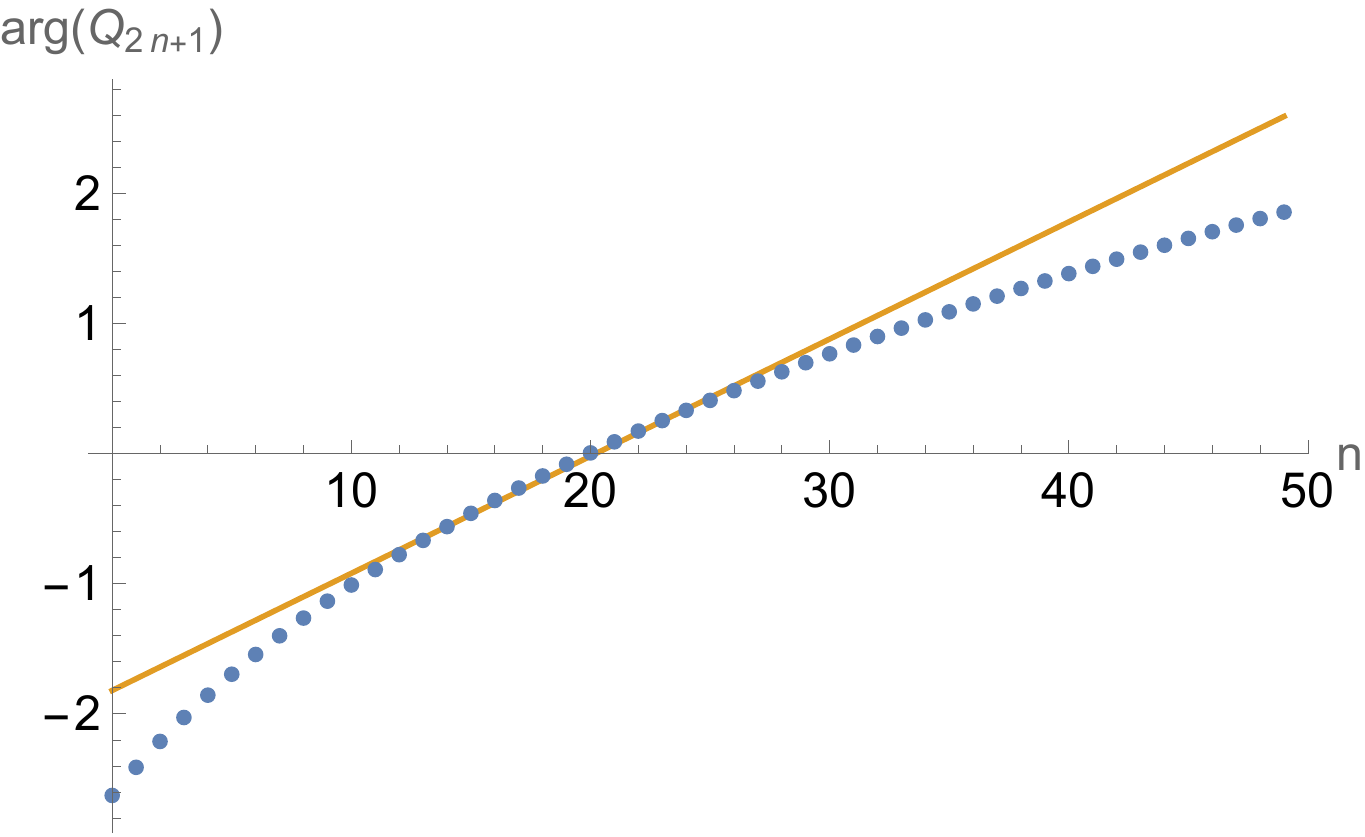}
\par\end{centering}
}\\
\subfloat[$\protect\b=1/3,t=-7$]{\begin{centering}
\includegraphics[height=2.3cm]{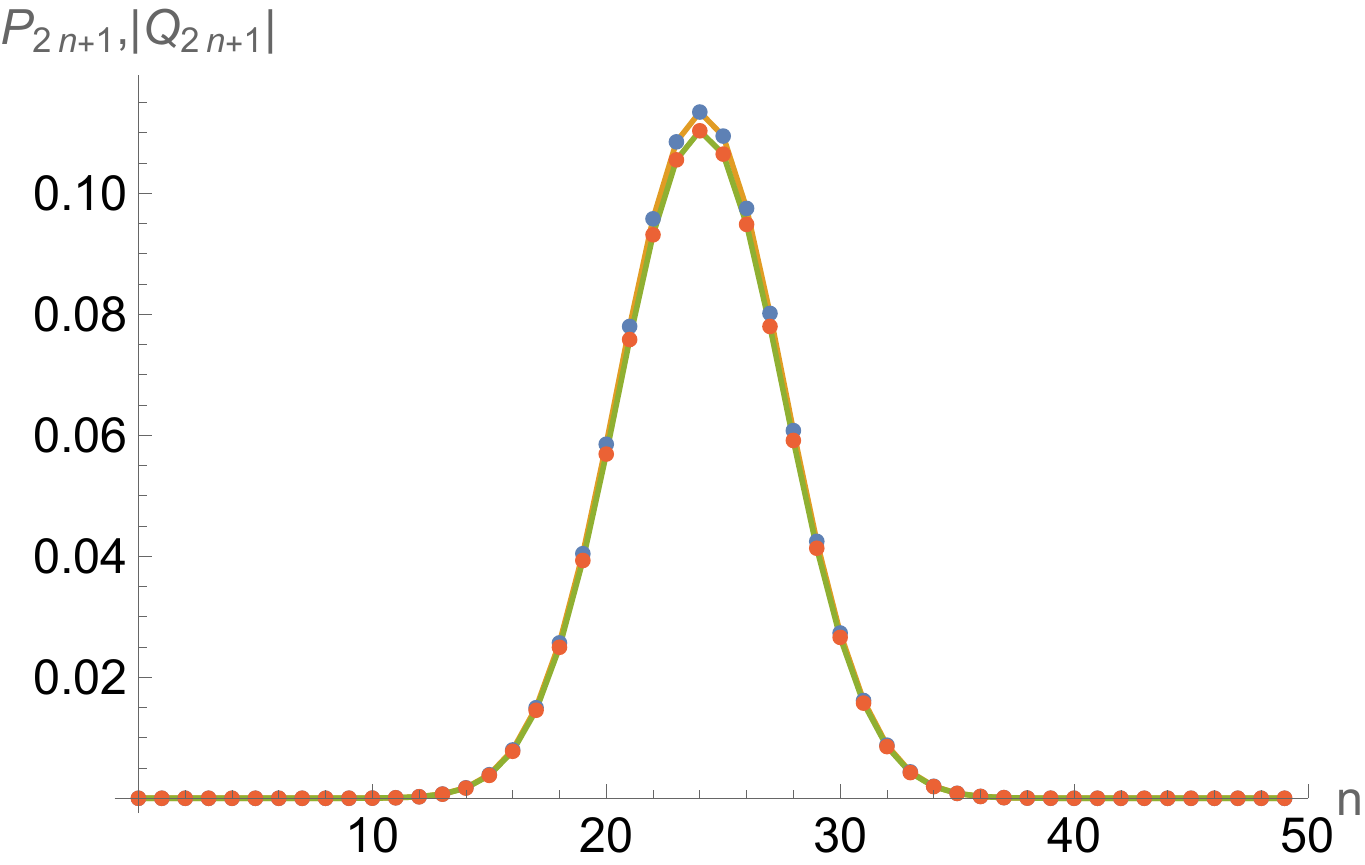}\includegraphics[height=2.3cm]{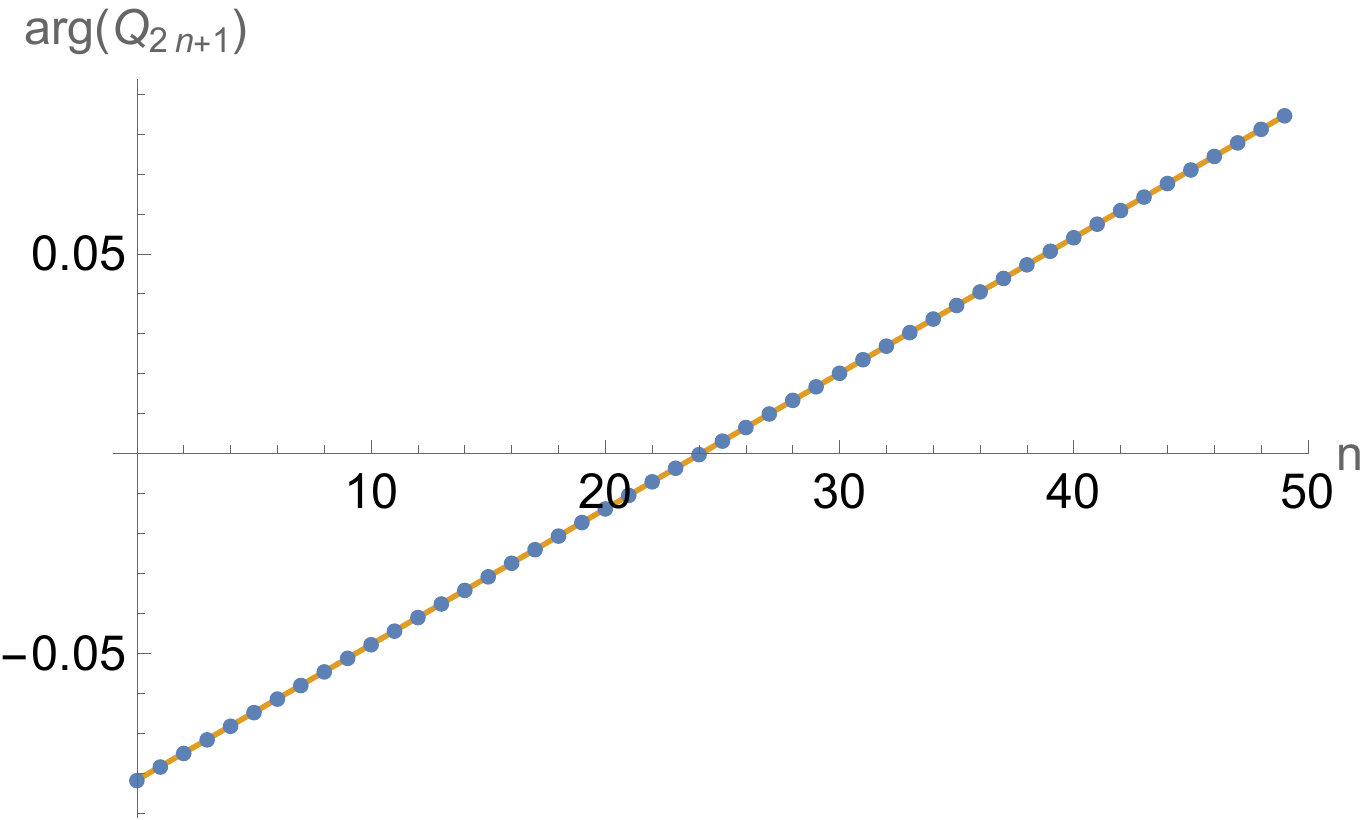}
\par\end{centering}
}\subfloat[$\protect\b=0.95,t=-7$]{\begin{centering}
\includegraphics[height=2.3cm]{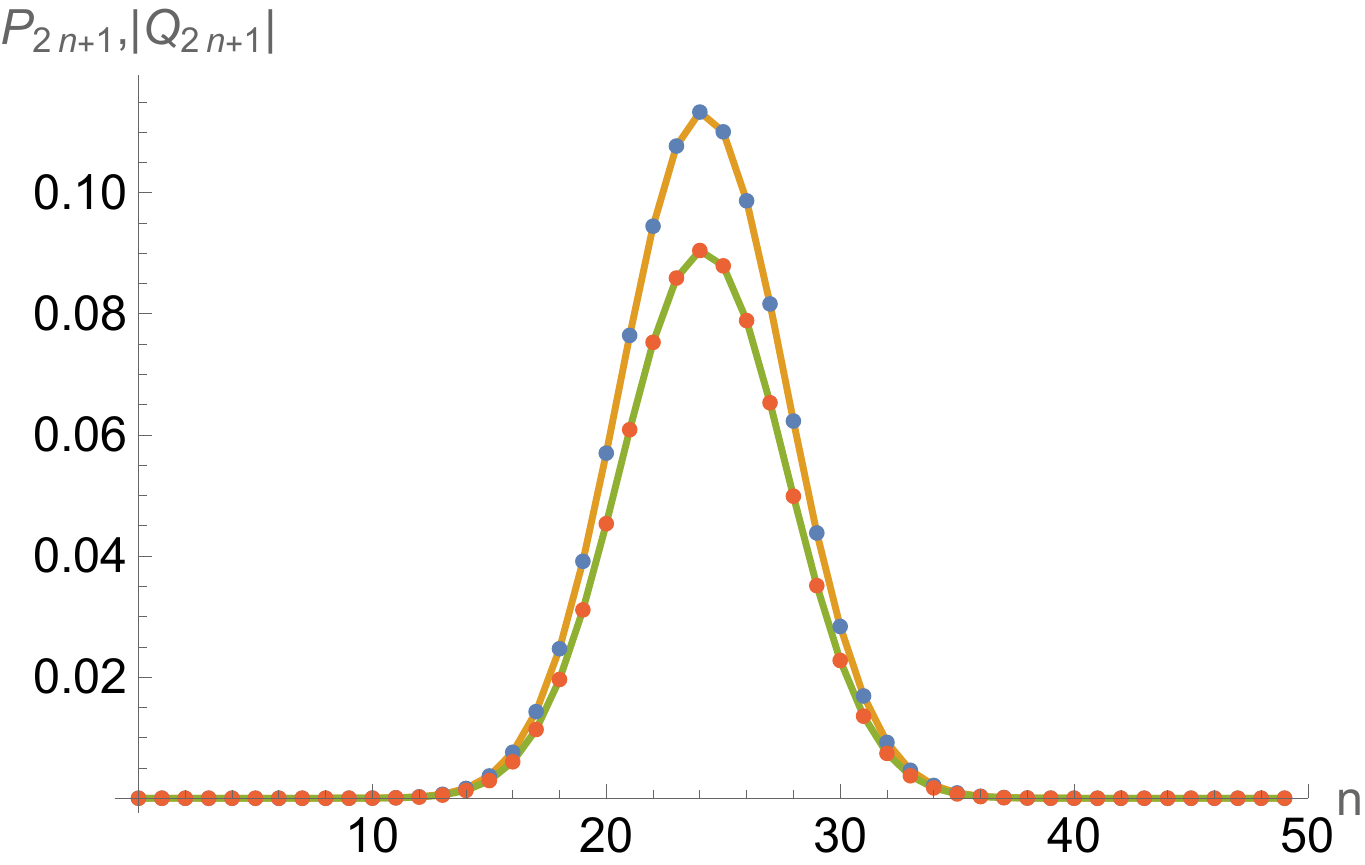}\includegraphics[height=2.3cm]{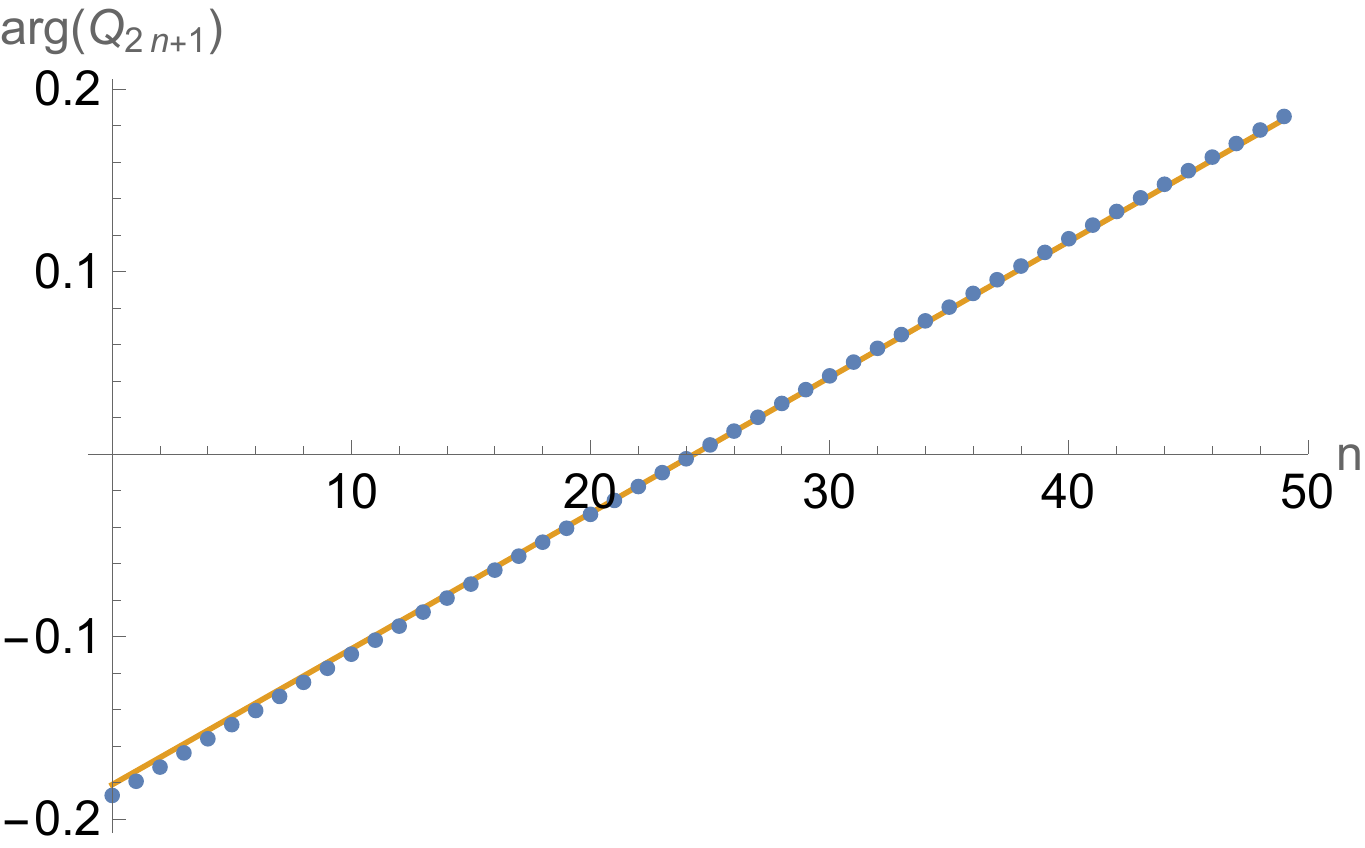}
\par\end{centering}
}
\par\end{centering}
\caption{The comparison between the simplified saddle approximation and exact
results with $N=100$ and for high temperature $\protect\b=1/3$ (a,c,e)
and intermediate temperature $\protect\b=0.95<\b_c$ (b,d,f). In the $P_{2n+1},|Q_{2n+1}|$
plots, the blue dots are exact values of $P_{2n+1}$ and the red dots
are exact values of $|Q_{2n+1}|$; the yellow joint lines are saddle
approximation of $P_{2n+1}$ and the green joint lines are saddle
approximation of $|Q_{2n+1}|$. In the $\arg Q_{2n+1}$ plots, the
blue dots are exact values of $\arg Q_{2n+1}$ and the yellow joint
lines are saddle approximation of $\arg Q_{2n+1}$. In (b) where the
simplified saddle approximation does not work very well, we plot the
saddle approximation using (\ref{eq:x0}). The purple joint line is
$P_{2n+1},$the gray joint line is $|Q_{2n+1}|$ in the left picture,
and the green joint line is $\arg Q_{2n+1}$ in the right picture.
For $n$ not too big, the saddle approximation is improved, but for
large $n$ it loses accuracy and we need to improve the saddle location
$x_{0}$ further in (\ref{eq:x0}). \label{fig:The-comparison-between}}
\end{figure}
Similarly, for size winding, we have 
\begin{align}
e^{-\mu N/2}\overline{g_{\mu}(\tau)} & =\f{e^{-\b^{2}/4}e^{-\mu N/2}\G(N/2)}{2\sqrt{a'\pi}}\int dx\sum_{i_{1}=0}^{N/2-1}\f{(\cosh\mu)^{N/2-i_{1}-1}(\sinh\mu)^{i_{1}}}{i_{1}!(N/2-1-i_{1})!}e^{-\f 1{a'}x^{2}+(2x+b'')i_{1}}\nonumber \\
 & =\f{e^{-\b^{2}/4}e^{-\mu}}{2^{N/2}\sqrt{a'\pi}}\int dxe^{-\f 1{a'}x^{2}}\left((1+e^{-2\mu})+(1-e^{-2\mu})e^{2x+b''}\right)^{N/2-1}
\end{align}
where 
\begin{equation}
b''=\f{(\b+4\tau)^{2}}{4(N/2-1)}-\b^{2}/4
\end{equation}
The saddle approximation is the same as before, which sets $x_{0}=0$
in the leading order of small $\mu$. This leads to
\begin{equation}
\overline{Q_{2n+1}(t)}=\f{e^{-\b^{2}/4}}{2^{N/2-1}}C_{N/2-1}^{n}(1+e^{-b_{q}(t)})^{N/2-1-n}(1-e^{-b_{q}(t)})^{n},\quad b_{q}(t)=\f{16t^{2}-\b^{2}-8i\b t}{4(N/2-1)}+\b^{2}/4\label{eq:120}
\end{equation}

Note that the phase of $\overline{Q_{2n+1}(t)}$ is perfectly proportional
to the size because $1+e^{-b_{q}(t)}$ has fixed phase $\phi_{1}$
and $1-e^{-b_{q}(t)}$ has fixed phase $\phi_{2}$, which together
leads to the phase of (\ref{eq:120}) as $(N/2-1)\phi_{1}+(\phi_{2}-\phi_{1})n$,
which is linear in size $2n+1$. By (\ref{eq:120}) the slope to $n$
of the phase is
\begin{equation}
\phi_{2}-\phi_{1}=\arg\f{1-e^{-b_{q}(t)}}{1+e^{-b_{q}(t)}}\app-\f{4\b t/N}{\sinh(\b^{2}/4+(16t^{2}-\b^{2})/(2N))}\label{eq:81}
\end{equation}
which increases with $t$ until the scrambling scale and then decreases.
A few numerics with $N=100$ in Figure \ref{fig:The-comparison-between}
shows that the simplified saddle approximation (\ref{eq:117}) and
(\ref{eq:120}) is quite good for high temperatures. One might complain
that the saddle approximation for the phase of $Q_{2n+1}$ is not
quite good for large and small $n$, especially at early time in Figure
\ref{fig:1a}. But this is not important because the dominant pieces
are around the peak of the magnitude where the linearity matches well.
As we decrease the temperature but still above the critical temperature $T_c$, the simplified saddle
approximation does not work very well in the early times but then improves as time increases. From thermal scale ($\sim\b$) to scrambling scale
($\sim\sqrt{N}$), the size distribution moves from small to large
and eventually stabilizes around $N/2$; the phase of $Q_{2n+1}$ organizes
itself from not quite linear to linear in $n$; and the slope of the
phase also decays, which is consistent with (\ref{eq:81}).

\begin{figure}
\begin{centering}
\subfloat[$\protect\b=0.1,t=-0.1$]{\begin{centering}
\includegraphics[height=2.3cm]{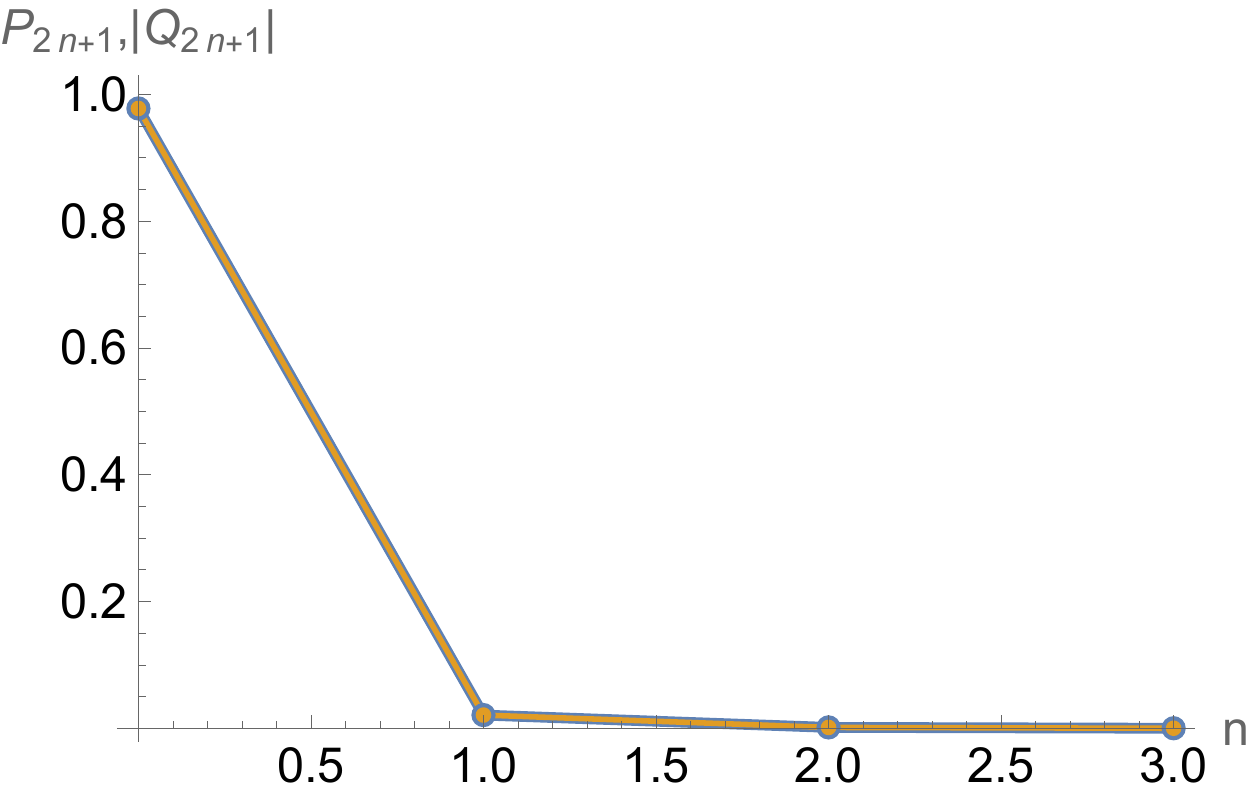}\includegraphics[height=2.3cm]{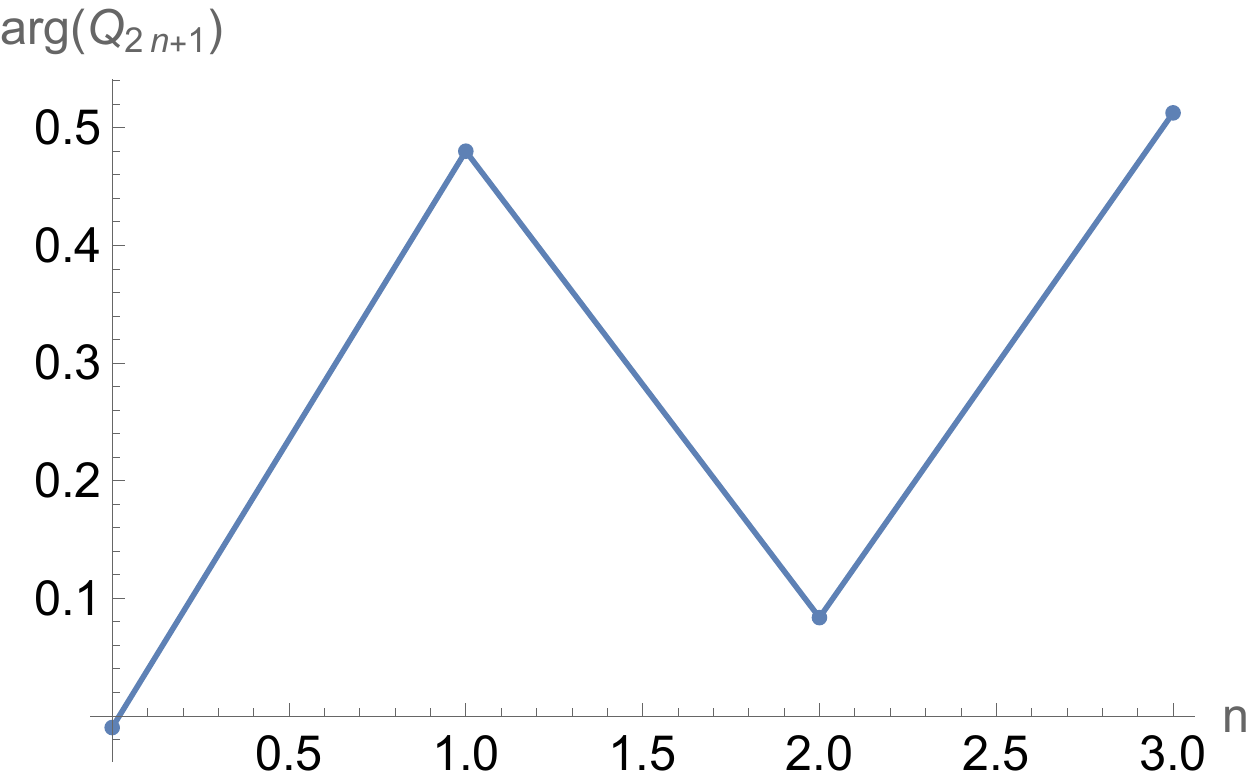}
\par\end{centering}
}\subfloat[$\protect\b=1,t=-0.1$]{\begin{centering}
\includegraphics[height=2.3cm]{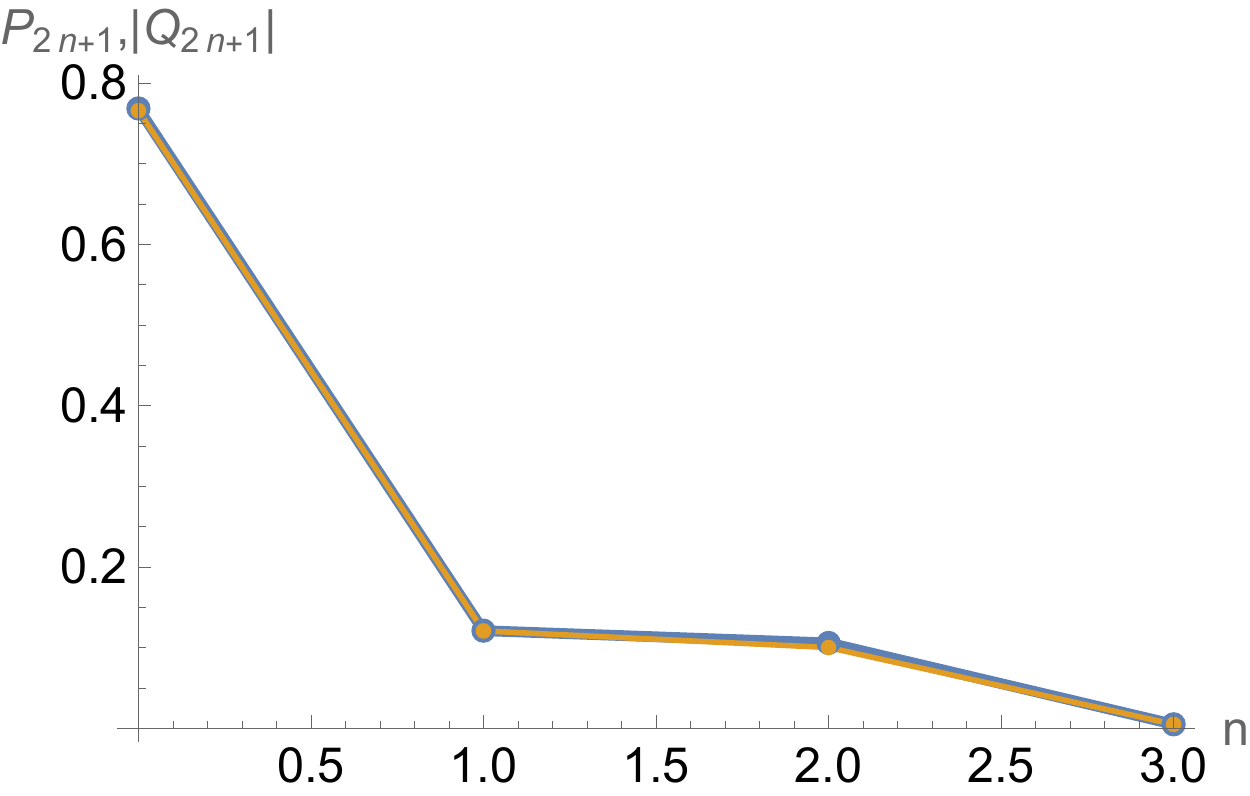}\includegraphics[height=2.3cm]{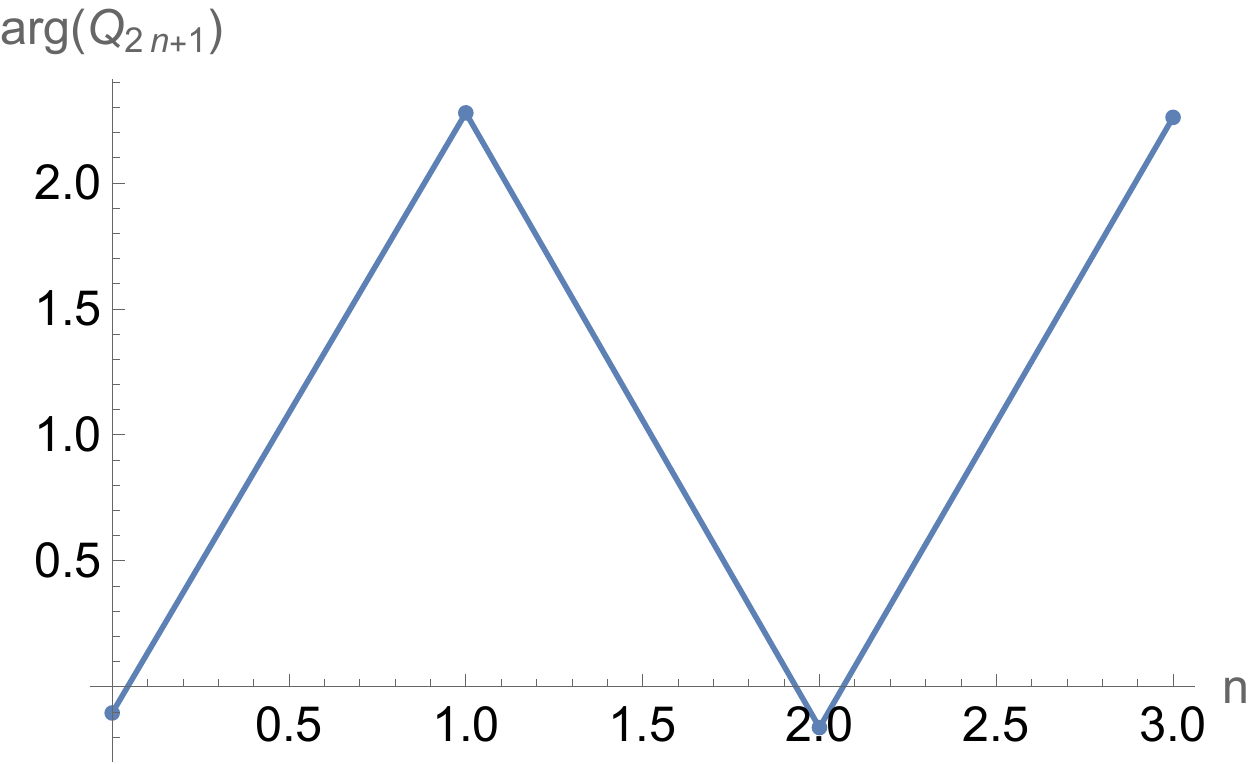}
\par\end{centering}
}\\
\subfloat[$\protect\b=0.1,t=-0.6$]{\begin{centering}
\includegraphics[height=2.3cm]{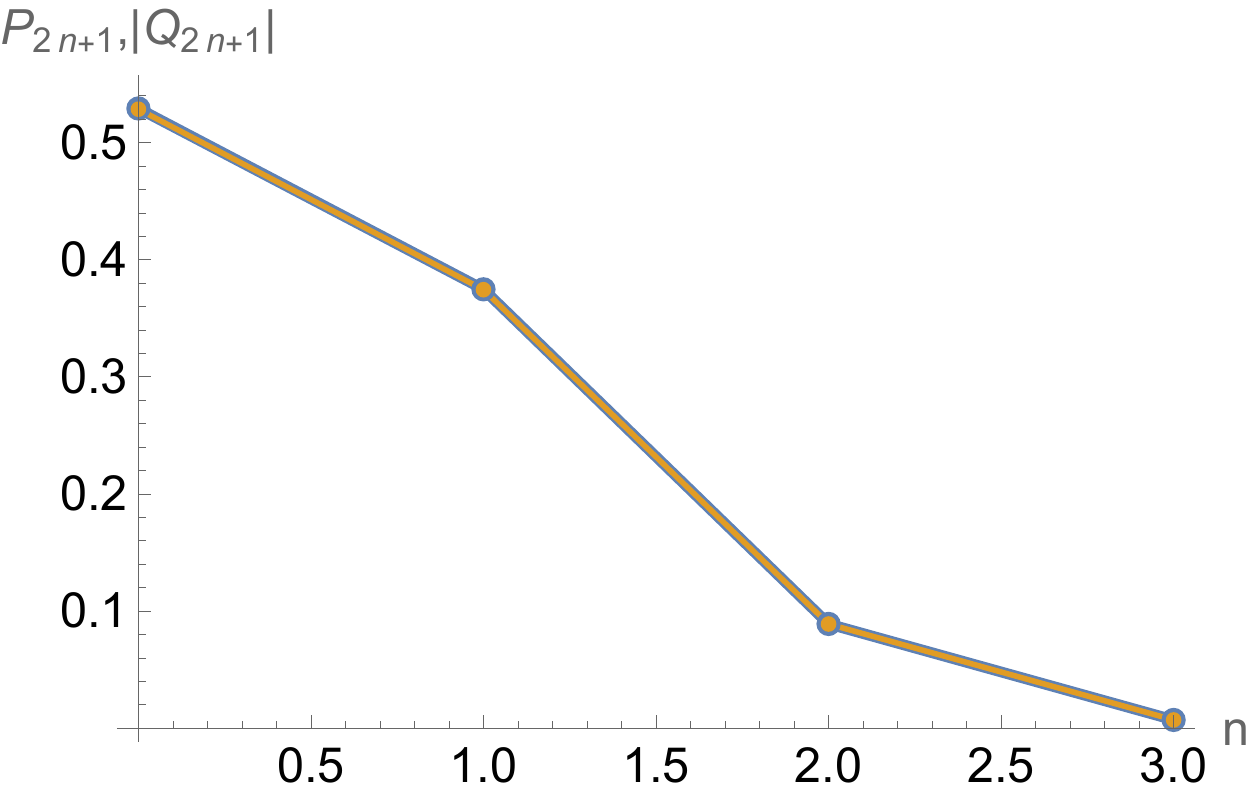}\includegraphics[height=2.3cm]{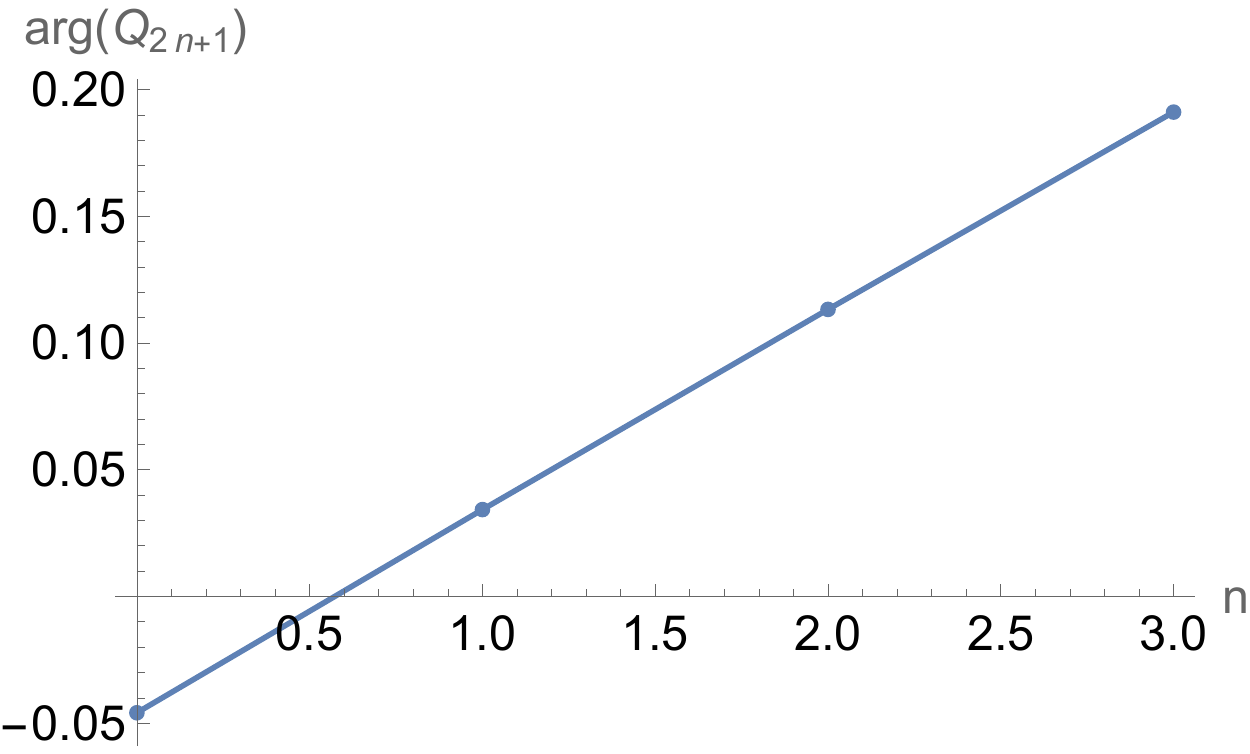}
\par\end{centering}
}\subfloat[$\protect\b=1,t=-0.8$\label{fig:2d}]{\begin{centering}
\includegraphics[height=2.3cm]{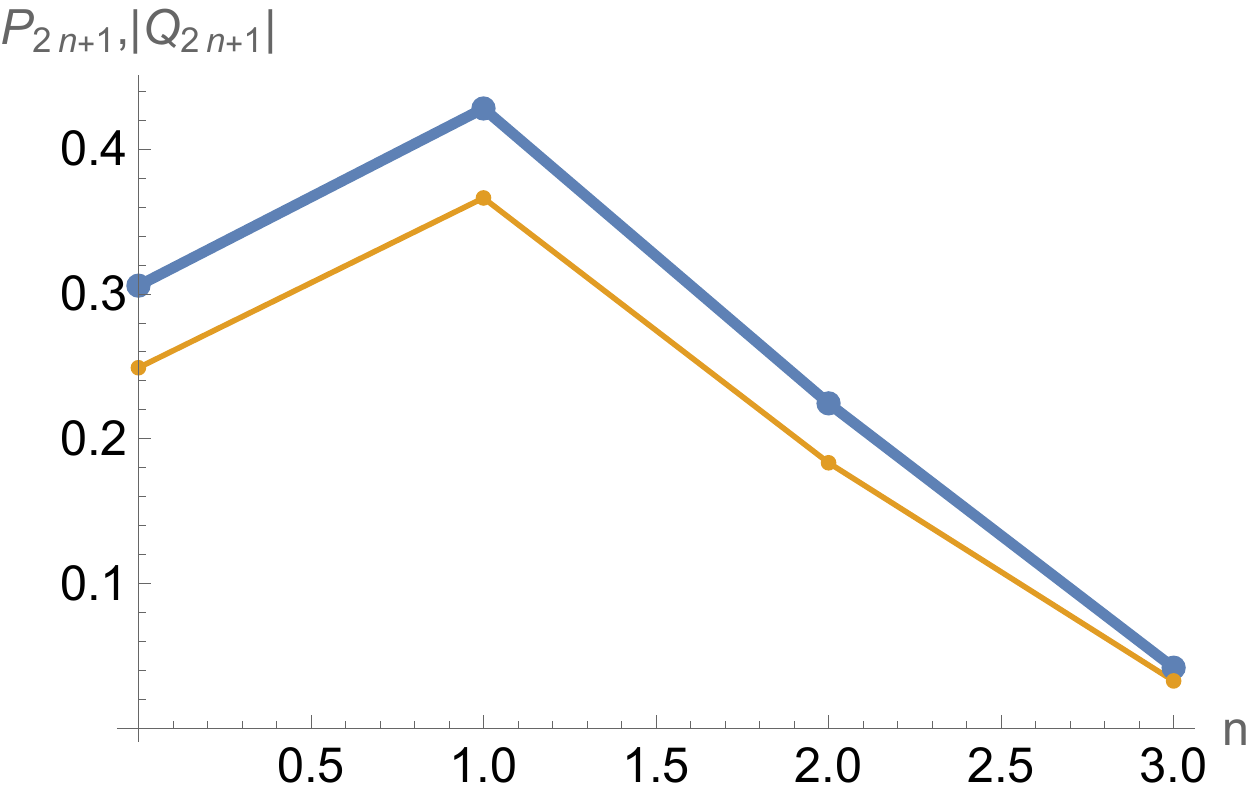}\includegraphics[height=2.3cm]{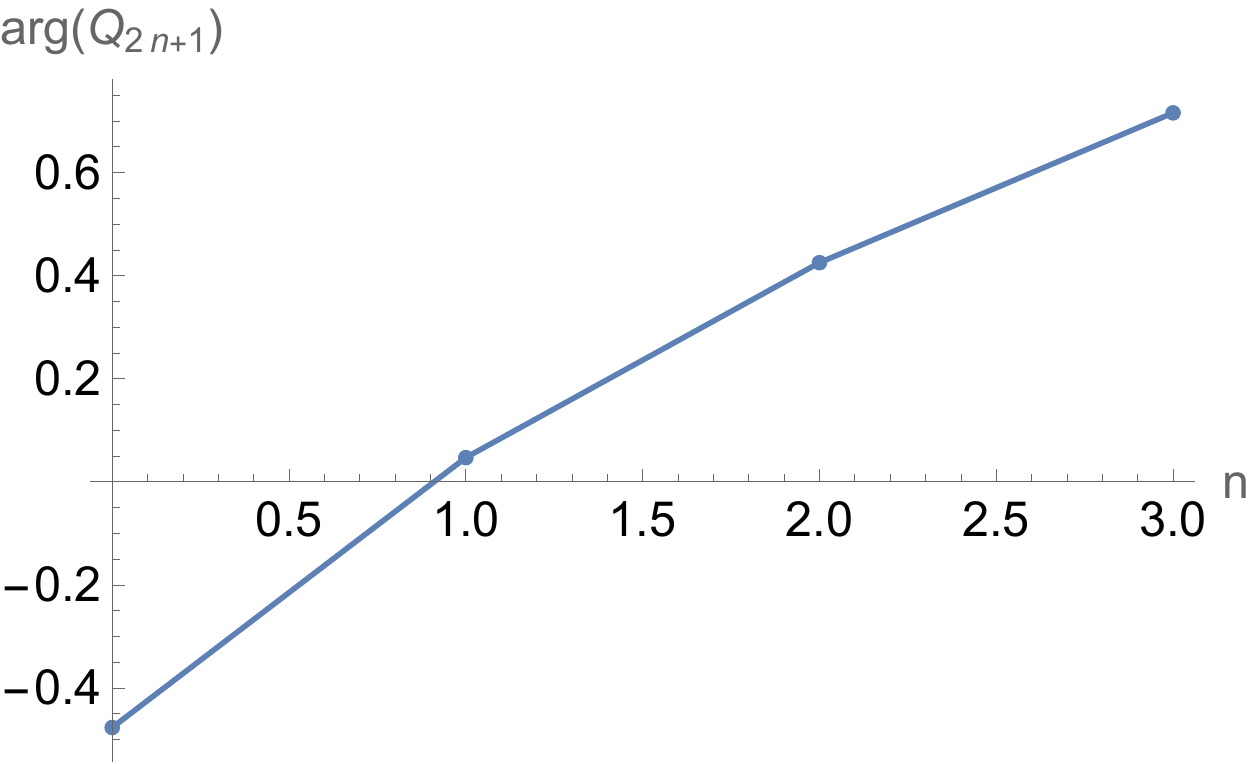}
\par\end{centering}
}\\
\subfloat[$\protect\b=0.1,t=-1.5$]{\begin{centering}
\includegraphics[height=2.3cm]{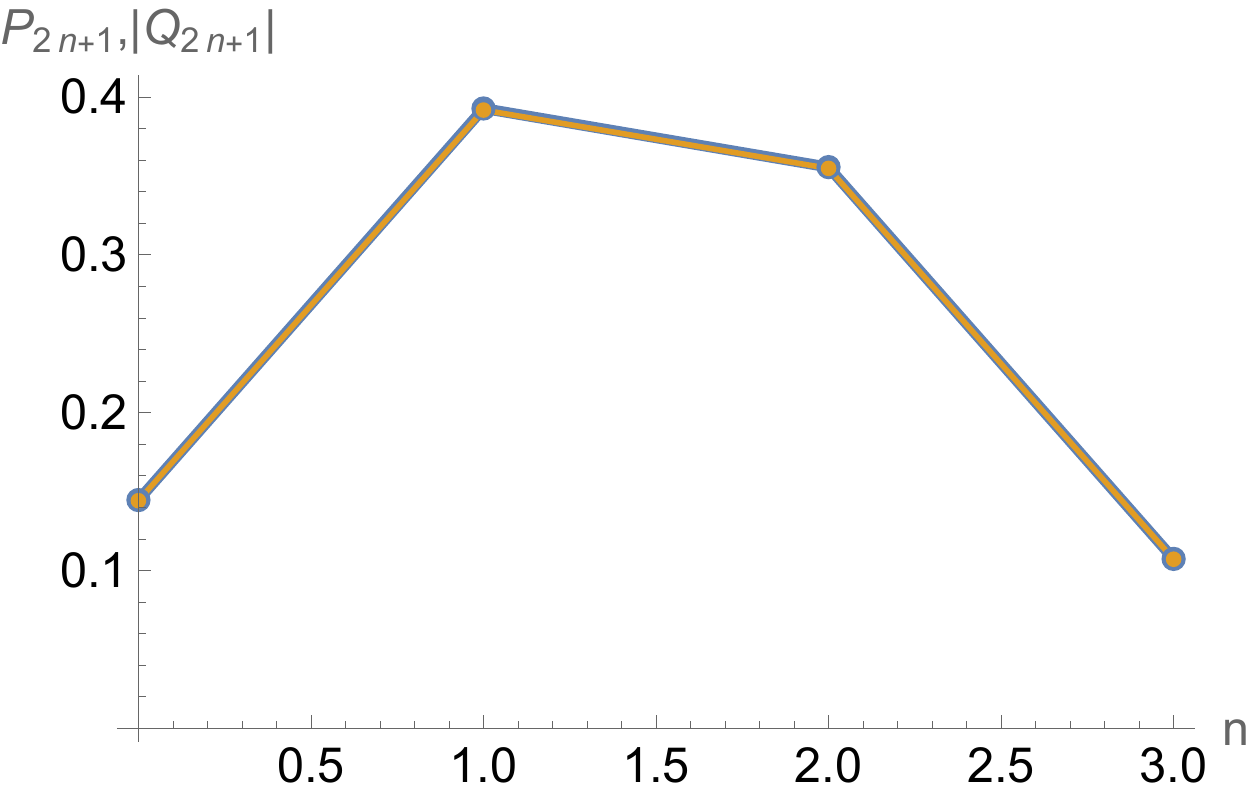}\includegraphics[height=2.3cm]{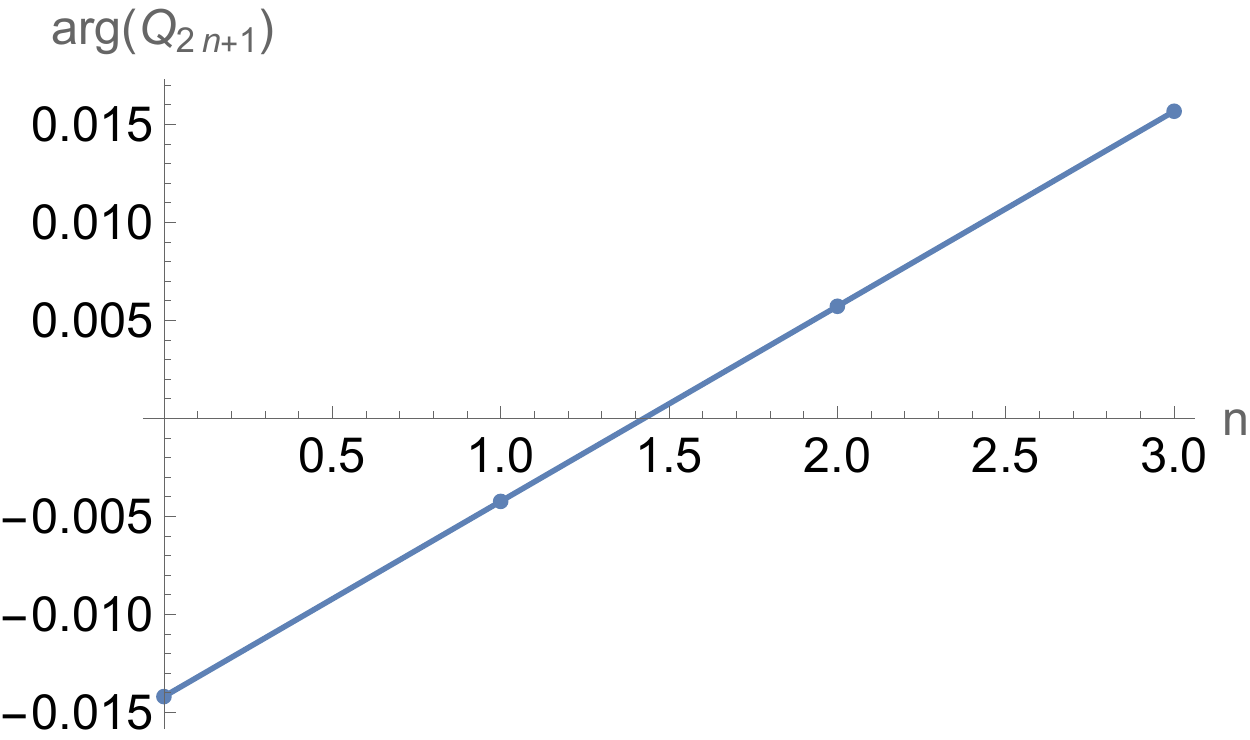}
\par\end{centering}
}\subfloat[$\protect\b=1,t=-1.5$]{\begin{centering}
\includegraphics[height=2.3cm]{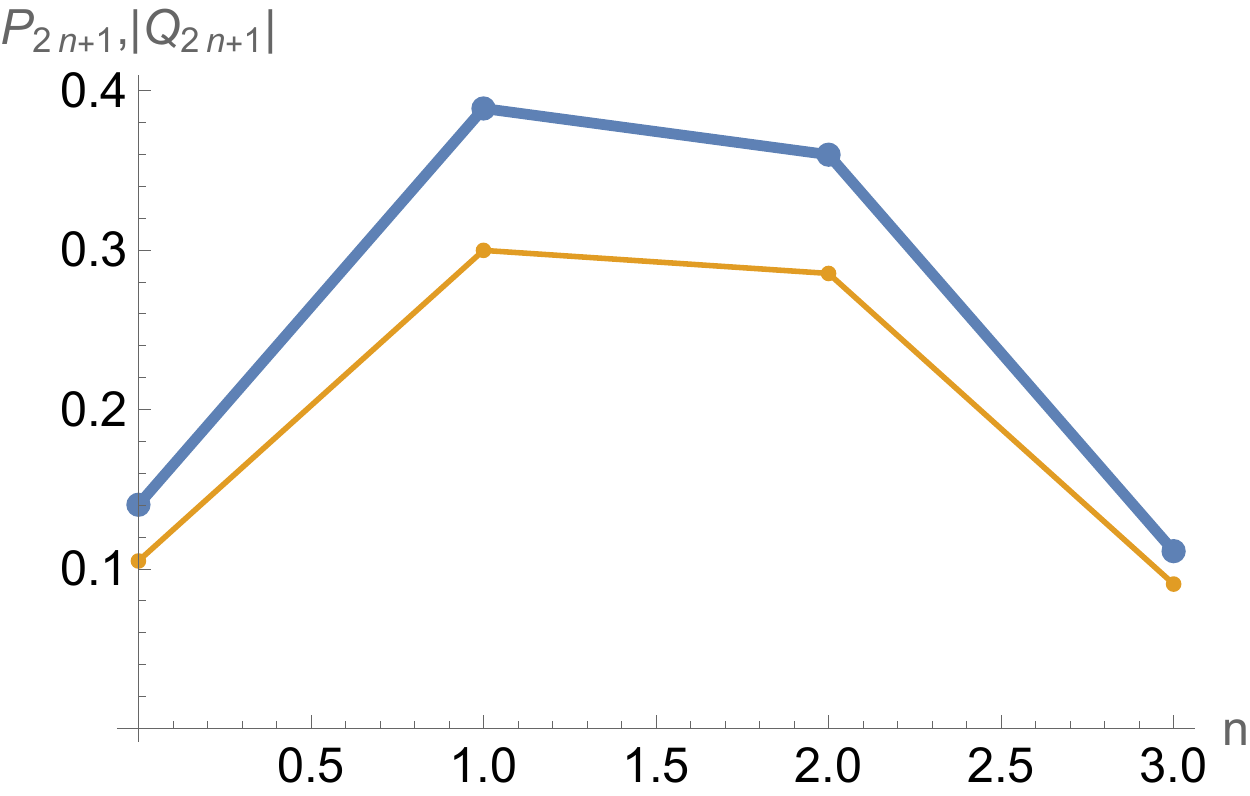}\includegraphics[height=2.3cm]{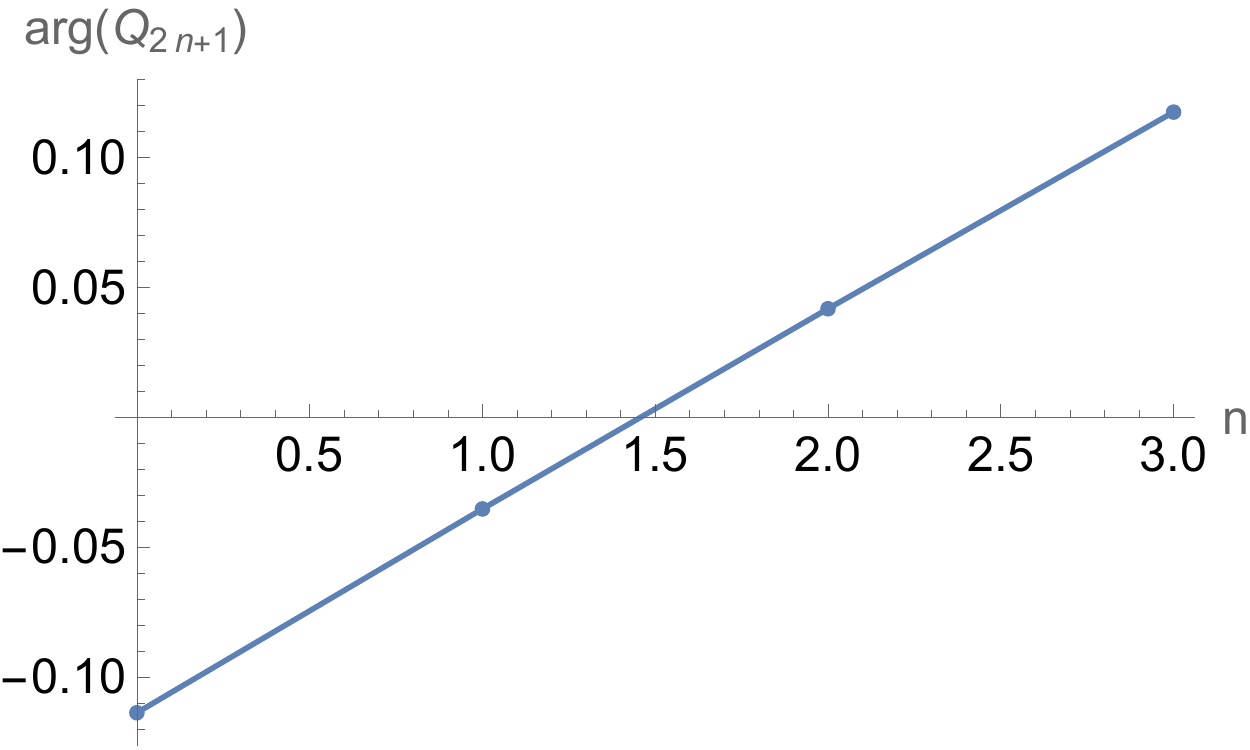}
\par\end{centering}
}
\par\end{centering}
\caption{The exact results with $N=8$ and for high temperature $\protect\b=0.1$
(a,c,e) and low temperature $\protect\b=1$ (b,d,f). In the $P_{2n+1},|Q_{2n+1}|$
plots, the joint blue dots are exact values of $P_{2n+1}$ and the
joint yellow dots are exact values of $|Q_{2n+1}|$. These two series
of joint dots are plotted with different thicknesses for clear comparison.
In the $\arg Q_{2n+1}$ plots, the joint blue dots are exact values
of $\arg Q_{2n+1}$. \label{fig:The-comparison-between-1}}
\end{figure}

Besides the linearity of the phase of $Q_{2n+1}$, we also need to
check if the phase of each individual size basis with the same size
aligns. This can be measured by how $r_{n}=|Q_{n}|/P_{n}$ is close
to one. With the numerical evidence in Figure \ref{fig:The-comparison-between},
in the following, we will use the simplified saddle approximation (\ref{eq:117})
and (\ref{eq:120}) to estimate $r_{n}$ for all temperatures. we
have
\begin{equation}
r_{2n+1}=e^{-\b^{2}/4}\left|\f{1+e^{-b_{q}(t)}}{1+e^{-b_{p}(t)}}\right|^{N/2-1-n}\left|\f{1-e^{-b_{q}(t)}}{1-e^{-b_{p}(t)}}\right|^{n}
\end{equation}
We approximate $e^{-b_{p}(t)}\app e^{-\b^{2}/4-8t^{2}/N}\left(1-\f{\b^{2}}{4(N/2-1)}\right)$
and $e^{-b_{q}(t)}\app e^{-\b^{2}/4-8t^{2}/N}\left(1+\f{\b^{2}+8i\b t}{4(N/2-1)}\right)$,
which leads to
\begin{equation}
r_{2n+1}\app\exp\left[-\f{\b^{2}(1-e^{-\b^{2}/4-8t^{2}/N})}{4(1+e^{-\b^{2}/4-8t^{2}/N})}-\f{\b^{2}n}{N\sinh(\b^{2}/4+8t^{2}/N)}\right]\label{eq:79-1}
\end{equation}
Note that the second term in the exponent of (\ref{eq:79-1}) is not
important in all scenarios. Since $\b\sim O(1)$, at late times this term is exponentially suppressed, and at early
times the dominant size is $n\sim O(1)$ and this term is of order
$O(1/N)$. Keeping only the first exponent in (\ref{eq:79-1}) leads
to the conclusion that $r_{2n+1}\sim e^{-\g\b^{2}/4}$ where
$\g\in[0,1]$, which implies that the size winding of commuting SYK
model is near-perfect in high temperatures but damped as we decrease the temperature.
When we cool down the system, the phase of the coefficients of the
size basis with the same size starts to spread out from the perfectly
alignment though their averaged phase is still proportional to the
size. This estimate can be verified by the numeric result in Figure
\ref{fig:The-comparison-between}. In high temperatures, the difference
between $P_{2n+1}$ and $|Q_{2n+1}|$ is quite small (a,c,e), but in lower temperature,
the difference is larger (b,d,f) due to the overall suppression $e^{-\g\b^{2}/4}$
as we just analyzed above. Therefore, we can confirm that the large
$N$ commuting SYK model has size winding in high temperatures. As
far as we know, this is the first nontrivial large $N$ non-holographic model that has near-perfect size winding.

As a comparison, we also show a case of small $N$ in Figure \ref{fig:The-comparison-between-1},
where we take $N=8$. Since there is no sharp phase transition for finite $N$, here we choose the temperatures for best exhibiting the features regardless of $T_c$. As we can see from these plots, the small $N$
is indeed qualitatively the same as the large $N$ case though the
thermal scale is not quite separable from the scrambling scale. The
main difference is that for small $N$ at early time ($t\lesssim\b$)
the phases are poorly lined up for both low and high temperatures,
but quickly reorganize themselves with linearity as time goes by.
However, phase linearity only guarantees near-perfect size winding
for high temperature because the magnitude of $Q_{2n+1}$ still matches
with $P_{2n+1}$ in later times, while the magnitude of $Q_{2n+1}$
starts to drop off after the thermalization scale from $P_{2n+1}$ for
lower temperature. 

\subsection{Peaked-size versus size-winding \label{subsec:Average-size-and}}

In \cite{Schuster:2021uvg}, there is another mechanism of teleportation in a generic
scrambling system in high temperatures called peaked-size teleportation.
This mechanism requires a narrow size distribution $P_{n}(t)$ of
the scrambled operator $\psi_{j}^{r}(t)\r_{r}^{1/2}$ around its average
size $\mS$. Such mechanism widely exists in many systems \cite{Schuster:2021uvg},
including the late-time regime of a generic scrambling system when
the dynamics can be approximated by Haar random unitaries \cite{Hayden:2007cs, Roberts:2016hpo},
random unitary circuits ($\geq1D$) with local gates, random unitary
circuits in $0D$ with all-to-all coupling and large $q$ SYK model
in infinite temperature. The last two require encoding the to-be-scrambled
qubit in terms of a large number of qubits. When the size distribution
is peaked, in the sense that the ratio between size fluctuation and
the size $\d\mS/\mS\ll1$, we can simply replace $e^{-\mu S}$ in
both $K_{\mu}(t)$ and $G_{\mu}(t)$ as $e^{-\mu\mS}$.

We can check if the commuting SYK model is peaked-size. The point
is to compare the average size $\mS$ and its variation $\d\mS$,
which can be easily computed by $K_{\mu}(t)$ 
\begin{equation}
\mS=-2\del_{\mu}K_{\mu}(t)|_{\mu=0},\quad(\d\mS)^{2}=2\del_{\mu}^{2}K_{\mu}(t)|_{\mu=0}-\mS^{2}
\end{equation}
where the additional coefficient 2 is due to the normalization of
$\psi_{j}^{r}(t_{r})\r_{r}^{1/2}$. Since $K_{\mu}=e^{-\mu N/2}k_{\mu}$,
we have
\begin{equation}
\mS=Nk_{0}-2k{}_{0}',\quad(\d\mS)^{2}=2k_{0}''-4(k_{0}')^{2}+2Nk_{0}'(2k_{0}-1)-\f 12N^{2}k_{0}(2k_{0}-1)
\end{equation}
where prime is derivative to $\mu$. From (\ref{eq:43-1}) we
have 
\begin{align}
k_{0} & =\f 12,\quad k_{0}'=\f 14(N-2)e^{-\b^{2}/4}e^{-8t^{2}/(N-2)}\\
k_{0}'' & =\f 14(N-2)+\f 18(N-2)(N-4)e^{-\f{N-4}{2(N-2)}\b^{2}}e^{-16t^{2}/(N-2)}
\end{align}
Let us consider the large $N$ limit. This leads to
\begin{align}
\mS & \app\f N2(1-e^{-\b^{2}/4}e^{-8t^{2}/N}) \label{eq:3.57}\\
\d\mS & \app\f{\sqrt{N}}2\left(2+(\b^{2}-2)e^{-\b^{2}/2}e^{-16t^{2}/N}\right)^{1/2}
\end{align}
where the ratio between size fluctuation and the size is $O(1/\sqrt{N})$.
This $O(1/\sqrt{N})$ ratio can also be understood as a common feature
of binary distribution (\ref{eq:117}) as long as $1\pm e^{-b_{p}(t)}\sim O(1)$.
This analysis implies that the commuting SYK model should follow the mechanism of peaked
size teleportation even though we see from
Section \ref{subsec:Saddle-approximation-for} that size winding is
also obeyed in high temperatures. There is no contradiction because
size winding requires $Q_{n}(t)\app P_{n}(t)e^{i(a_{1}+a_{2}n)}$
but does not impose any restriction on the distribution of $P_{n}(t)$.
It is a little bit surprising that in the past work \cite{Schuster:2021uvg} (and
also \cite{Brown:2019hmk, Nezami:2021yaq}), they are regarded as two distinct mechanisms, which are found in different (holographic versus non-holographic) models or exclusive regimes (e.g. low-temperature versus high-temperature SYK model).

Despite the ratio $\d\mS/\mS$ is of order $O(1/\sqrt{N})$, as we
will discuss in the next section, the teleportation protocol follows
the peaked-size mechanism only for long time scale $t\sim O(\sqrt{N})$.
For short time scale $t\sim O(1)$, the teleportation protocol works
quite differently is related to thermalization.

\section{Traversable wormhole teleportation protocol in commuting SYK}\label{sec4}

As the commuting SYK model shows near-perfect size winding in high
temperature, it is natural to consier how it behaves in the traversable
wormhole teleportation protocol \cite{Gao:2019nyj}. It has been shown in \cite{Brown:2019hmk, Nezami:2021yaq}
that the size winding is the microscopic mechanism for the teleportation
through a traversable wormhole in the ordinary SYK model. 

\begin{figure}
\begin{centering}
\includegraphics[width=4cm]{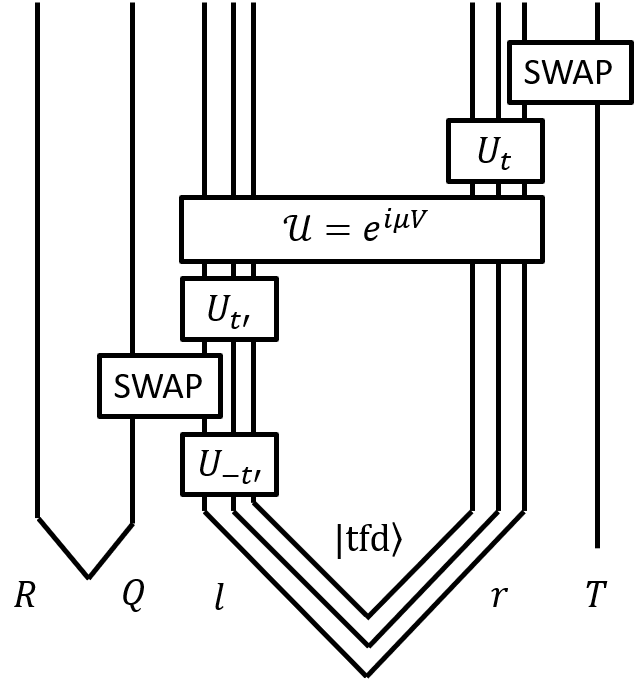}
\par\end{centering}
\caption{The traversable wormhole teleportation protocol (copied from \cite{Gao:2019nyj})
\label{fig:The-traversable-wormhole}}
\end{figure}
The quantum circuit of the traversable wormhole teleportation protocol
is given by Figure \ref{fig:The-traversable-wormhole}, in which $U_{t}=e^{-iHt}$
is the evolution operator and SWAP works as an injection and extraction
of the qubit being teleported. However, to study its effectiveness,
computing a left-right causal correlator is sufficient. For fermionic
system, this is defined as 
\begin{equation}
C(t_{l},t_{r})=\avg{\{e^{i\mu V}\psi_{j}^{l}(t_{l})e^{-i\mu V},\psi_{j}^{r}(t_{r})\}}_{TFD}=-2\Im H_{i\mu}(t_{l},t_{r})\label{eq:anticom}
\end{equation}
where we define the perturbed two-point function 
\begin{equation}
H_{i\mu}(t_{l},t_{r})=-i\avg{0|\r_{r}^{1/2}e^{i\mu V}\psi_{j}^{l}(t_{l})e^{-i\mu V}\psi_{j}^{r}(t_{r})\r_{r}^{1/2}|0}\label{eq:84}
\end{equation}
For convenience, we compute the Euclidean version
\begin{equation}
h_{\mu}(\tau_{1},\tau_{2})=-i\avg{0|\r_{r}^{1/2}e^{\mu V}\psi_{j}^{l}(\tau_{1})e^{-\mu V}\psi_{j}^{r}(\tau_{2})\r_{r}^{1/2}|0}\label{eq:79}
\end{equation}

The computation is similar to Section \ref{sec:Commuting-SYK-has}
but more involved because we need to expand two exponentials $e^{\pm\mu V}$
in (\ref{eq:79}) and counting the number of terms with index conditions
needs a finer analysis. We leave the computations ($q=4$) in Appendix
\ref{sec:Computation-of} and just present the result here. We have 
\begin{equation}
h_{\mu}(\tau_{1},\tau_{2})=h_{\mu}^{1}(\tau_{1},\tau_{2})+h_{\mu}^{2}(\tau_{1},\tau_{2})
\end{equation}
where 
\begin{align}
h_{\mu}^{1} & =\f{e^{-\b^{2}/4+(\tau_{1}+\tau_{2})^{2}}\cosh\mu}{2\sqrt{a\pi}}\int dxe^{-\f 1ax^{2}}\left(\cosh^{2}\mu-e^{-b_{1}}\sinh^{2}\mu-\sinh2\mu\sinh\f{\tau_{1}(2\tau_{2}+\b)}{N/2-1}e^{2x-b_{2}}\right)^{N/2-1}\label{eq:h1}\\
h_{\mu}^{2} & =-\f{e^{(\tau_{1}+\tau_{2})(\tau_{1}+\tau_{2}+\b)}\sinh\mu}{2\sqrt{a\pi}}\int dxe^{-\f 1ax^{2}}\left(\cosh^{2}\mu-e^{-c_{1}}\sinh^{2}\mu-\sinh2\mu\sinh\f{2\tau_{1}\tau_{2}}{N/2-1}e^{2x-c_{2}}\right)^{N/2-1}\label{eq:h2}
\end{align}
and a few parameters are defined as 
\begin{align}
a & =\f{\b^{2}}{4(N/2-1)},\quad b_{1}=\f{4\tau_{1}\tau_{2}}{N/2-1}\quad b_{2}=\f{\tau_{2}(2\tau_{1}-\b)+(N-4)\b^{2}/8}{N/2-1}\\
c_{1} & =\f{2\tau_{1}(2\tau_{2}+\b)}{N/2-1},\quad c_{2}=\f{2\tau_{1}\tau_{2}+(\tau_{1}+\tau_{2})\b+N\b^{2}/8}{N/2-1}
\end{align}

\subsection{Saddle approximation}

Similar to Section \ref{subsec:Saddle-approximation-for}, for large
$N$ we can do the saddle approximation for the $x$ integral in $h_{\mu}^{i}$.
In order to discuss the $\mu$ dependence, we will not restrict $\mu$
to be small. We will solve the saddle after analytic continuation
$\mu\ra i\mu$, $\tau_{1}\ra it_{l}$ and $\tau_{2}\ra it_{r}$. Up
to a sign flip, from (\ref{eq:h1}) and (\ref{eq:h2}), we find that
$C$ has $\mu$ periodicity of $\pi$. Therefore, we can restrict
$\mu$ to the range of $[-\pi/2,\pi/2]$. 

For $H_{i\mu}^{1}$ in large $N$ limit, we have
\begin{equation}
H_{i\mu}^{1}=\f{e^{-\b^{2}/4-(t_{l}+t_{r})^{2}}\cos\mu}{2\sqrt{a\pi}}\int dxe^{-\f 1aF_{1}(x)}\label{eq:110}
\end{equation}
where $a=\b^{2}/(2N)$ and 
\begin{equation}
F_{1}(x)=x^{2}-\f{\b^{2}}4\log\left(\cos^{2}\mu+e^{\f{4t_{l}t_{r}}{N/2}}\sin^{2}\mu+ie^{-\b^{2}/4}\sin2\mu\sinh\f{2t_{l}t_{r}-i\b t_{l}}{N/2}e^{\f{2t_{l}t_{r}+i\b t_{r}}{N/2}}e^{2x}\right)
\end{equation}
In this equation, we have kept some $N$ dependence for later consideration
for scales of $t_{l,r}$ up to $O(N)$. Taking derivative to $x$,
we have the saddle equation
\begin{equation}
x=\f{\b^{2}}4\f{ie^{-\b^{2}/4}\sin2\mu\sinh\f{2t_{l}t_{r}-i\b t_{l}}{N/2}e^{\f{2t_{l}t_{r}+i\b t_{r}}{N/2}}e^{2x}}{\cos^{2}\mu+e^{\f{4t_{l}t_{r}}{N/2}}\sin^{2}\mu+ie^{-\b^{2}/4}\sin2\mu\sinh\f{2t_{l}t_{r}-i\b t_{l}}{N/2}e^{\f{2t_{l}t_{r}+i\b t_{r}}{N/2}}e^{2x}}\label{eq:112}
\end{equation}
For teleportation, we consider $t_{l}>0$ and $t_{r}<0$, which leads
to $t_{l}t_{r}<0$. Due to the overall exponential suppression factor
$e^{-(t_{l}+t_{r})^{2}}$ in (\ref{eq:110}), we only need to consider
the case $t_{l}+t_{r}\sim O(1)$. In the large $N$ limit, we have
\begin{align}
i\sinh\f{2t_{l}t_{r}-i\b t_{l}}{N/2}e^{\f{2t_{l}t_{r}+i\b t_{r}}{N/2}} & \app\begin{cases}
\f{2it_{l}t_{r}+\b t_{l}}{N/2} & t_{l}\sim-t_{r}\sim O(1)\\
\f i2\left(e^{\f{4t_{l}t_{r}}{N/2}}-1\right) & t_{l}\sim-t_{r}\gtrsim O(\sqrt{N})
\end{cases}\label{eq:113}
\end{align}
where we considered two scales of $t_{l}\sim-t_{r}$. For the short
time scale $t_{l}\sim-t_{r}\sim O(1)$, the saddle of $x$ is of order
$1/N$ and we can approximate $e^{2x}\sim1$ in leading order, which
leads the solution to (\ref{eq:112}) as
\begin{equation}
x_{saddle}\app\f{\b^{2}}4e^{-\b^{2}/4}\sin(2\mu)\f{2it_{l}t_{r}+\b t_{l}}{N/2}
\end{equation}
At this saddle point, we have in large $N$ limit
\begin{align}
-\f 1aF_{1}(x_{saddle}) & \app4t_{l}t_{r}\sin^{2}\mu+e^{-\b^{2}/4}\sin(2\mu)(2it_{l}t_{r}+\b t_{l})\\
F''_{1}(x_{saddle}) & \app2
\end{align}
which leads to
\begin{equation}
H_{i\mu}^{1}=\f{e^{-\b^{2}/4-(t_{l}+t_{r})^{2}}\cos\mu}2\exp\left(4t_{l}t_{r}\sin^{2}\mu+e^{-\b^{2}/4}\sin(2\mu)(2it_{l}t_{r}+\b t_{l})\right)\label{eq:h1-short}
\end{equation}

For the long time scale where $t_{l}\sim-t_{r}$ is the same or higher
order than $O(\sqrt{N})$ in (\ref{eq:113}), the saddle of $x$ is
an $O(1)$ complex number, which gives an $O(1)$ value for $F_{1}(x_{saddle})$.
These saddles do not have an analytic expression but we can easily
find their values numerically. One can show numerically that the saddle
leads to an $O(1)$ positive real part of $F_{1}(x_{saddle})$ for
most choices in the parameter space. Due to large $1/a$ coefficient,
$H_{i\mu}^{1}$ is exponentially suppressed in large $N$. An interesting
exception is $\mu$ close to zero and scales as $1/N$. Let us assume
$\mu=\mu_{0}/N$, and the saddle of $x$ in (\ref{eq:112}) is again
of order $1/N$ 
\begin{equation}
x_{saddle}\app\f{i\mu_{0}\b^{2}}{4N}e^{-\b^{2}/4}\left(e^{\f{4t_{l}t_{r}}{N/2}}-1\right)
\end{equation}
At this saddle point, we have in large $N$ limit
\begin{align}
-\f 1aF(x_{saddle}) & \app\f i2e^{-\b^{2}/4}\mu_{0}\left(e^{\f{4t_{l}t_{r}}{N/2}}-1\right)\\
F''(x_{saddle}) & \app2
\end{align}
which leads to
\begin{equation}
H_{i\mu}^{1}=\f{e^{-\b^{2}/4-(t_{l}+t_{r})^{2}}}2\exp\left(\f{i\mu_{0}}2e^{-\b^{2}/4}\left(e^{\f{4t_{l}t_{r}}{N/2}}-1\right)\right)\label{eq:122-1}
\end{equation}

Similarly, for $H_{i\mu}^{2}$ in large $N$ limit, we have
\begin{equation}
H_{i\mu}^{2}=-\f{e^{i\b(t_{l}+t_{r})-(t_{l}+t_{r})^{2}}i\sin\mu}{2\sqrt{a\pi}}\int dxe^{-\f 1aF_{2}(x)}\label{eq:110-1}
\end{equation}
where 
\begin{equation}
F_{2}(x)=x^{2}-\f{\b^{2}}4\log\left(\cos^{2}\mu+e^{\f{4t_{l}t_{r}-2i\b t_{l}}{N/2}}\sin^{2}\mu+ie^{-\b^{2}/4}\sin2\mu\sinh\f{2t_{l}t_{r}}{N/2}e^{\f{2t_{l}t_{r}-i\b(t_{l}+t_{r})}{N/2}}e^{2x}\right)
\end{equation}
Taking the derivative to $x$, we have the saddle equation
\begin{equation}
x=\f{\b^{2}}4\f{ie^{-\b^{2}/4}\sin2\mu\sinh\f{2t_{l}t_{r}}{N/2}e^{\f{2t_{l}t_{r}-i\b(t_{l}+t_{r})}{N/2}}e^{2x}}{\cos^{2}\mu+e^{\f{4t_{l}t_{r}-2i\b t_{l}}{N/2}}\sin^{2}\mu+ie^{-\b^{2}/4}\sin2\mu\sinh\f{2t_{l}t_{r}}{N/2}e^{\f{2t_{l}t_{r}-i\b(t_{l}+t_{r})}{N/2}}e^{2x}}\label{eq:112-1}
\end{equation}
Again due to the overall exponential suppression factor $e^{-(t_{l}+t_{r})^{2}}$
in (\ref{eq:110-1}), we only need to consider the case $t_{l}+t_{r}\sim O(1)$.
In the large $N$ limit, we have 
\begin{align}
i\sinh\f{2t_{l}t_{r}}{N/2}e^{\f{2t_{l}t_{r}-i\b(t_{l}+t_{r})}{N/2}} & \app\begin{cases}
\f{2it_{l}t_{r}}{N/2} & t_{l}\sim-t_{r}\sim O(1)\\
\f i2\left(e^{\f{4t_{l}t_{r}}{N/2}}-1\right) & t_{l}\sim-t_{r}\gtrsim O(\sqrt{N})
\end{cases}\label{eq:113-1}
\end{align}
For the short time scale $t_{l}\sim-t_{r}\sim O(1)$, the saddle of
$x$ is of order $1/N$
\begin{equation}
x_{saddle}\app\f{\b^{2}}4e^{-\b^{2}/4}\sin(2\mu)\f{2it_{l}t_{r}}{N/2}
\end{equation}
At this saddle point, we have in large $N$ limit
\begin{align}
-\f 1aF_{2}(x_{saddle}) & \app(4t_{l}t_{r}-2i\b t_{l})\sin^{2}\mu+e^{-\b^{2}/4}\sin(2\mu)(2it_{l}t_{r})\\
F''_{2}(x_{saddle}) & \app2
\end{align}
which leads to
\begin{equation}
H_{i\mu}^{2}=-\f{e^{i\b(t_{l}\cos2\mu+t_{r})-(t_{l}+t_{r})^{2}}i\sin\mu}2\exp\left(4t_{l}t_{r}\sin^{2}\mu+e^{-\b^{2}/4}\sin(2\mu)(2it_{l}t_{r})\right)\label{eq:h2-short}
\end{equation}

For the long time scale $t_{l}\sim-t_{r}\gtrsim O(\sqrt{N})$, the
saddle is a complex $O(1)$ number, which can be shown numerically
leading to positive real part of $F_{2}(x_{saddle})$ for most choices
in the parameters space, which results in an exponentially suppressed
$H_{i\mu}^{2}$ in large $N$ limit. The interesting exception is
to consider $\mu=\mu_{0}/N$, which gives order one value saddle approximation
for the integral (\ref{eq:110-1}). However, the $\sin\mu$ factor
will be order $1/N$ and $H_{i\mu}^{2}$ is suppressed relative to
$H_{i\mu}^{1}$. 

Putting all together, we have for the short time scale $t_{l}\sim-t_{r}\sim O(1)$
\begin{align}
H_{i\mu}= & \f{e^{-(t_{l}+t_{r})^{2}+4t_{l}t_{r}\sin^{2}\mu}}2\left[\cos\mu\exp\left(-\b^{2}/4+e^{-\b^{2}/4}\sin(2\mu)(2it_{l}t_{r}+\b t_{l})\right)\right.\nonumber \\
 & \left.-i\sin\mu\exp\left(e^{-\b^{2}/4}\sin(2\mu)(2it_{l}t_{r})+i\b(t_{l}\cos2\mu+t_{r})\right)\right]\label{eq:131}
\end{align}
and for the long time scale $t_{l}\sim-t_{r}\gtrsim O(\sqrt{N})$
\begin{equation}
H_{i\mu}=\f{e^{-\b^{2}/4-(t_{l}+t_{r})^{2}}}2\exp\left(\f{i\mu_{0}}2e^{-\b^{2}/4}\left(e^{\f{4t_{l}t_{r}}{N/2}}-1\right)\right)\label{eq:132}
\end{equation}
Note that (\ref{eq:131}) has exponential decay as we send a signal
earlier and receive the signal later, namely $t_{l}\sim-t_{r}\gg1$,
due to the factor $e^{-(t_{l}+t_{r})^{2}+4t_{l}t_{r}\sin^{2}\mu}$,
while (\ref{eq:132}) tends to an $O(1)$ constant in the regime $t_{l}\sim-t_{r}\gg\sqrt{N}$.
One should not be confused by this because in the long time scale
(\ref{eq:132}), $\mu$ is been rescaled to $\mu_{0}/N$ that compensates
the decaying effect of large time in the short time scale in the term
$4t_{l}t_{r}\sin^{2}\mu$.

\subsection{Sign of $\mu$\label{subsec:Sign-of}}

There is a crucial feature of the traversable wormhole teleportation
that only one sign of $\mu$ allows the information sent through \cite{Gao:2019nyj}.
In the semiclassical picture, the sign of $\mu$ is proportional to
the stress tensor of the injected matter that supports the traversable
wormhole. The throat of a traversable wormhole opens only when the averaged
null energy of the matter, which in turn is proportional to $\mu$,
is negative \cite{Gao:2016bin}. It is interesting to check if the teleportation
in the commuting SYK model follows the same rule.

Since we have two distinct time scales, we need to discuss the dependence
on the sign of $\mu$ separately. For the long time scale $t_{l}\sim-t_{r}\gtrsim O(\sqrt{N})$. It is interesting that the sign of $\mu$ does not affect the leading
order magnitude of $C$. This can be seen easily using \eqref{eq:132}, which leads to
\begin{equation}
C(t_{l},t_{r})=e^{-\b^{2}/4-(t_{l}+t_{r})^{2}}\sin\left(\f{\mu_{0}}2e^{-\b^{2}/4}\left(1-e^{\f{4t_{l}t_{r}}{N/2}}\right)\right)\label{eq:113C}
\end{equation}
It is noteworthy that this formula holds for any temperature $T>T_c$ and scrambling time scale $O(\sqrt{N})$.
This is very different from the large $N$ limit of the ordinary SYK model
at scrambling time scale $O(\log N)$, which prefers positive $\mu$
in low temperature but has indifference in the sign of $\mu$ only
in high temperature \cite{Gao:2019nyj}. 

As we discussed in Section \ref{subsec:Average-size-and} that the
distribution is peaked in the sense $\d\mS/\mS\ll 1$, we can find that (\ref{eq:113C})
follows directly from the peaked-size mechanism \cite{Schuster:2021uvg}. The result
of the peaked-size teleportation is quite simple that the $e^{-i\mu V}$
in (\ref{eq:84}) measures the averaged size of $\psi_{j}^{r}(t_{r})\r_{r}^{1/2}$
and becomes an exponential factor $e^{-i\mu(\mS-N/2)}$ due to the
narrow size distribution. It follows that
\begin{equation}
H_{i\mu}(t,-t)\app G_{lr}(t,-t)\exp\left(-i\mu(\mS+NG_{lr}(0,0)-N/2)\right)\label{eq:122}
\end{equation}
where the first exponential $e^{i\mu V}$ simply factorizes as $e^{i\mu\avg V}=e^{-i\mu G_{lr}(0,0)}$
because it is a TOC. Taking (\ref{eq:122}) into (\ref{eq:anticom}),
the left-right causal correlator has the form
\begin{equation}
C(t,-t)=2G_{lr}(t,-t)\sin\mu(\mS+NG_{lr}(0,0)-N/2)\label{eq:123}
\end{equation}
Note that the magnitude of $C$ is bounded by $G_{lr}(t_{l},t_{r})$
which decays as we drop the temperature. Therefore, the peaked-size
teleportation works better in high temperatures. From (\ref{eq:21}),
we can easily find that the left-right correlator in the commuting
SYK model is given by 
\begin{equation}
G_{lr}(t_{l},t_{r})\equiv-i\avg{0|\r_{r}^{1/2}\psi_{j}^{l}(t_{l})\psi_{j}^{r}(t_{r})\r_{r}^{1/2}|0}=\f 12e^{-\b^{2}/4-(t_{l}+t_{r})^{2}}\label{eq:124}
\end{equation}
Using (\ref{eq:124}) and \eqref{eq:3.57}, we see that (\ref{eq:113C}) exactly matches
with (\ref{eq:123}). For late time $t\ra\infty$, we have $C(t,-t)=2G_{lr}(t,-t)\sin\mu NG_{lr}(0,0)$,
which is the universal late-time behavior of ``quantum traversable
wormhole'' due to the interference effect \cite{Gao:2018yzk}, in which OTOCs
simply vanish and TOCs factorize.

\begin{figure}
\begin{centering}
\par\end{centering}
\begin{centering}
\subfloat[$\protect\b=0.95$\label{fig:9b}]{\begin{centering}
\includegraphics[height=3cm]{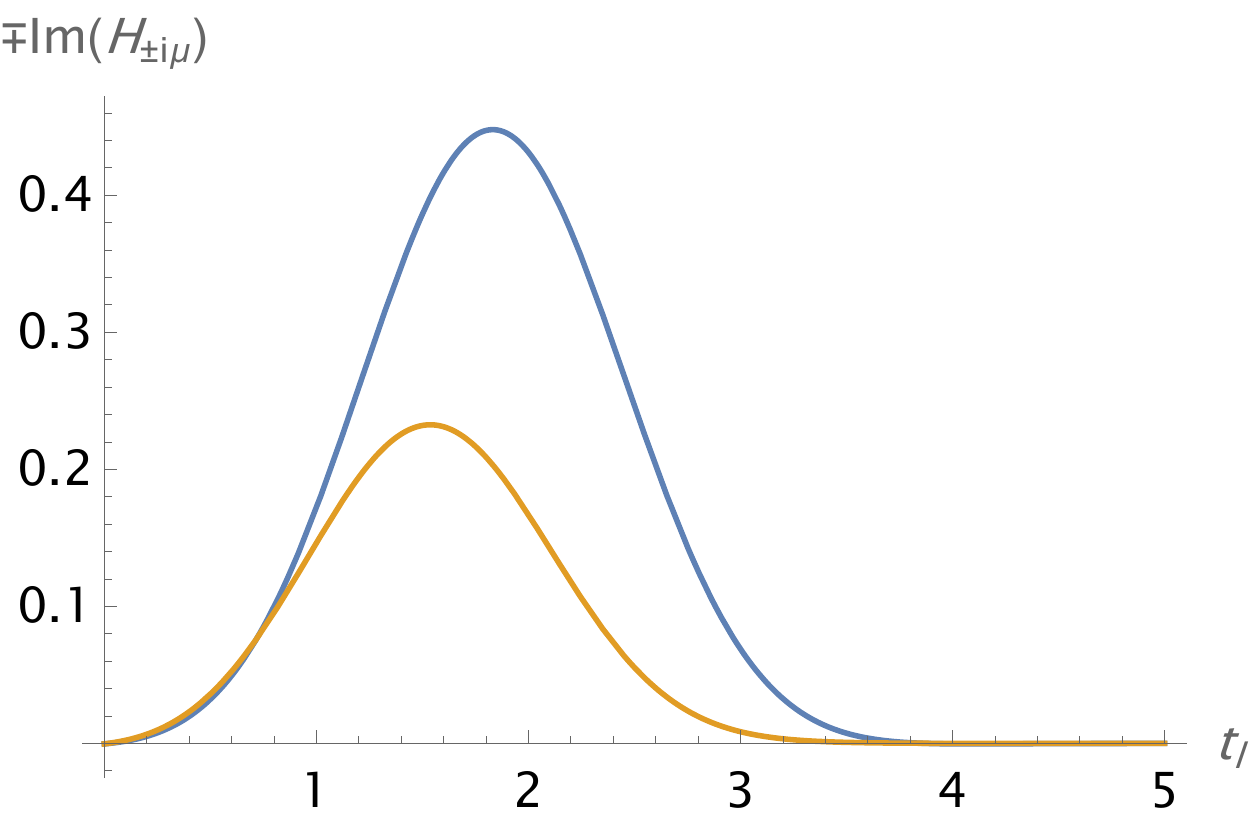}
\par\end{centering}
}\subfloat[$\protect\b=0.1$\label{fig:9c}]{\begin{centering}
\includegraphics[height=3cm]{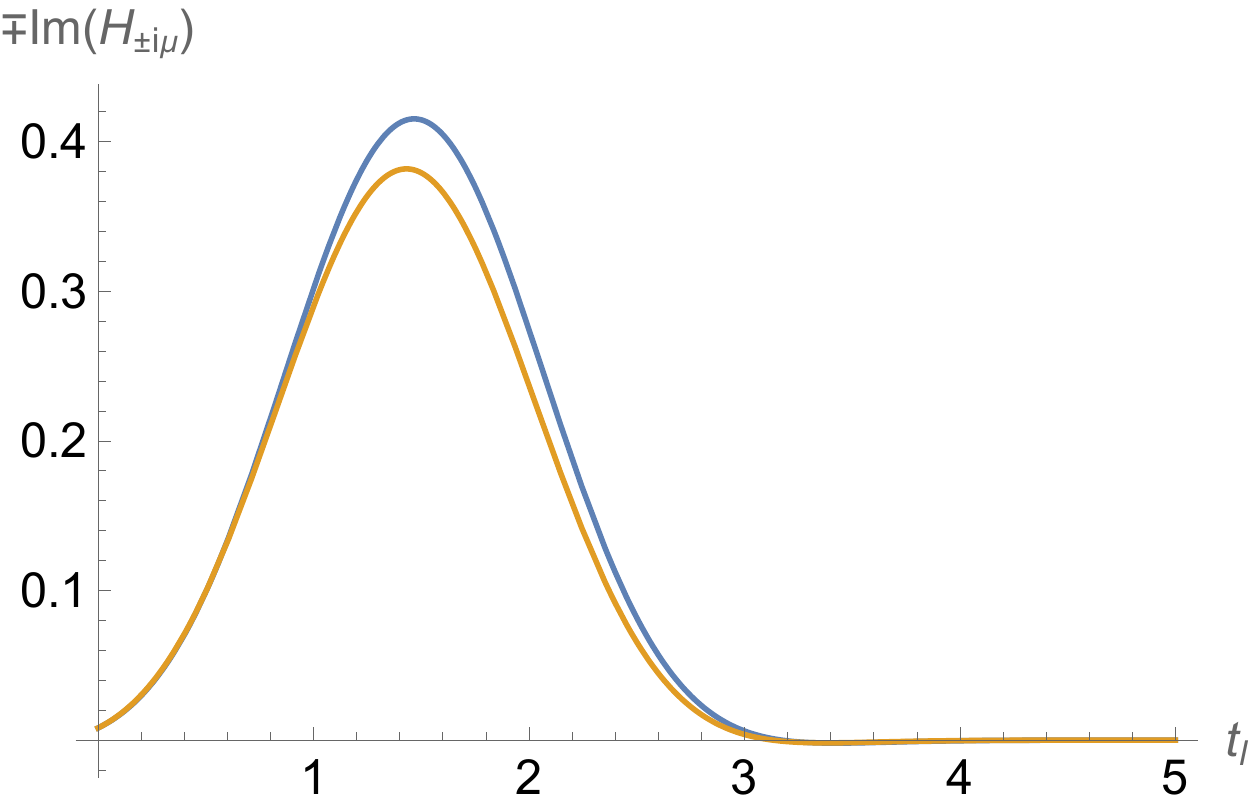}
\par\end{centering}
}
\par\end{centering}
\caption{The optimized $\max|\Im H_{i\mu}|-\max|\Im H_{-i\mu}|$ for the scale
$t_{l}\sim-t_{r}\sim O(1)$. The blue curves are $-\Im H_{i\mu}$
and the yellow curves are $\Im H_{-i\mu}$. (a) Lower temperature
$\protect\b=0.95$, the optimal $\mu=0.0468\pi$ and the injection time
is at $t_{r}=-1.791$. (d) High temperature $\protect\b=0.1$, the
optimal $\mu=0.0509\pi$ and the injection time is at $t_{r}=-1.507$.
\label{fig:The-optimized-}}
\end{figure}

For the short time scale $t_{l}\sim-t_{r}\sim O(1)$,
there is an obvious asymmetry for the flip $\mu\ra-\mu$ from (\ref{eq:131}).
To have a good exhibition of the sign effect, we can optimize the
peak difference between $|\Im H_{i\mu}|$ and $|\Im H_{-i\mu}|$ for
different temperatures in the range of $\mu\in[0,\pi/2]$ and $t_{r}<0$. The result is shown in Figure \ref{fig:The-optimized-}. From the
plots, we find that in high temperature $\b=0.1$, the sign difference
is not large. Then as we decrease the temperature to an intermediate
level $\b=0.95$ but still in non-spin glass phase, the sign difference becomes large. 
In the both cases, we see that positive $\mu$
leads to a higher peak in $|H_{i\mu}|$ while negative $\mu$ gives
a lower peak. Though this is qualitatively compatible with the analysis
in large $q$ SYK model \cite{Gao:2019nyj} that positive $\mu$ generates a negative
energy shockwave into a black hole, we would like to emphasize again
that the commuting SYK is not holographic. In particular, we do not
see strong suppression of $\Im H_{i\mu}$ for negative $\mu$, which was observed in the large $q$ SYK model in the low temperature limit \cite{Gao:2019nyj}. This means that for any nonzero $\mu$, there is always an $O(1)$ fidelity of teleportation. 

Note that the sign difference in $C$ in the short time scale is completely
different from the size-winding mechanism. Let us first recall why
size winding prefers a specific sign (and indeed the value) of $\mu$
for teleportation. For the time scale larger than the thermalization
scale, the exponential $e^{i\mu V}$ factorizes because it is a TOC
piece \cite{Maldacena:2017axo}
\begin{equation}
H_{i\mu}(t_{l},t_{r})\app-ie^{i\mu\avg V}\avg{0|\r_{r}^{1/2}\psi_{j}^{l}(t_{l})e^{-i\mu V}\psi_{j}^{r}(t_{r})\r_{r}^{1/2}|0}\label{eq:121}
\end{equation}
where $\avg V$ is the expectation value in the thermofield double
state. Given the size winding assumption $\psi_{j}^{r}(t)\r_{r}^{1/2}=\f 1{\sqrt{2}}\sum_{I}|c_{I}(t)|e^{i(a_{1}(t)+a_{2}(t)|I|)}\G_{I}$,
(\ref{eq:121}) becomes
\begin{equation}
H_{i\mu}(t_{l},t_{r})\app\f 12e^{i\mu(\avg V+N/2)}\sum_{I}|c_{I}(t_{r})c_{I}(-t_{l})|e^{i(a_{1}(t_{r})+a_{1}(-t_{l})+(a_{2}(t_{r})+a_{2}(-t_{l})-\mu)|I|)}\label{eq:121-1}
\end{equation}
If we take $t_{l}=-t_{r}$ and $\mu=2a_{2}(t_{r})$, we have $H_{i\mu}=\f 12e^{i\t}$
for a pure phase $\t$. This guarantees the success of teleportation
only for the specific sign (and also the value as a function of $t_{r}$)
of $2a_{2}(t_{r})$ with an $O(1)$ imaginary part of $H_{i\mu}$.
If we choose the opposite sign of $\mu$, each term in (\ref{eq:121-1})
will have a nonzero phase and the sum will be highly suppressed by
the cancellation among terms, which leads to the failure of teleportation.
From (\ref{eq:81}) we know that the slope to $n$ of the phase is
$O(t/N)$ and $\mu$ should have been the same magnitude if the sign
difference of $\mu$ were caused by size winding. However, in the
short time scale and Figure \ref{fig:The-optimized-}, we have taken
$N\ra\infty$ but the optimal $\mu$ is still of order one.

To understand the sign difference of $\mu$ for finite $\b$, let us first consider the high temperature limit $\b\ra0$ in (\ref{eq:131}). We have
\begin{equation}
H_{i\mu}(t_{l},t_{r})=\f 12e^{-(t_{l}+t_{r})^{2}+4t_{l}t_{r}\sin^{2}\mu}e^{i(2t_{l}t_{r}\sin2\mu-\mu)}
\end{equation}
which changes to its complex conjugate under $\mu\ra-\mu$ and is
consistent with Figure \ref{fig:9c}. As we decrease the temperature from $\b = 0$, consider the small $\mu$ expansion in \eqref{eq:131} with $t_l\sim-t_r$. The second line of \eqref{eq:131} in leading order of $\mu$ is $-i\mu \exp(e^{-\b^2/4}(2\mu)(-2i t_l^2))$, which does not contribute to the sign difference of $\mu$ of $\Im H_{i\mu}$. For the first line of \eqref{eq:131}, in leading order of small $\b$, we can rewrite it approximately as
\be  
H_{i\mu}\app \f 1 2 e^{i\mu(-4t_l^2-2i\b t_l)}\app -i \avg{\psi^l_j(t_l)\psi^r_j(-t_l)}e^{i\mu\avg{[V,\psi^l_j(t_l)]\psi^r_j(-t_l)}/\avg{\psi^l_j(t_l)\psi^r_j(-t_l)}+i\mu} \label{eq:4.34}
\ee
where the $\avg{\cdots}$ is evaluated in the thermofield double state at high temperature. In this equation, we have $\avg{\psi^l_j(t_l)\psi^r_j(-t_l)}\app i/2$ (for $\b=0$) and the exponent is the difference between a pair of OTOC and TOC
\begin{align}
\avg{[V,\psi^l_j(t_l)]\psi^r_j(-t_l)}/\avg{\psi^l_j(t_l)\psi^r_j(-t_l)}&\app \sum_{i=1}^N-\f 1 2-2 \overline{W_{ji}}(it_l+\b/2,\b/2,it_l,0) \nn\\
&\app -4t_l^2-2i\b t_l-1 \label{eq:4.35}
\end{align}
where $\overline{W_{ji}}$ is the OTOC in \eqref{eq:2.28}, in which we expanded the exponent in leading order. The exponentiation of the difference between TOC and OTOC in \eqref{eq:4.34} also exists in the regenesis phenomenon in 2d CFT \cite{Gao:2018yzk}, though the latter is valid in a wider regime. The commutator \eqref{eq:4.35} can be roughly understood as the ``scattering" between $V$ and the pair of $\psi_j^l(t_l)\psi_j^r(-t_l)$. From \eqref{eq:4.34}, we see that the term linear in $\b$  in the exponent contributes to the sign difference of $\mu$ because it gives an enhancement for positive $\mu$ and a suppression for negative $\mu$. As we discussed below \eqref{eq:thG}, this linear in $\b$ term is exactly the effective characteristic frequency $\b\mJ^2$ by thermalization and reflects the underlying integrability of the commuting SYK model. As we decrease the temperature, the relative enhancement/suppression is stronger because the effective characteristic frequency is proportional to $\b$. Though this analysis is for small $\b$, it is compatible with the observation in Figure \ref{fig:The-optimized-} for finite temperatures as long as we are still in non spin-glass phase. It is interesting to examine how the sign difference behaves if we drop the temperature below $T_c$ and enter the spin-glass phase. We leave this investigation to future work.

We can also compare the $O(1)$ time and high temperature result with the peaked-size formula \eqref{eq:122}. In early times, the size is $\mS\app 4t^2$ by \eqref{eq:3.57}. Comparing \eqref{eq:122} with \eqref{eq:4.34} at high temperature, it is easy to see that the quadratic term in the exponent in \eqref{eq:4.34} is exactly the size $\mS$. The piece beyond the peaked-size mechanism in this regime completely comes from the effective characteristic frequency $\b\mJ^2$.

\begin{figure}
\begin{centering}
\subfloat[$\protect\b=4$\label{fig:8a}]{\begin{centering}
\includegraphics[height=3cm]{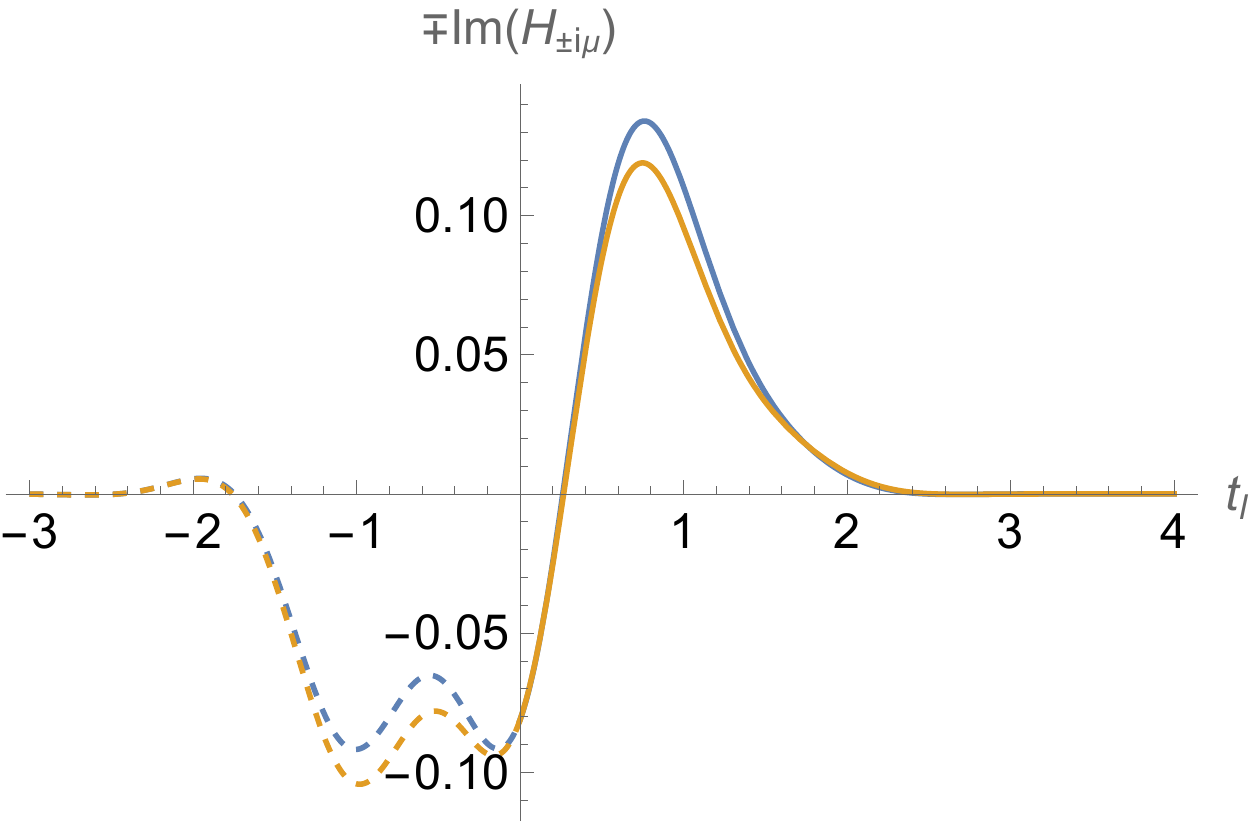}
\par\end{centering}
}\subfloat[$\protect\b=2.5$\label{fig:8b}]{\begin{centering}
\includegraphics[height=3cm]{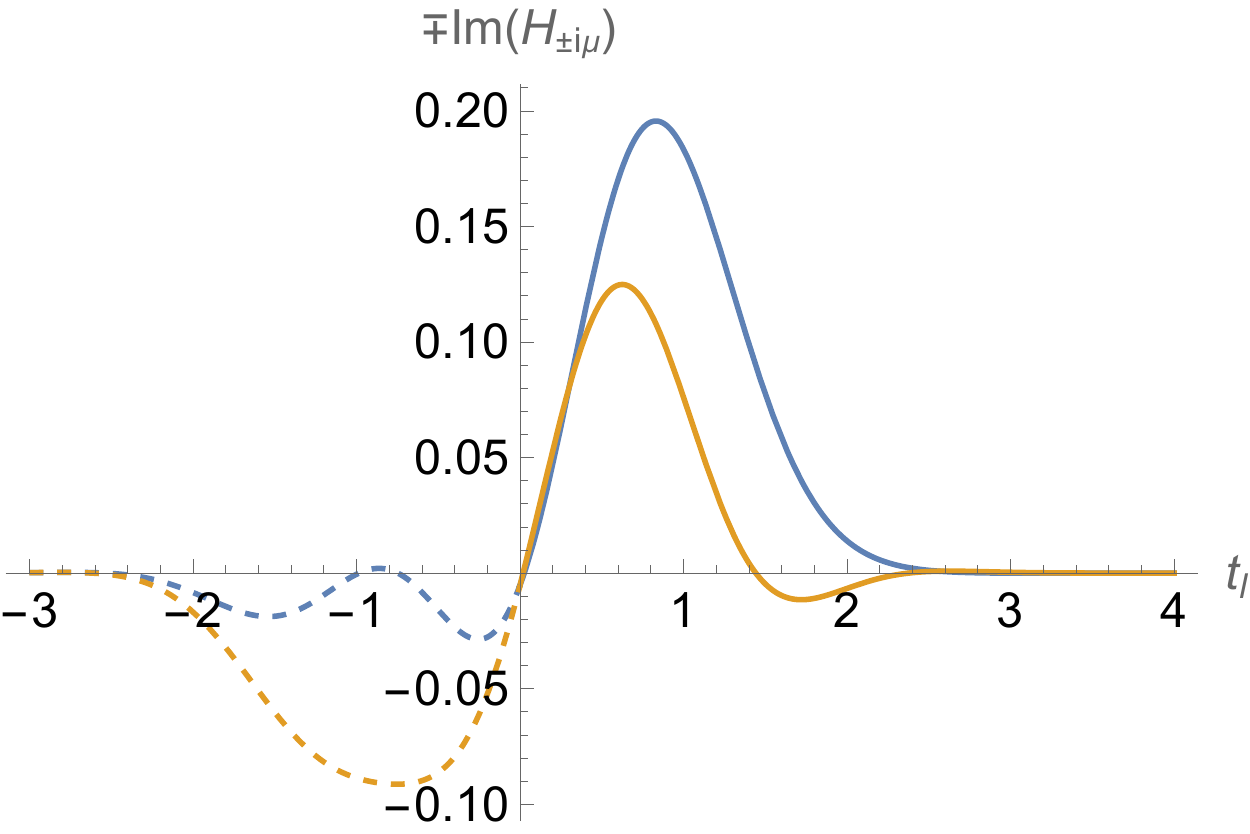}
\par\end{centering}
}\\
\subfloat[$\protect\b=1$\label{fig:8c}]{\begin{centering}
\includegraphics[height=3cm]{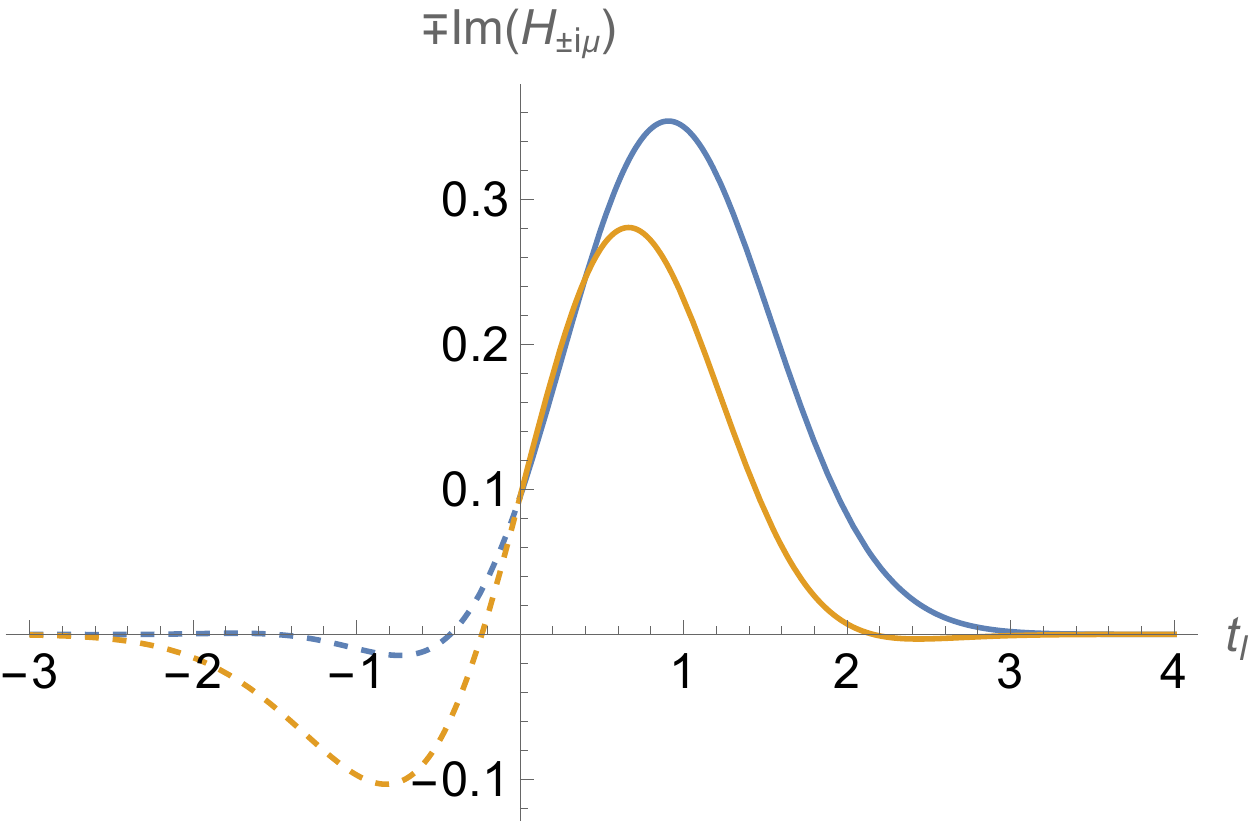}
\par\end{centering}
}\subfloat[$\protect\b=0.1$\label{fig:8d}]{\begin{centering}
\includegraphics[height=3cm]{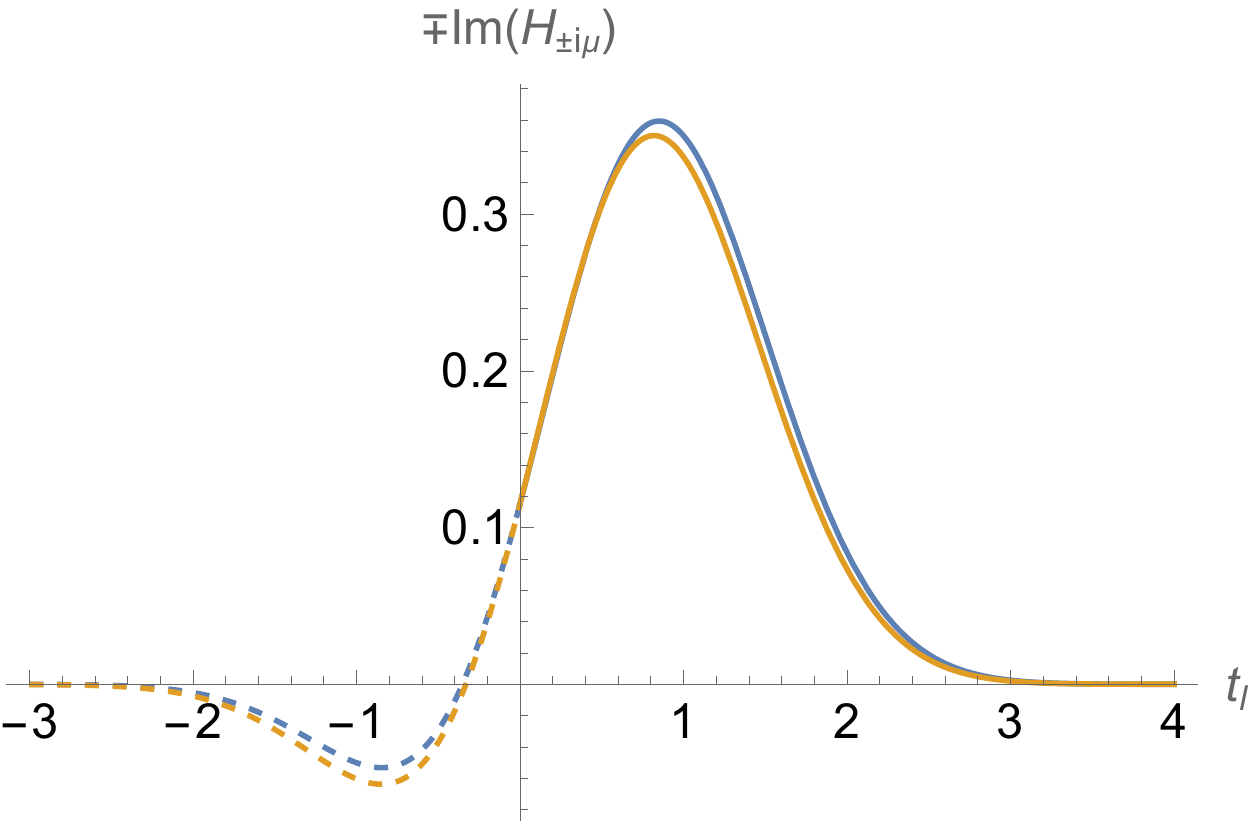}
\par\end{centering}
}
\par\end{centering}
\caption{The optimized $\max|\Im H_{i\mu}|-\max|\Im H_{-i\mu}|$ for $N=8$.
The blue curves are $-\Im H_{i\mu}$ and the yellow curves are $\Im H_{-i\mu}$.
The dashed lines are extended plots for $t_{l}<0$, which is an unphysical
regime. (a) Even-lower temperature $\protect\b=4$, the optimal $\mu=0.221\pi$
and the injection time is at $t_{r}=-0.474$. (b) Low temperature
$\protect\b=2.5$, the optimal $\mu=0.195\pi$ and the injection time
is at $t_{r}=-0.614$. (c) Intermediate temperature $\protect\b=1$,
the optimal $\mu=0.139\pi$ and the injection time is at $t_{r}=-0.720$.
(d) High temperature $\protect\b=0.1$, the optimal $\mu=0.131\pi$
and the injection time is at $t_{r}=-0.742$. \label{fig:The-optimized--1}}
\end{figure}

On the other hand, one might be confused that why the short time scale does not follow
the peaked-size mechanism given that $\d\mS/\mS\sim O(1/\sqrt{N})$
even in early times. Indeed, the simple criterion $\d\mS/\mS\ll1$
is not always enough to guarantee peaked-size teleportation \cite{Schuster:2021uvg}
and sometimes we need a much finer criterion. For the current case,
a simple explanation is that the peaked-size teleportation in \cite{Schuster:2021uvg}
assumes $\mu\sim O(1/K)$ for $K$ being a large number ($K$ could
equal to $N$). Then the size fluctuation $\d\mS\sim O(\sqrt{K})$
only affects the phase of $e^{i\mu\mS}$ by a negligible $O(1/\sqrt{K})$
amount. However, when we choose $\mu\sim O(1)$, the size fluctuation
affects the phase by a large $O(\sqrt{K})$ amount and the peaked-size
criterion must be much tighter.

As a comparison, we can also study the sign difference effect for small $N$
systems, which might be related to the recent simulation of traversable
wormhole dynamics on the Sycamore quantum processor \cite{Jafferis:2022crx}.
For small $N$ the saddle approximation is poorly behaved, but the
explicit expression of $H_{i\mu}$ can be easily written down by expanding
the $N/2-1$ power term in both (\ref{eq:h1}) and (\ref{eq:h2}).
We take $N=8$ and the numerics are straightforward. The optimized
curves of $|\Im H_{i\mu}|-$$|\Im H_{-i\mu}|$ for a few choices
of temperatures are shown in Figure \ref{fig:The-optimized--1}.

For small $N$, different time scales are not separable and we only
need to consider $t_{l,r}\sim O(1)$. Surprisingly, we find that the
behavior of small $N$ in Figure \ref{fig:The-optimized--1} is qualitatively
similar to the behavior of large $N$ in the short time scale in Figure
\ref{fig:The-optimized-}. In all temperatures checked in Figure \ref{fig:The-optimized--1},
positive $\mu$ leads to a higher maximum value than the negative
$-\mu$. It is noteworthy that the dependence of the sign
difference of $\mu$ on temperature follows a similar pattern as Figure
\ref{fig:The-optimized-}, where we when cool down the system from high temperature, the sign difference becomes more visible until some critical temperature. If we continue to decrease the temperature, the sign difference is again diminished. This critical temperature seems to be the same order as the critical temperature $T_c=1$ for spin glass phase in the large $N$ case. Though we do not have a clean analysis of effective characteristic frequency for finite $N$ case, this similarity of sign difference suggest that the thermalization process of the commuting SYK model should play a crucial role. Moreover, as we drop the temperature
from high to low, $\Im H_{i\mu}$ develops more and more peaks. To show the generation of new peaks, we
extend the plot to an unphysical negative $t_{l}$ regime in Figure
\ref{fig:The-optimized--1}. For very large $\b$ if we inject the signal
around or earlier than $-\b$, the plot of $\Im H_{i\mu}$ will have
many wiggles, and the sign difference is not visible (which is not plotted here because the self-average result may not be reliable for too low temperatures).

\subsection{Peak location and signal ordering}

Another feature of a holographic model that is dual to a semiclassical
traversable wormhole is the signal ordering \cite{Gao:2019nyj}. If two signals
are sent consecutively from the right side early enough to go through
a traversable wormhole, they should be received on the left side with
the same ordering of signals due to the smooth geometry in the traversable
wormhole. This signal ordering preserving feature is unusual because
the thermofield double state of two entangled black holes has maximal
correlation at $t_{l}=-t_{r}$ due to the boost symmetry. This maximal
correlation at opposite times on two sides indicates that the signal
received time $t_{l}$ is approximately around $-t_{r}$, which leads
to opposite signal ordering before scrambling time. Only when we send
the signal in the time window for the semiclassical traversable wormhole
throat, which is around scrambling time, the effects of two-sided
instant coupling at $t=0$ generates a large enough backreaction that
alters the signal ordering. This feature was verified in the large
$q$ SYK model in low temperatures \cite{Gao:2019nyj}. 

\begin{figure}
\begin{centering}
\subfloat[$\protect\b=0.95$\label{fig:10b}]{\begin{centering}
\includegraphics[width=3cm]{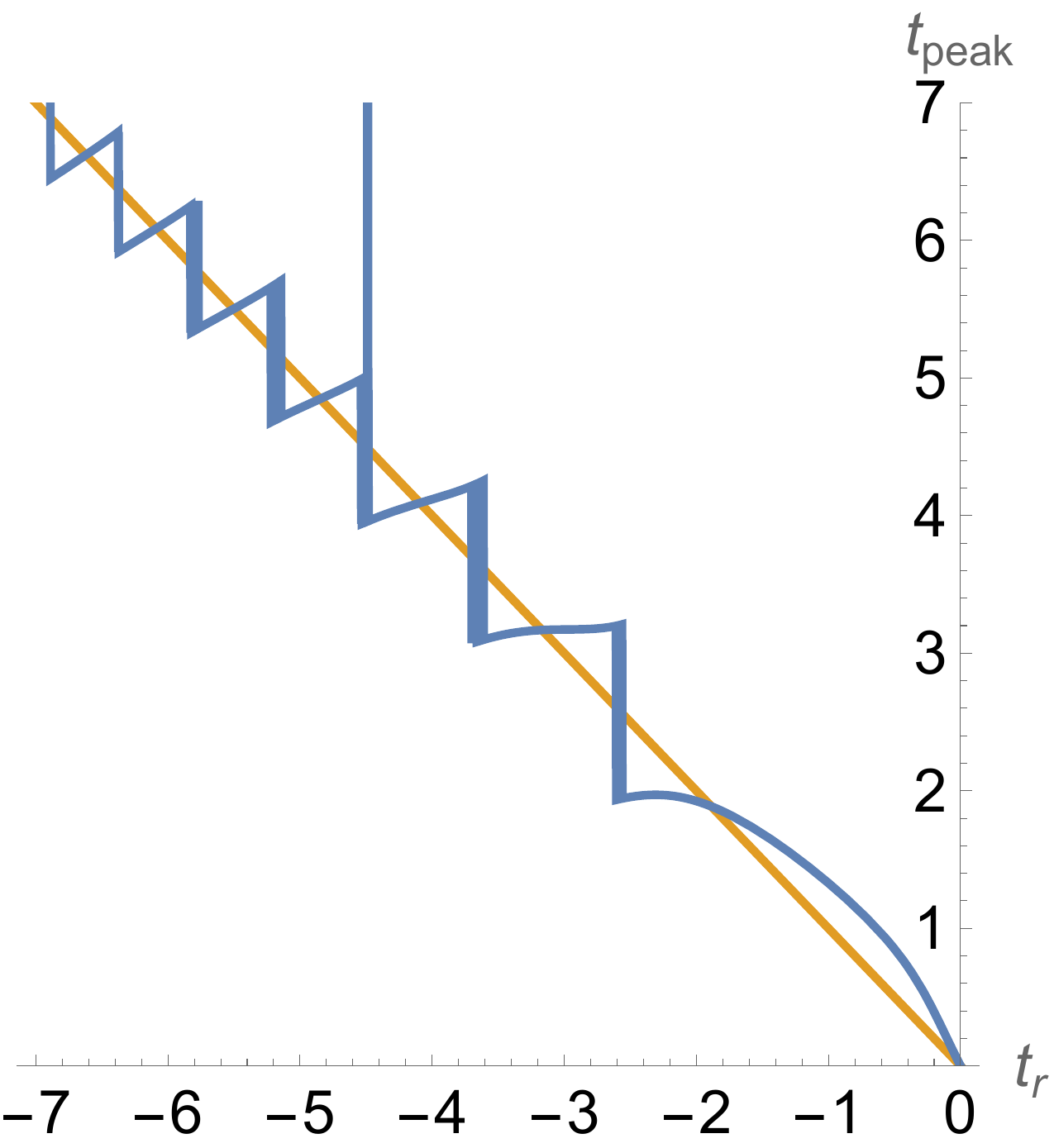}
\par\end{centering}
}\subfloat[$\protect\b=0.1$\label{fig:10c}]{\begin{centering}
\includegraphics[width=3cm]{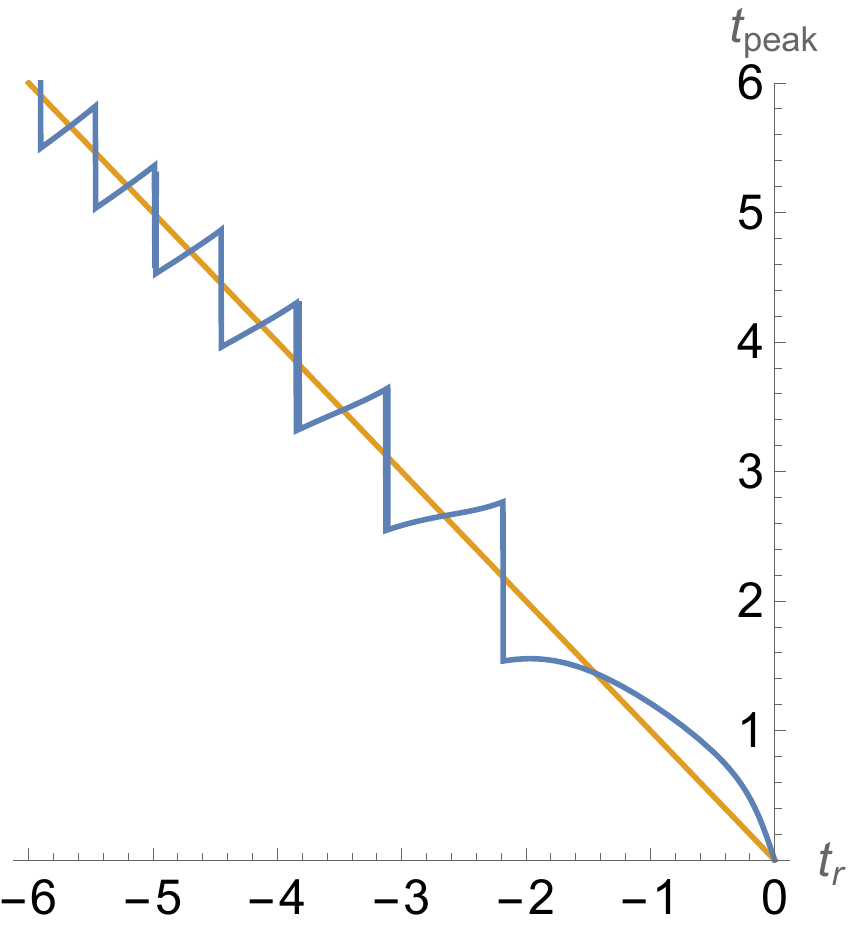}
\par\end{centering}
}\\
\par\end{centering}
\begin{centering}
\subfloat[$t_{r}\sim-1.6$\label{fig:10d}]{\begin{centering}
\includegraphics[width=3cm]{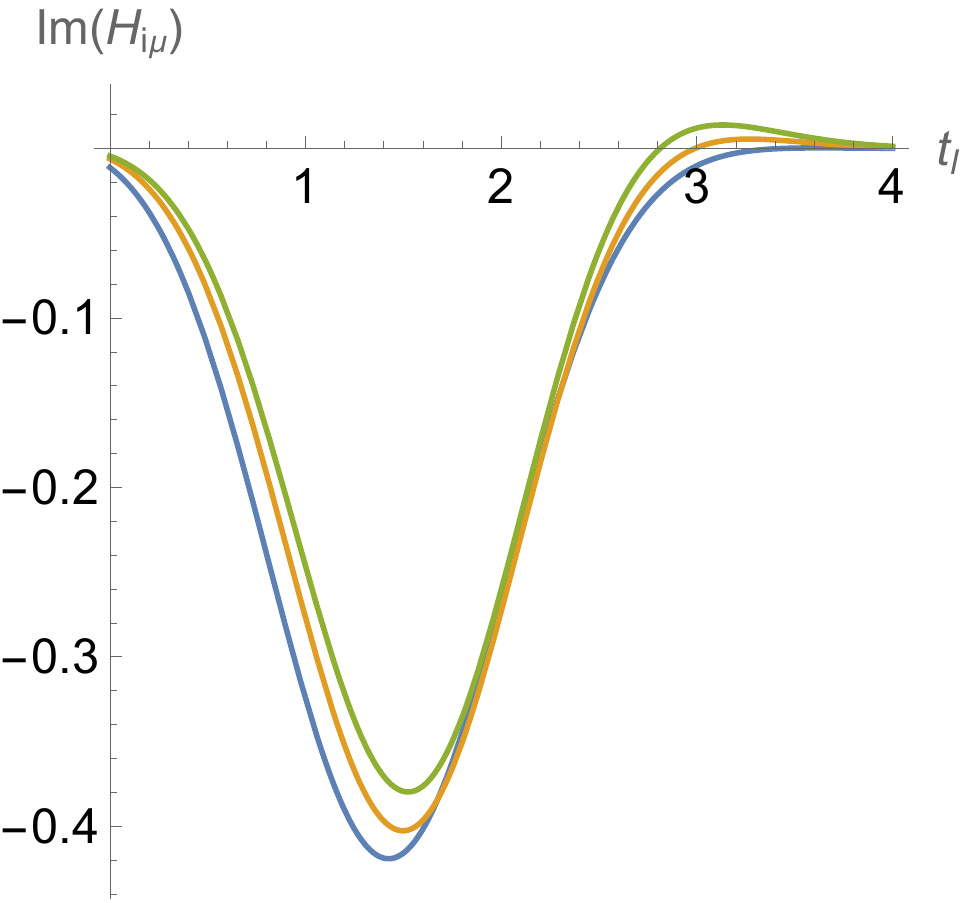}
\par\end{centering}
}\subfloat[$t_{r}\sim-2.2$\label{fig:10e}]{\begin{centering}
\includegraphics[width=3cm]{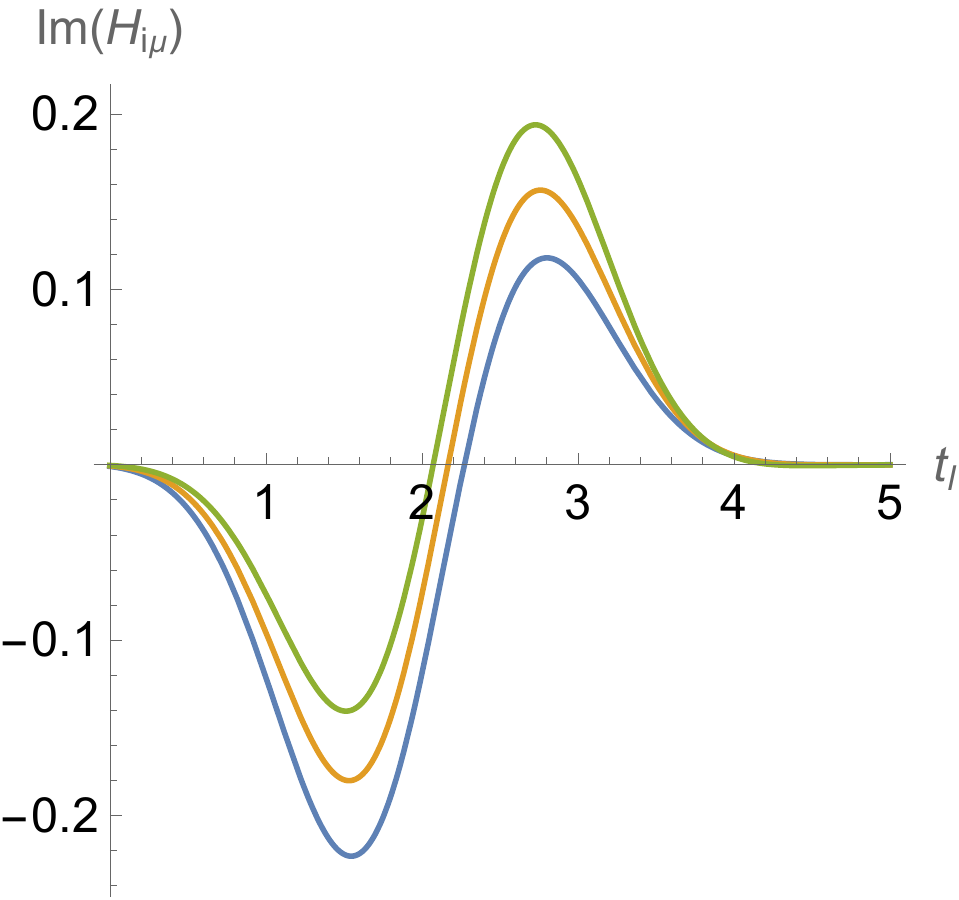}
\par\end{centering}
}\subfloat[$t_{r}\sim-2.8$\label{fig:10f}]{\begin{centering}
\includegraphics[width=3cm]{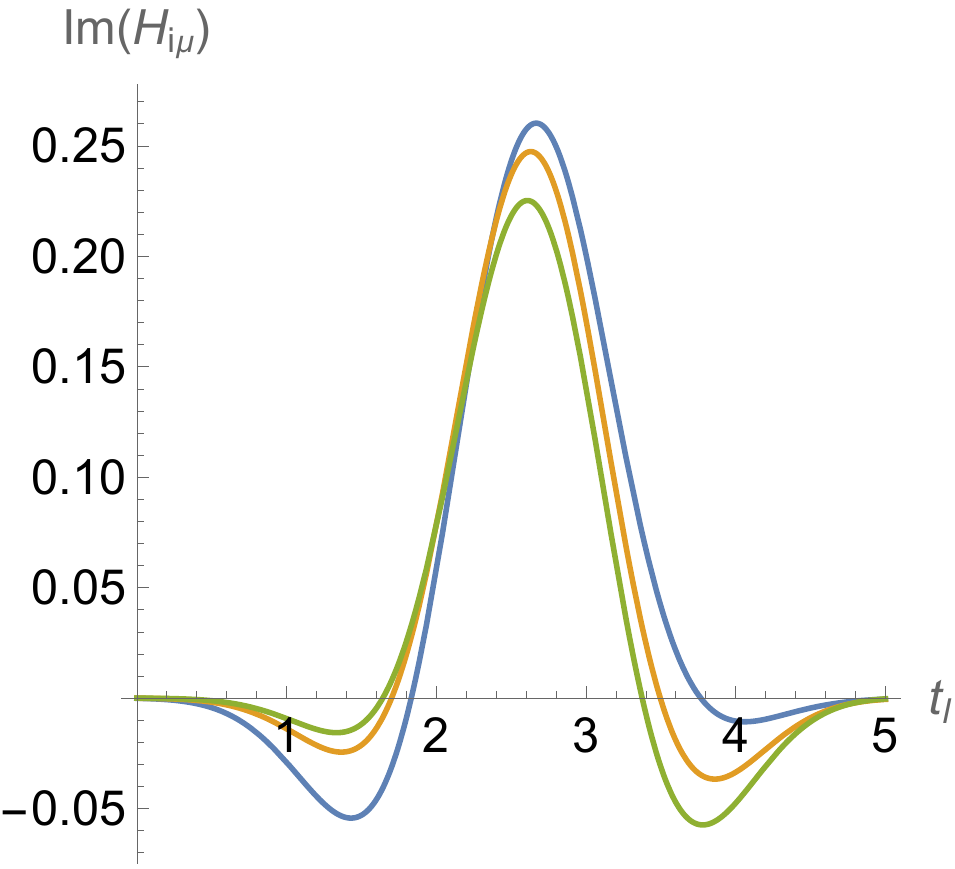}
\par\end{centering}
}
\par\end{centering}
\caption{Signal ordering for the scale $t_{l}\sim-t_{r}\sim O(1)$ in large
$N$ limit. The first row is the signal receiving time $t_{peak}$
as a function of signal sending time $t_{r}$ for different temperatures.
The yellow straight line is the reference $t_{peak}=-t_{r}$. The
choices of $\mu$ are the same as Figure \ref{fig:The-optimized-}
for both temperatures. The second row is the plot of $\Im H_{i\mu}$
as a function of $t_{l}$ for three consecutive signals sent around
different times $t_{r}$ at high temperature $\protect\b=0.1$. In
each figure, the blue, yellow and green curves are for the latest,
middle and earliest signals respectively around $t_{r}$.\label{fig:The-first-row}}
\end{figure}
It is interesting that for the non-holographic commuting SYK model,
we could also find some time regime, in which the signal ordering
is preserved. We define the signal receiving time as the $t_{l}=t_{peak}$
at the highest peak of $H_{i\mu}$. Let us first take the large $N$
case. For short time scale, the solution to $\del_{t_{l}}H_{i\mu}=0$
for (\ref{eq:131}) can be numerically solved and is shown in Figure
\ref{fig:The-first-row}. The first row is the signal receiving time
$t_{peak}$ as a function of signal sending time $t_{r}$ for different
temperatures. By definition, if the slope of the function $t_{peak}(t_{r})$
is positive, this means that the signal ordering is preserved at $t_{r}$.
We also draw a yellow straight line as the reference $t_{peak}=-t_{r}$
to show the asymptotic reverse signal ordering. For convenience, we
choose $\mu$ to be the same as Figure \ref{fig:The-optimized-} for
all three temperatures. 

A few interesting features can be readily observed from the first
row of Figure \ref{fig:The-first-row}. First, the function $t_{peak}(t_{r})$
is split into many intervals, in which the signal orderings could
be different. Second, as we increase the temperature, we see more
times preserving signal ordering. But in all temperatures, the first
time interval with the largest $t_{r}$ does not preserve the signal ordering.
For the intermediate ($\b=0.95$) and high ($\b=0.1$) temperatures, starting
from the second largest time interval of $t_{r}$, the signal ordering
is preserved. Third, the function $t_{peak}(t_{r})$ is ambiguous
between two neighboring intervals. This is because at these times
multiple comparable peaks emerge and the highest peak has a discontinuous
jump. 

To have an intuitive picture of the signal peak, we draw the second
row of Figure \ref{fig:The-first-row} for $\Im H_{i\mu}$ as a function
of $t_{l}$ for three consecutive signals sent around different times
$t_{r}$ at high temperature $\b=0.1$. In each figure, the blue,
yellow, and green curves are for the latest, middle and earliest signals
respectively around $t_{r}$. In Figure \ref{fig:10d}, the signal
sending time $t_{r}\sim-1.6$ is chosen from the first interval of
Figure \ref{fig:10c}, which shows reverse signal ordering. In Figure
\ref{fig:10e}, the signal sending time $t_{r}\sim-2.2$ is chosen
from the bordering regime between the first and second interval of Figure
\ref{fig:10c}, which shows comparable two peaks. In Figure \ref{fig:10f},
the signal sending time $t_{r}\sim-2.8$ is chosen from the second
interval of Figure \ref{fig:10c}, which shows signal ordering is
preserved. 

This observation is indeed similar to the semiclassical traversable
wormhole in the large $q$ SYK model in low temperature in the sense that
the signal ordering will be preserved only when you send it early
enough. However, the transition time occurs at scrambling time for
large $q$ SYK with $\mu\sim O(1/N)$ but only at $O(1)$ time for
the commuting SYK model with a much large $\mu\sim O(1)$. Furthermore,
the preservation of signal ordering occurs in many intervals and only
for intermediate and high temperatures in the commuting SYK model.

For the long time scale, we always have reverse signal ordering due
to the simple formula (\ref{eq:113C}). Let us define
\begin{equation}
t_{l}=-\sqrt{N/2}T+t,\quad t_{r}=\sqrt{N/2}T
\end{equation}
which in large $N$ limit leads to
\begin{equation}
C=e^{-\b^{2}/4-t^{2}}\sin\left[\f{\mu_{0}}2e^{-\b^{2}/4}\left(1-e^{-4T^{2}}\right)\right]\label{eq:135}
\end{equation}
for a given $T<0$. The signal receiving time is the peak as a function
of $t$, which is always at $t=0$. From (\ref{eq:135}), we can infer
that the peak is in the form of Gaussian. If $\mu_{0}$ is large enough,
the sending time is split into a few intervals that are separated
by $T_{n}=-\f 12\sqrt{\log\f 1{1-2\pi ne^{\b^{2}/4}/\mu_{0}}}$ for
$n\in\Z_{+}$, at which the signal vanishes $C=0$. The threshold
of $\mu_{0}$ for the existence of such multiple intervals is $\mu_{0}\geq2\pi e^{\b^{2}/4}$.

\begin{figure}
\begin{centering}
\subfloat[\label{fig:12a}]{\begin{centering}
\includegraphics[height=3cm]{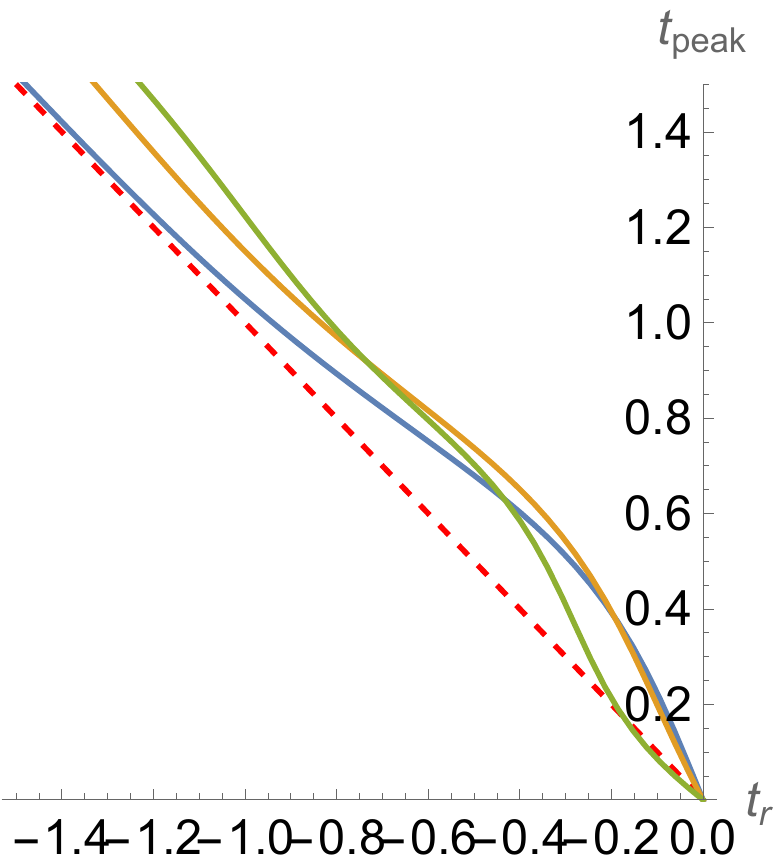}
\par\end{centering}
}\subfloat[\label{fig:12b}]{\begin{centering}
\includegraphics[height=3cm]{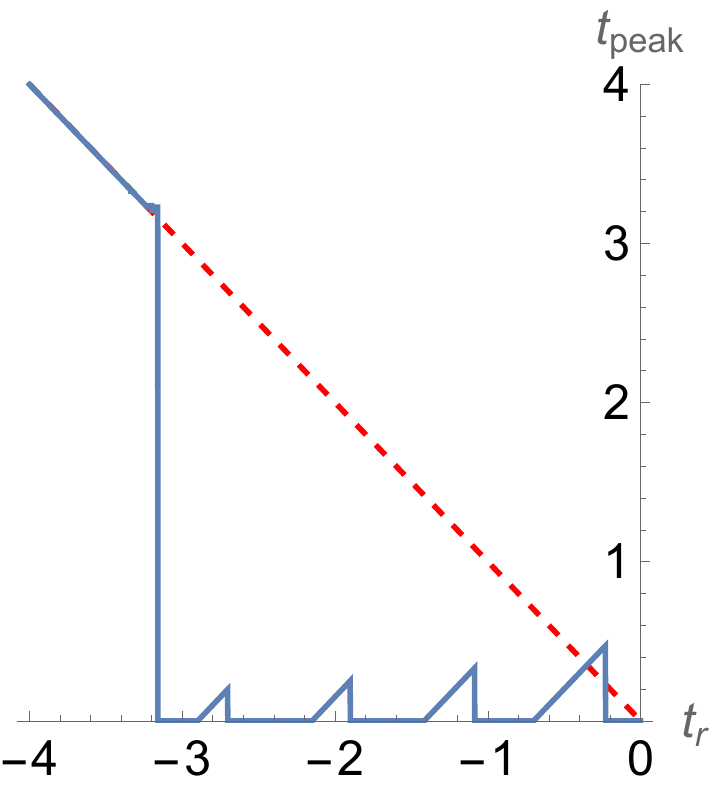}
\par\end{centering}
}\subfloat[$t_{r}\sim-0.35$\label{fig:12c}]{\begin{centering}
\includegraphics[height=3cm]{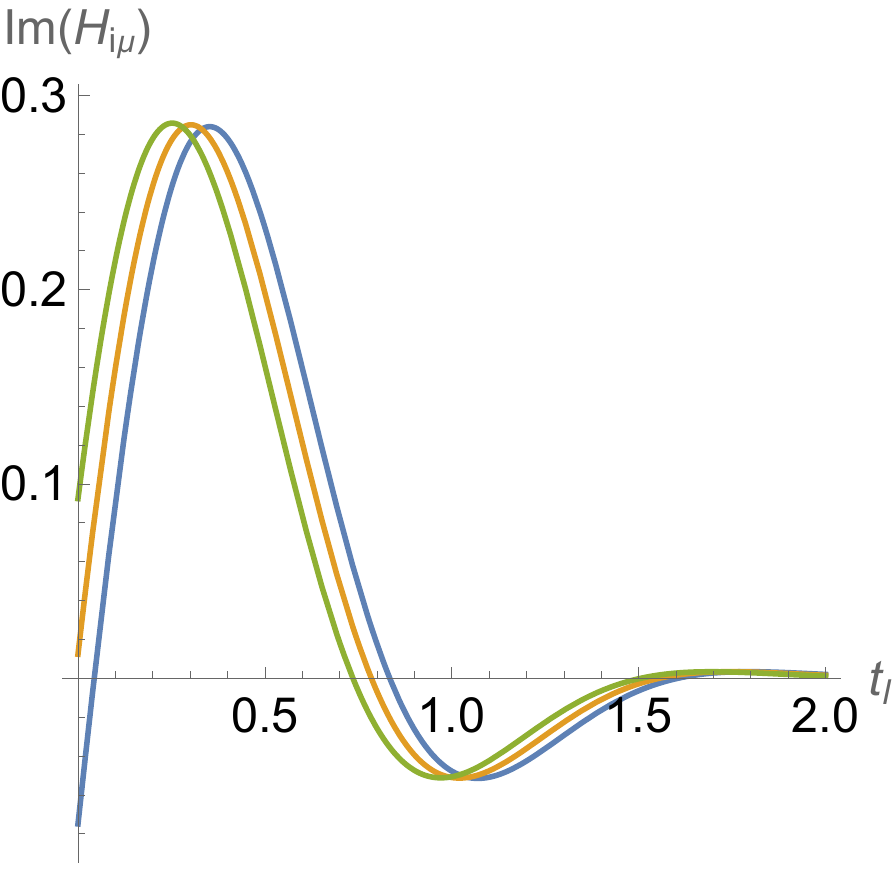}
\par\end{centering}
}\subfloat[$t_{r}\sim-3.35$\label{fig:12d}]{\begin{centering}
\includegraphics[height=3cm]{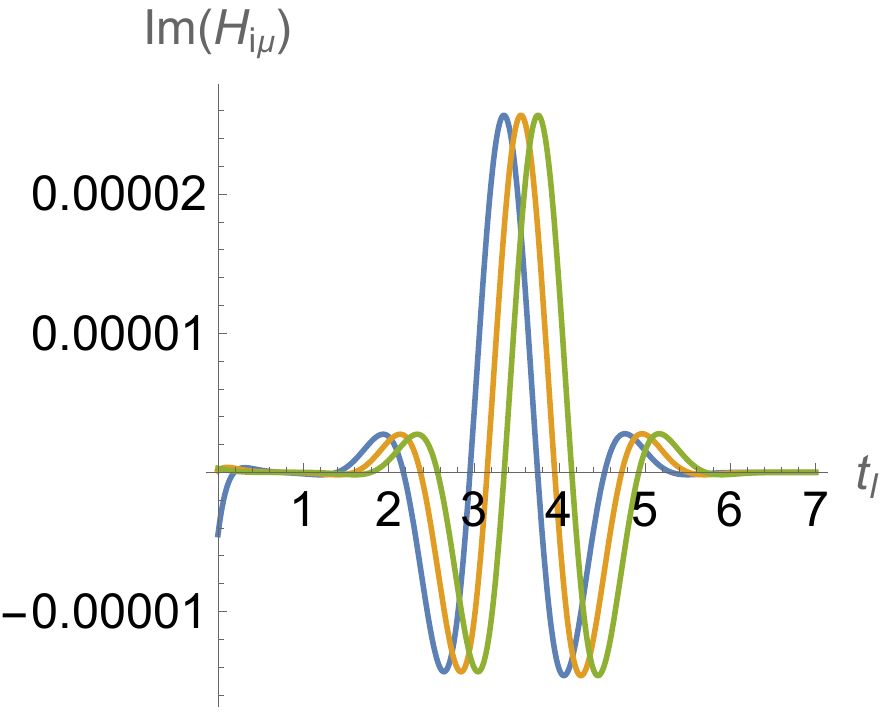}
\par\end{centering}
}
\par\end{centering}
\caption{Signal ordering for $N=8$. (a) The signal receiving time $t_{peak}$
as a function of signal sending time $t_{r}$ for different temperatures.
The blue, yellow, and green curves are for $\protect\b=0.1,1,2.5$
respectively. The red dashed line is the reference $t_{peak}=-t_{r}$.
(b) Choose large $\mu=0.473\pi$ and $\protect\b=4$ leads to signal
ordering preserved in early times. (c) $\Im H_{i\mu}$ in the signal
ordering preserved regime. (d) $\Im H_{i\mu}$ in the signal ordering
reversed regime. In both (c) and (d), blue, yellow, and green curves
are for the latest, middle and earliest signals in the consecutive
sequence of signals sent around $t_{r}$ with the same $\protect\b$
and $\mu$ as (b).}
\end{figure}
As a comparison, let us check the signal ordering for small $N$.
We again take $N=8$ and the same parameters in Figure \ref{fig:The-optimized--1}.
The result is in Figure \ref{fig:12a}, which shows that the signal
ordering is reversed for all high ($\b=0.1)$, intermediate ($\b=1$)
and low temperatures ($\b=2.5$). This is compatible with the observation
in \cite{Jafferis:2022crx}  that a one-time two-sided coupling is not strong
enough to preserve the signal ordering for a learned commuting Hamiltonian
with small $N$. While \cite{Jafferis:2022crx} finely tuned the model by trotterized
the one-time coupling into three times to achieve the preservation
of signal ordering in some time range, here we can instead tune a
large enough $\mu$ or $\b$ to see a similar effect. Choosing $\mu$
close to $\pi/2$ and large $\b$, we find that the signal ordering
is preserved in early times as shown in Figure \ref{fig:12b}. An
illustration of $\Im H_{i\mu}$ in this regime for three consecutive
signals sent around $t_{r}\sim-0.35$ is given by Figure \ref{fig:12c}.
However, such a regime is smaller than the thermal scale and quite transient,
which is in essence quite different from the case for large $N$ and the
short time scale case in Figure \ref{fig:The-first-row}. The preservation
of signal ordering is also piecewise in Figure \ref{fig:12b} because
there are multiple peaks competing as we send a signal toward earlier
times. Roughly after the thermal scale $\b$, we see from \ref{fig:12b}
that the signal ordering is reversed and obeys $t_{peak}=-t_{r}$
quite well. In this regime, we plot $\Im H_{i\mu}$ for three consecutive
signals sent around $t_{r}\sim-3.35$ in Figure \ref{fig:12d} as
an example.

\section{Conclusion and discussion\label{sec:Conclusion-and-discussion}}

In this work, we studied the large $N$ limit of a variant of the SYK
model whose Hamiltonian contains only commutative $q$-local interaction
terms. There are many different ways to define such a commuting SYK-like
Hamiltonian and we choose the simplest one by constructing each term
in the Hamiltonian by a $q/2$ product of commutative ingredients
$X_{i}=\psi_{2i-1}\psi_{2i}$ with a random coupling $\mJ_{I}$ that
is drawn from a Gaussian ensemble. Since this model has infinite numbers
of conserved charges in the large $N$ limit, it is integrable and completely
solvable. It turns out that this model is non-holographic by checking
its spectrum, two-point functions, and out-of-time-ordered correlators. Due to the large numbers of degrees of freedom in this model, an excitation $\psi_i$ thermalizes but in a way different from holographic models. In particular, its thermalization has two features: it has a non-holographic Gaussian tail decay in two-point function, and an oscillation with effective characteristic frequency $\b\mJ^2$, which is the typical energy of the excitation $\psi_i$ on a thermal state. The existence of this effective characteristic frequency reflects the underlying integrability of the commuting SYK model. This effective characteristic frequency also appears in four-point functions.

In spite of this, this model has some holography-like features, especially
the near-perfect size-winding in high temperatures. It has been shown
 and also briefly reviewed in Section \ref{subsec:Size-winding}
that size-winding is a feature of holographic models but it is not
known that any non-holographic models with size winding before this
work. Nevertheless, the size winding in the commuting SYK model is
quite different from the ordinary SYK model because it simultaneously
has peaked-size distribution, to which the teleportation in the long time
(scrambling) scale $t\gtrsim O(\sqrt{N})$ is attributed. Because
of this, we would like to call the non-holographic commuting SYK model
as \textit{pseudo-holographic}.

Applying the traversable wormhole teleportation protocol to this commuting
SYK model, we also find some similarities and differences with the
ordinary SYK model. For large $N$, we find two different behaviors
in short time scale $t\sim O(1)$ and long time scale $t\gtrsim O(\sqrt{N})$.
In the short time scale, the sign of $\mu$ matters more and more as we decrease from infinite temperature but still above critical temperature $T_c=1$. A positive $\mu$ leads to a stronger signal
transmission than the negative $-\mu$, which is compatible with
the expectation of holography. However, the mechanism for the sign difference is neither size-winding nor peaked-size teleportation. Though the size is peaked in the sense that $\d \mS/\mS\ll 1$, there is an important correction from the effective characteristic frequency $\b \mJ^2$ that leads to the relative enhancement/suppression for the sign choice of $\mu$. This is a special feature of the thermalization of this model. On
the other hand, the distinctions are also obvious that the time scale
is much shorter than scrambling time $O(\sqrt{N})$, and it does not
prohibit teleportation with negative $\mu$ in any temperatures, unlike
the large $q$ SYK model in low temperatures. For the long time scale,
we must take $\mu\sim O(1/N)$ and the sign of $\mu$ does not matter
for teleportation efficiency for all $T>T_c$. In this case, the teleportation undergoes the peaked-size mechanism.

Besides the sign of $\mu$, the preservation of signal ordering also
has interesting features. In the short time scale, there are many
intervals that preserve the signal ordering with $O(1)$ value of
$\mu$ turned on for intermediate and high temperatures above $T_c$. This is different
from the semiclassical picture of the large $q$ SYK model, which preserves
the signal ordering in low temperature with $\mu\sim O(1/N)$ and
in scrambling time scale. On the other hand, the commuting SYK model
always has reversed signal ordering in the long time scale. 

As a comparison, we also numerically studied the small $N$ case and
take $N=8$ as an example. Since $N$ is finite, there is only one time scale, in which the sign of $\mu$ matters as we drop from high temperature, which is qualitatively similar
to the large $N$ case. As we decrease the temperature further, the sign difference disappears. The signal ordering is mostly reversed
unless we tune large enough $\mu$ and $\b$, in which we see quite
transient reversed signal ordering in early times that are smaller than the thermal scale.

Below we end with a few discussions and comparisons with other related
works.

\paragraph{Geometric picture of size winding }

In the last section of \cite{Nezami:2021yaq}, the authors suggested that the
existence of size winding alone could potentially provide a geometrical
picture for teleportation because one can heuristically identify the
winding size distribution $Q_{n}(t)$ as the momentum wave function
of a one-dimensional particle. However, the non-holographic essence
of the commuting SYK model would give a strong constraint on the interpretation
of such a geometric picture if it could be defined explicitly. Another
aspect to note is that the argument for size winding in holographic
systems \cite{Nezami:2021yaq} is based on the near-AdS$_{2}$ isometry, which
exists near the horizon of a semiclassical near-extremal black hole,
which usually occurs in low temperature.\footnote{An $SL(2)$ symmetry still exists for large
$q$ SYK model in finite temperature. But the bulk dual of this $SL(2)$
is not well understood.} But in the commuting SYK model, the size winding occurs at high temperatures
and becomes damped when we decrease the temperature. This suggests that the size winding of the two types is from two different microscopic origins. This is a very interesting direction that we leave for future
investigation.

\paragraph{Comparison with the learned SYK Hamiltonian}

It is interesting to compare the $N=8$ ensemble averaged size winding
result in Section \ref{subsec:Saddle-approximation-for} with the
learned commuting SYK model in \cite{Jafferis:2022crx}  (and also follow-ups \cite{Kobrin:2023rzr,Jafferis:2023moh}).
The learnt Hamiltonian in \cite{Jafferis:2022crx}  consists of five terms
\begin{equation}
H=-0.36\psi_{1}\psi_{2}\psi_{4}\psi_{5}+0.19\psi_{1}\psi_{3}\psi_{4}\psi_{7}-0.71\psi_{1}\psi_{3}\psi_{5}\psi_{6}+0.22\psi_{2}\psi_{3}\psi_{4}\psi_{6}+0.49\psi_{2}\psi_{3}\psi_{5}\psi_{7}\label{eq:77}
\end{equation}
Note that this commuting Hamiltonian can not be written in terms of
(\ref{eq:3}) because the five terms do not share the same set of
commuting ingredients $X_{i}$. However, it can be separated into
two commuting groups, each of which shares a set of commuting ingredients.\footnote{$\psi_{1}\psi_{2}\psi_{4}\psi_{5}$, $\psi_{2}\psi_{3}\psi_{5}\psi_{7}$
and $\psi_{1}\psi_{3}\psi_{4}\psi_{7}$ can be constructed by $\psi_{1}\psi_{4}$,
$\psi_{2}\psi_{5}$ and $\psi_{3}\psi_{7}$; $\psi_{1}\psi_{3}\psi_{5}\psi_{6}$
and $\psi_{2}\psi_{3}\psi_{4}\psi_{6}$ can be constructed by $\psi_{3}\psi_{6}$,
$\psi_{2}\psi_{4}$ and $\psi_{1}\psi_{5}$.} Nevertheless, comparing (\ref{eq:77}) with our $N=8$ model is still
qualitatively reasonable. The main reason is that multiple groups
of shared commuting ingredients just splits one $F$ into multiple
commutative factors, and the counting in Section \ref{subsec:Size-distribution}
and \ref{subsec:Size-winding} mostly care about how many commuting
terms of different overlap types in the Hamiltonian. Since $N=8,q=4$
allows at most $C_{N/2}^{q/2}=6$ terms, which is close to 5, we should
expect qualitatively close behavior.

Let us simply fit the five coefficients of (\ref{eq:77}) with Gaussian
distribution by comparing the moments, which gives $-0.034$ mean
value and fluctuation $\s=0.436$. Comparing with (\ref{eq:4}) for $N=8$ and
$q=4$, we have $\mJ=0.267$. In \cite{Jafferis:2022crx} , the size winding is
mainly checked at $\b=4$ and $t_{r}=-2.8$ with $\mu=12/(4\times7)=0.429$,
which in our notation is $\b_{\text{\text{here}}}=\b_{\text{there}}\mJ=1.07$
and $t_{r,\text{here}}=t_{r,\text{there}}\mJ=-0.748$. The size winding
of close parameters is demonstrated in Figure \ref{fig:2d}, which
shows that the phases are almost lined up and the difference between
$P_{2n+1}$ and $|Q_{2n+1}|$ is not quite large. The alignment of
phases with the same size can be measured by the ratio $r_{2n+1}=|Q_{2n+1}|/P_{2n+1}$,
which on the four data points in Figure \ref{fig:2d} is $r_{2n+1}\gtrsim0.8$.
In Figure \ref{fig:8c}, we optimized the difference between the peak
of $|\Im H_{i\mu}|$ and the peak of $|\Im H_{-i\mu}|$ for $\b=1$
by tuning parameters $\mu=0.139\pi\app0.437$ and $t_{r}=-0.72$.
Note that the specific Hamiltonian (\ref{eq:77}) with $N=7$ was
learnt in \cite{Jafferis:2022crx}  also by maximizing the sign difference of $\mu$
but for another closely related quantity, the mutual information between
the reference system and the qubit receiver in the traversable wormhole
teleportation protocol. An interesting coincidence is that the optimal
parameters $t_{r}$ and $\mu$ are quite close between our commuting
SYK model and \cite{Jafferis:2022crx}. 

\begin{figure}
\begin{centering}
\subfloat[\label{fig:13a}]{\begin{centering}
\includegraphics[width=3.5cm]{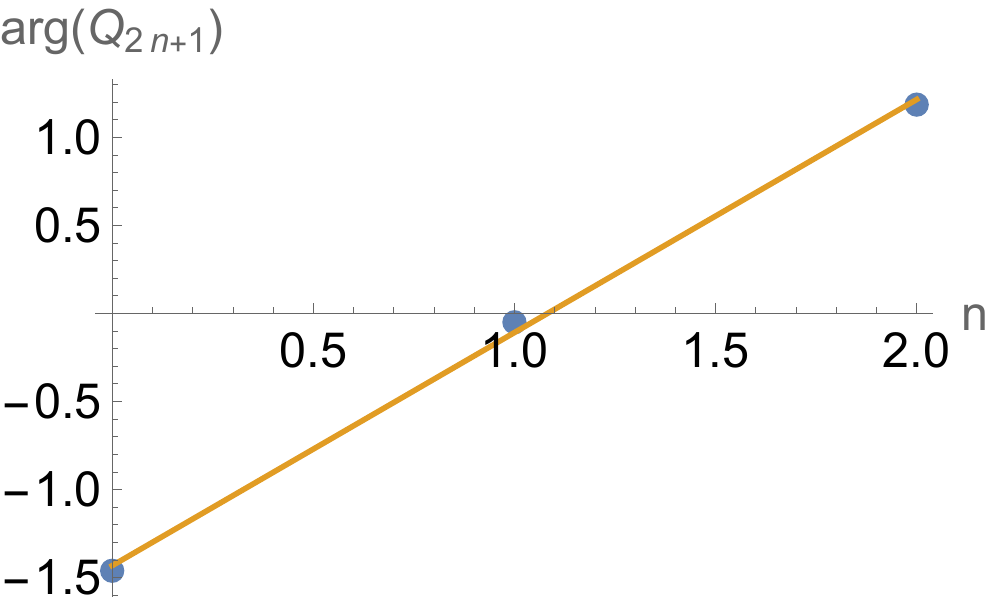}
\par\end{centering}
}\subfloat[\label{fig:13b}]{\begin{centering}
\includegraphics[width=3.5cm]{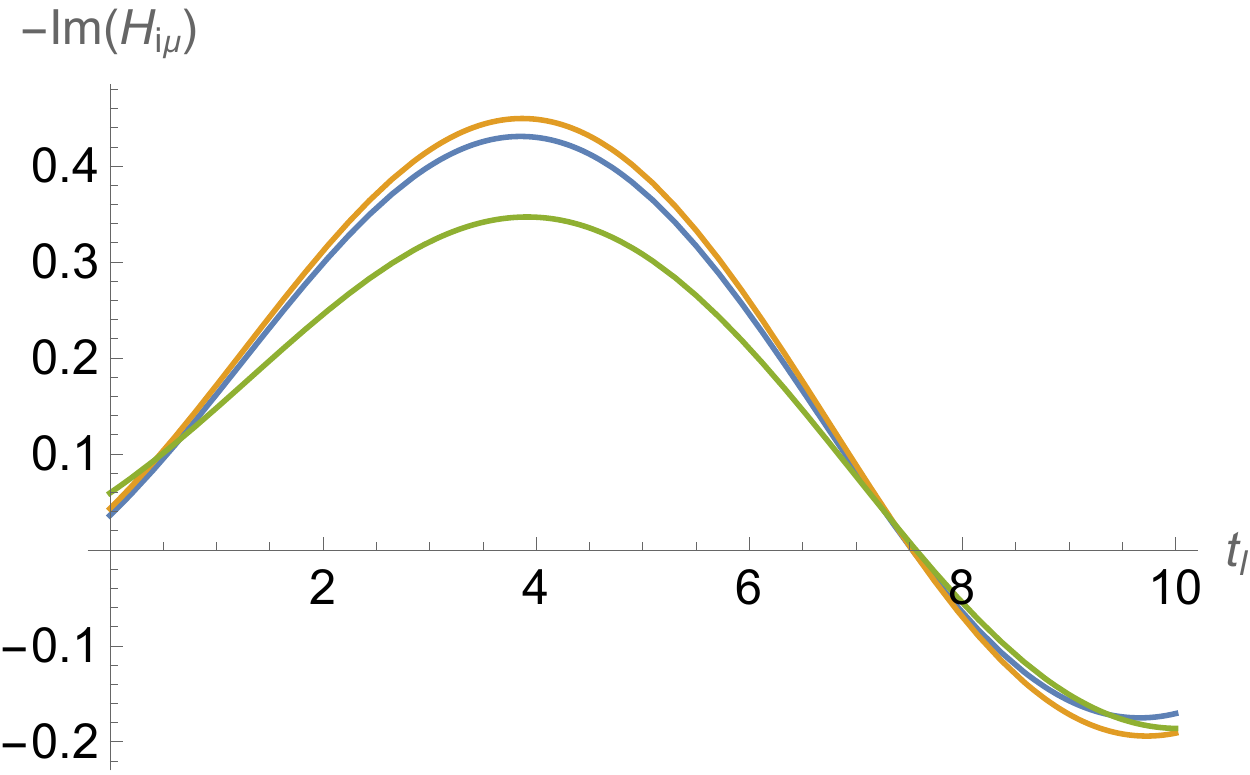}
\par\end{centering}
}\subfloat[\label{fig:13c}]{\begin{centering}
\includegraphics[width=3.5cm]{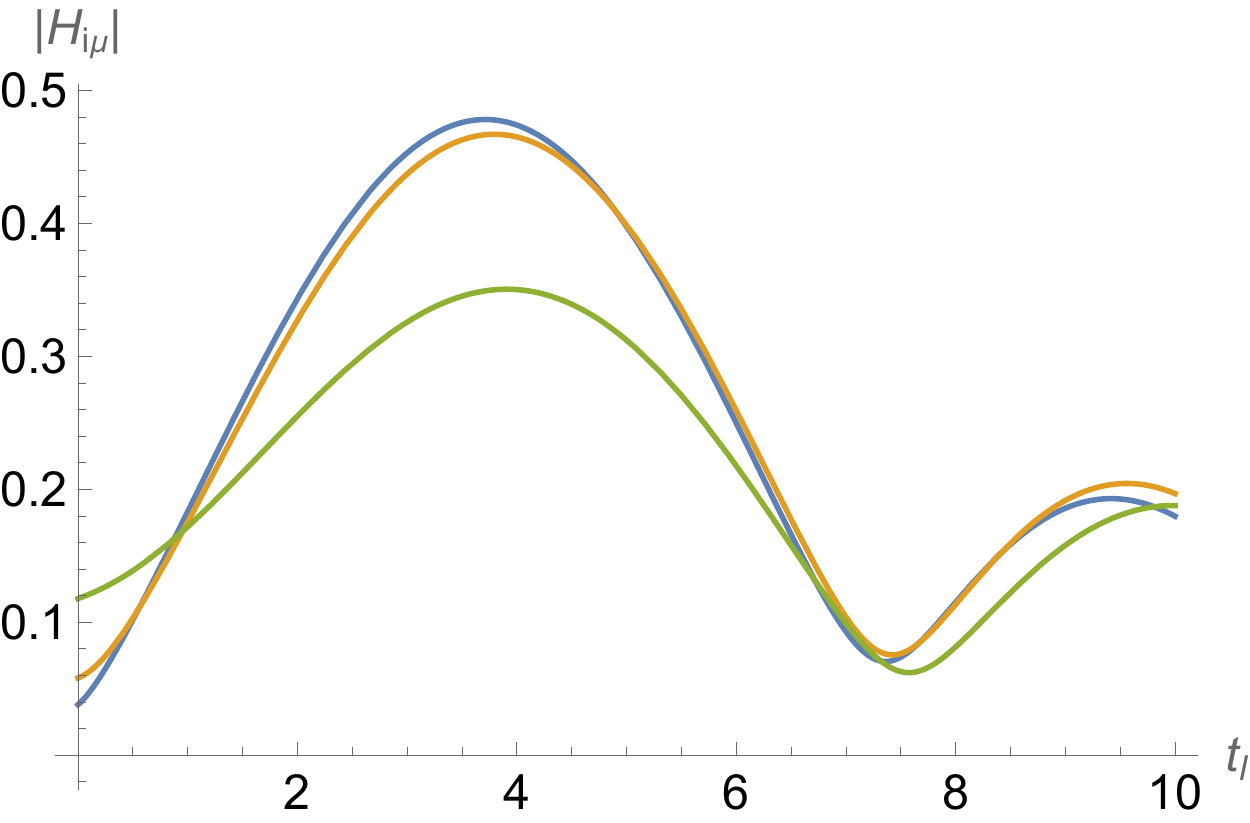}
\par\end{centering}
}
\par\end{centering}
\begin{centering}
\subfloat[\label{fig:13a-1}]{\begin{centering}
\includegraphics[width=3.5cm]{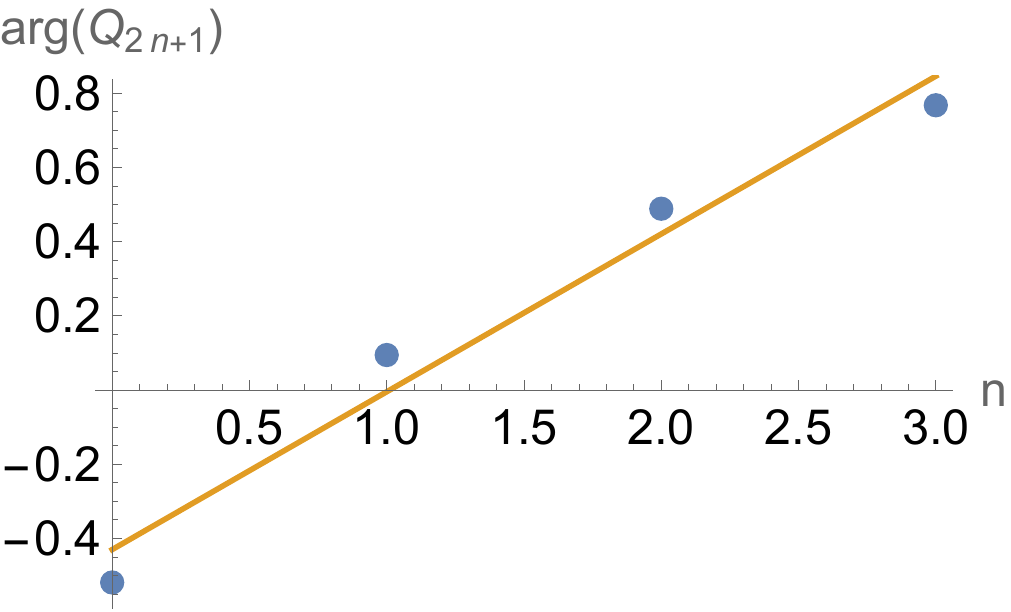}
\par\end{centering}
}\subfloat[\label{fig:13b-1}]{\begin{centering}
\includegraphics[width=3.5cm]{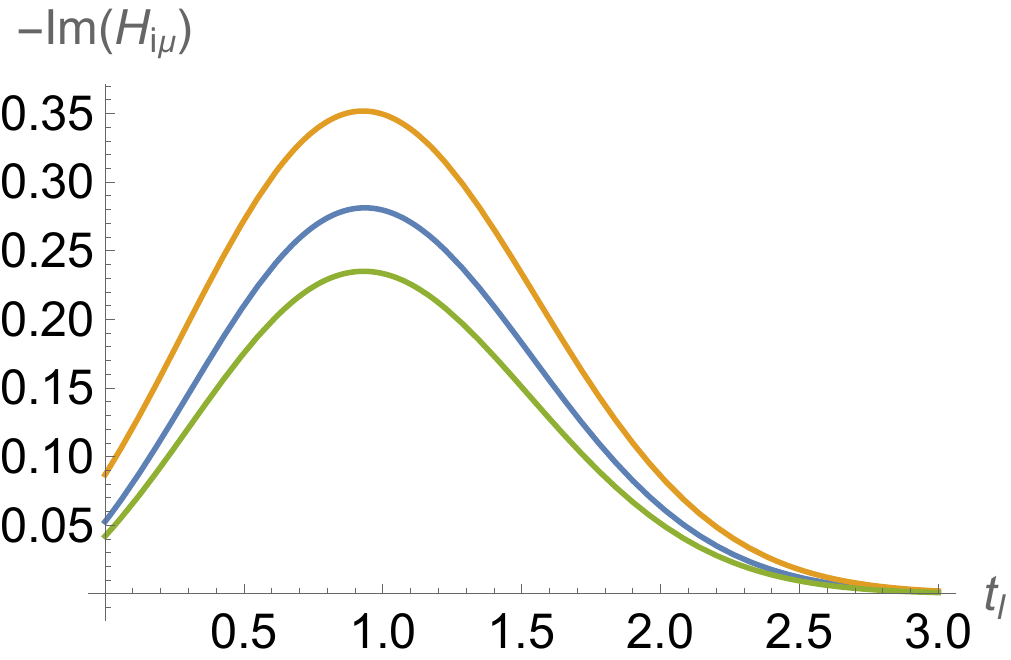}
\par\end{centering}
}\subfloat[\label{fig:13c-1}]{\begin{centering}
\includegraphics[width=3.5cm]{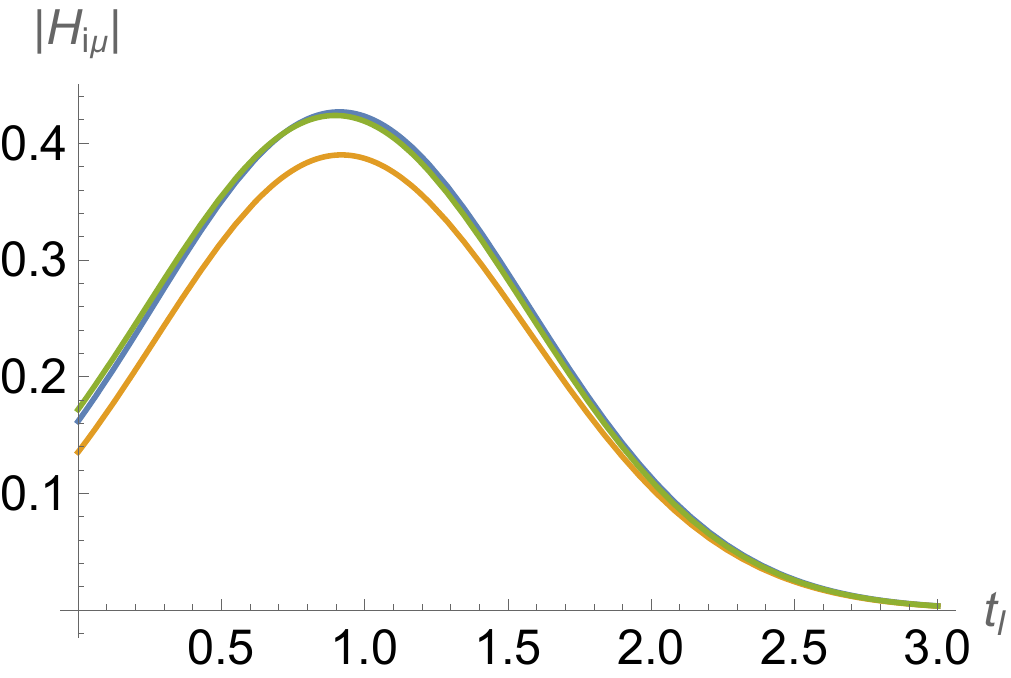}
\par\end{centering}
}\\
\subfloat[\label{fig:13a-1-1}]{\begin{centering}
\includegraphics[width=3.5cm]{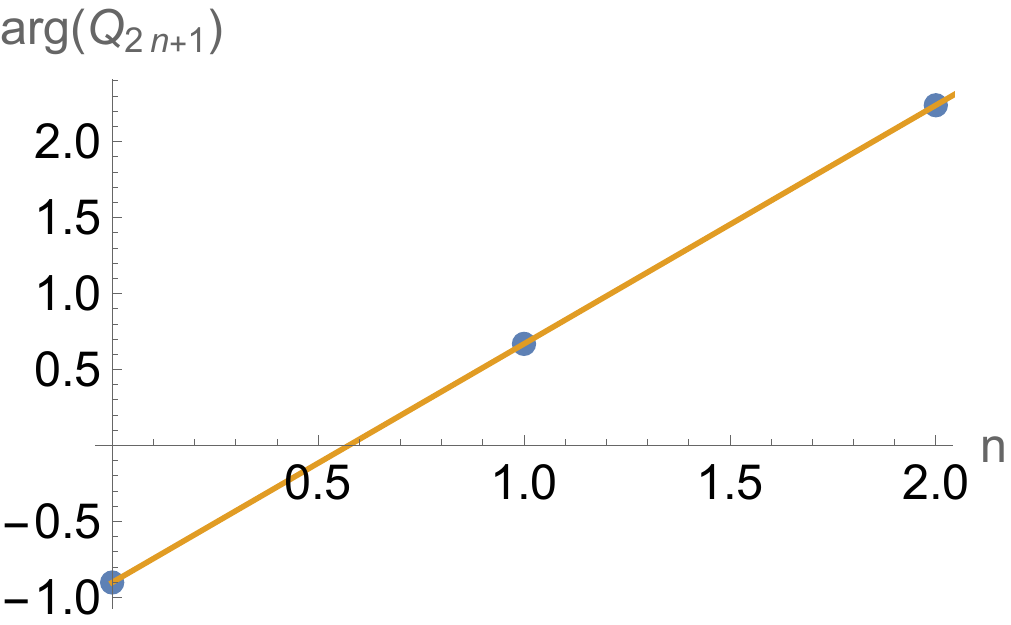}
\par\end{centering}
}\subfloat[\label{fig:13b-1-1}]{\begin{centering}
\includegraphics[width=3.5cm]{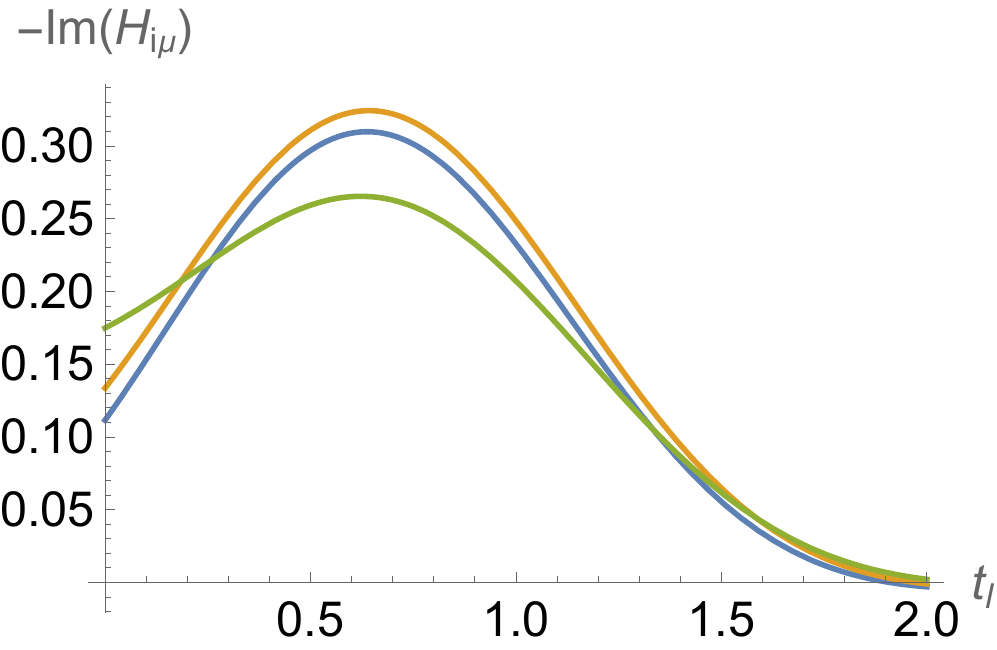}
\par\end{centering}
}\subfloat[\label{fig:13c-1-1}]{\begin{centering}
\includegraphics[width=3.5cm]{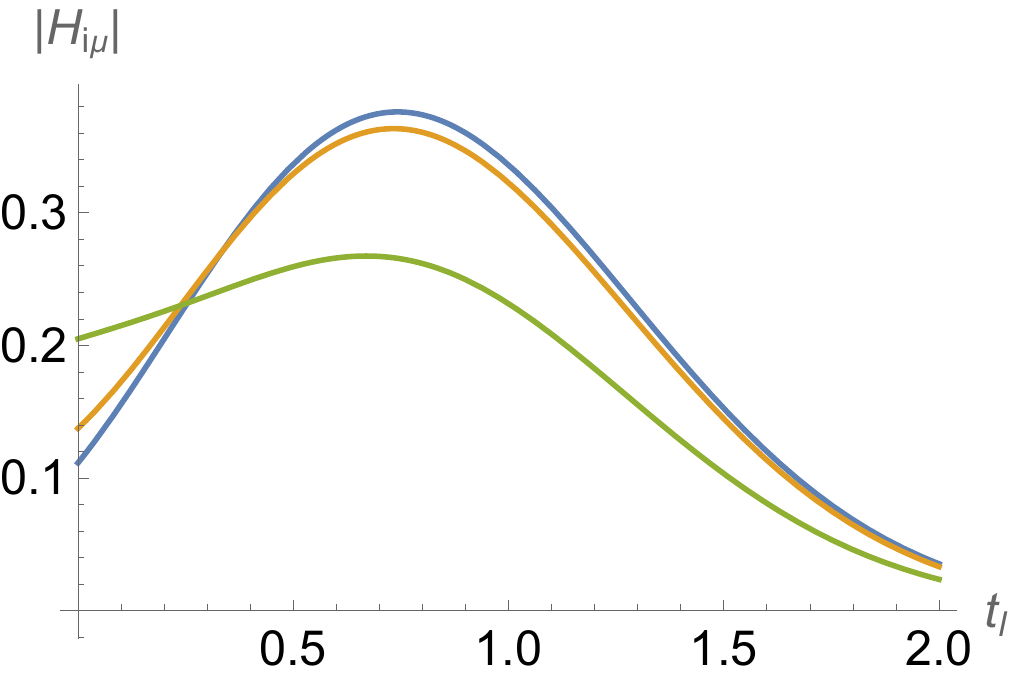}
\par\end{centering}
}
\par\end{centering}
\caption{The first row is for the learned Hamiltonian in \cite{Jafferis:2022crx} ; the second
row is for the ensemble-averaged commuting SYK model with $N=8$,
$\protect\b=1.07$ and $t=-0.748$ ($\protect\mJ=1$); the third row
is for the ensemble-averaged commuting SYK model with $N=6$, $\protect\b=1.9$
and $t=-0.475$ ($\protect\mJ=1$). The left column is the phase of
$Q_{2n+1}$ as a function of $n$. The yellow line is the linear fit
of the three/four data points. The middle column is $-\Im H_{i\mu}$
for $\mu_{|H|}$ (blue), $\mu_{\Im H}$ (yellow) and $\mu_{\text{SW}}$
(green). The right column is $|H_{i\mu}|$ for $\mu_{|H|}$ (blue),
$\mu_{\Im H}$ (yellow) and $\mu_{\text{SW}}$ (green). \label{fig:The-first-row-1}}
\end{figure}
However, if we look more carefully into these two models, we will
see that the underlying mechanism is not completely due to size winding.
Since the Hamiltonian (\ref{eq:77}) is given explicitly and each
term is commutative to each other, it is quite straightforward to
compute $H_{i\mu}$ and $Q_{2n+1}$ with (\ref{eq:77}) in Mathematica.
We first compute the phase of $Q_{2n+1}(t)$ for $t=-2.8$ in Figure
\ref{fig:13a}, which is equivalent to Fig. 3d (or Fig. S15) in \cite{Jafferis:2022crx} 
after shifting the first data point to the origin. We see that the
three data points are aligned pretty well. The yellow straight line
is the linear fit for these three points read as $Q_{2n+1}(-2.8)=-1.432+1.323n$.
By the argument of size winding below (\ref{eq:121-1}), we should
choose $\mu$ equal to half of the slope, namely $\mu_{\text{SW}}=0.662$,
for best fidelity.\footnote{By the definition of perfect size-winding, this $\mu_{\text{SW}}$ should also maximize $|G_{i\mu}(-2.8)|$. But because the linearity of phase to size is not perfect, we will instead get a very close $\mu=0.659$ if we maximize $|G_{i\mu}(-2.8)|$.} On the other hand, we can maximize $|\Im H_{i\mu}|$
by finding the appropriate $t_{l}$ and $\mu$ given $t_{r}=-2.8$.
It turns out that the optimal choice is $\mu_{\Im H}=0.463$ with
$t_{l}=3.87$. By (\ref{eq:121-1}), if $e^{i\mu V}$ can be approximated
by $e^{i\mu\avg V}$, the optimal $\mu$ for size winding should also
give the maximal $|H_{i\mu}|$. Maximizing $|H_{i\mu}|$ by scanning
$t_{l}$ and $\mu$ given $t_{r}=-2.8$, we find the optimal choice
is $\mu_{|H|}=0.381$ with $t_{l}=3.71$.\footnote{One can also search the maximal $|H_{i\mu}|$ with $t_{l}=2.8$ fixed.
The optimal result is $\mu=0.346$, which deviate even more from $\mu_{\text{SW}}$. } We plot the $-\Im H_{i\mu}$ and $|H_{i\mu}|$ in Figure \ref{fig:13b}
and \ref{fig:13c} respectively for these three choices of $\mu$. 

Clearly, these three optimal choices of $\mu$ are different. In particular,
$\mu_{|H|}$ deviates from $\mu_{\text{SW}}$ quite much and is just
$58\%$ of the latter. The explanation behind this distinction is
that the size-winding mechanism is compatible with traversable wormhole
teleportation only when the time scale is much larger than the thermal
scale so that we can factorize out a pure phase in (\ref{eq:121-1})
by the argument of time-order-correlator. In this case $H_{i\mu}\propto G_{i\mu}$
and perfect size winding implies maximal $|H_{i\mu}|$. However, in
the above example, $t_{r}=-2.8$, which is even shorter than the thermal
scale $\b=4$. Therefore we should expect the thermalization process
to have an important interplay with the size winding and tunes the optimal
$\mu$. The difference between $\mu_{\Im H}$ and $\mu_{|H|}$ is
much subtler. Even for the factorized case (\ref{eq:121}), these
two are not necessarily the same because the phase $e^{i\mu(\avg V+N/2)}$
exerts an additional rotation to $G_{i\mu}$, which will affect the
imaginary part in a nontrivial way. However, in the known holographic
large $q$ SYK model, one can check from \cite{Gao:2019nyj} that these three
$\mu$'s coincide in low temperature and equal to $-4\pi Ne^{-2\pi t/\b}/(\b\mJ)$
(with the notation of \cite{Gao:2019nyj}).

If we check the same aspects with the ensemble-averaged commuting
SYK model, we will find that thermalization also plays an important
role. It has been discussed in Section \ref{subsec:Sign-of} that
in the short time scale and large $N$, thermalization is responsible
for the sign difference of $\mu$ rather than size winding. In small
$N$ cases in Figure \ref{fig:The-optimized--1}, we see similar sign
difference effect only when $t$ is comparable with $\b$ for intermediate
$\b$. Though for small $N$ we have size winding with order one phase
slope as shown in Figure \ref{fig:The-comparison-between}, we now
show that the size winding is not the complete mechanism for the sign
difference. Let us take $N=8$, $\b=1.07$ and $t=-0.748$ as an example.\footnote{These numbers are in $\mJ=1$ unit and are equivalent to the parameters
of \cite{Jafferis:2022crx}  as we discussed earlier.} The phase of $Q_{2n+1}(-0.748)$ is shown in Figure \ref{fig:13a-1},
where the four data points are mostly aligned. The yellow straight
line is the linear fit for these three points read as $Q_{2n+1}(-0.748)=-0.430+0.425n$,
which by size-winding assumption leads to $\mu_{\text{SW}}=0.213$.
Maximizing $|\Im H_{i\mu}|$ leads to the optimal choice $\mu_{\Im H}=0.451$
with $t_{l}=0.929$.\footnote{If we maximize the peak difference between $|\Im H_{i\mu}|$ and $|\Im H_{-i\mu}|$,
we will get a slightly different optimal choice that is close to Figure
\ref{fig:8c}.} Maximizing $|H_{i\mu}|$ leads to the optimal choice $\mu_{|H|}=0.267$
with $t_{l}=0.912$. We plot the $-\Im H_{i\mu}$ and $|H_{i\mu}|$
in Figure \ref{fig:13b-1} and \ref{fig:13c-1} respectively for these
three choices of $\mu$. 

Here we see that these three $\mu$'s are different and the minimum
$\mu_{\text{SW}}$ is just about 47\% of the maximum $\mu_{\Im H}$.
Compare the learned SYK model and the ensemble-averaged one, we see
that the optimal $\mu$'s are different, the size-winding phase aligns
better, and the peak of $\Im H_{i\mu}$ is slightly higher in the learned
model. This is reasonable because the Hamiltonian (\ref{eq:77}) is especially learned to achieve the best teleportation fidelity for a specific
operator $\psi_{1}$. For other operators $\psi_{i}$, the size-winding
quality is worse at the same time \cite{Jafferis:2022crx,Kobrin:2023rzr}. On the other hand, the
ensemble-averaged model works for all operators $\psi_{i}$ equally
and shows an average level of teleportation efficiency. 

As we argued before, the large deviation between $\mu_{|H|}$ and
$\mu_{\text{SW}}$ for the learned Hamiltonian implies that the system
is undergoing thermalization. For the ensemble-averaged model with
equivalent parameter $\b=1.07$, we see that $\mu_{|H|}$ is not far
from $\mu_{\text{SW}}$, which means the thermalization is close to
complete. This suggests that the effective temperature for the learned
Hamiltonian is even lower than $1/\b$, which strengthens the conclusion
that thermalization plays a crucial role besides size-winding.
To justify this, we tune the parameters in the ensemble-averaged model
with lower temperature in the third row of Figure \ref{fig:The-first-row}.
We find that for $N=6$, $\b=1.9$, and $t_{r}=-0.475$, the linearity of size-winding
phase is pretty good\footnote{The averaged $r_n$ ratio weighted by probability is $r=\sum_n |Q_n(t)|\app 0.58$ \cite{Kobrin:2023rzr}, which shows the phases of the same size are not quite aligned (due to the low temperature) but not too bad. As a comparison, the learned Hamiltonian has an averaged $r\app 0.95$ for all $\psi_i$.} with a large phase slope that gives $\mu_{\text{SW}}=0.785$
as shown in Figure \ref{fig:13a-1-1}. At the same time, we find that
$\mu_{\Im H}=0.570$ and $\mu_{|H|}=0.468$. On one hand, this is
quite similar to the learned Hamiltonian case (the first row of Figure
\ref{fig:The-first-row-1}), including the relative ordering of three
$\mu$'s ($\mu_{\text{SW}}>\mu_{\Im H}>\mu_{|H|}$ with $\mu_{|H|}/\mu_{\text{SW}}\app59.6\%$).
On the other hand, the ratio $|t_{r}|/\b=25\%$ suggests that the
size-winding of the learned Hamiltonian occurs at the early stage of
thermalization.

\paragraph{Non-commutative terms}

The non-existence of non-commutative terms was discussed in \cite{Kobrin:2023rzr,Jafferis:2023moh},
where the main focus was on their effects on size winding. However,
based on the observation of the commuting SYK model, we find that
size winding is not a unique feature of holographic models. It can
exist in the non-holographic commuting SYK model, though in long time
scale and large $N$ it also has peaked size distribution. Clearly,
we need to introduce non-commutative terms to save it from peaked
size distribution in large $N$ limit and keep the size-winding feature
at the same time. For small $N$ cases, there is no peaked size distribution
and the size-winding also exists. However, teleportation does
not completely follow the rule of size-winding, and thermalization
becomes important.

Recent simulation of the dynamics of traversable wormholes in \cite{Jafferis:2022crx} is an excellent start toward the project ``quantum gravity in the lab". However, given the holography-like features in the commuting SYK model, it is not obvious that the simulation in \cite{Jafferis:2022crx} with $N=7$ is for the non-holographic commuting SYK model or the holographic full SYK model. To have a better simulation for the dynamics of traversable wormhole that is mostly due to non-peaked size winding, we must increase $N$, separate
the thermal scale with scrambling scale, and consider times after
thermalization. This indicates that we will not learn commuting Hamiltonian
as we scale up the system.\footnote{The possible exception is a learnt large $N$ commuting Hamiltonian
that has teleportation-like behavior as Figure \ref{fig:The-optimized-},
which has $t$ in the thermal scale with $\mu\sim O(1)$. } Otherwise, it will be peaked-size and non-holographic. On the other
hand, non-commutative terms will bring challenges because the implementation
steps of simulation will be more complicated. Therefore, to understand
how to minimally introduce non-commutative terms while preserving
essential holographic features (e.g. non-peaked size winding distribution,
very close three optimal $\mu$'s, etc.) is crucial for future simulation
studies.

On the theoretical side, understanding the non-commutative terms is
equally important. As we know that the commuting SYK model has large
numbers of conserved charges that are related to overly abundant symmetries
of the system. However, in quantum gravity, we should have much fewer
symmetries and most of them need to be explicitly broken by introducing
the non-commutative terms. It is extremely interesting to construct a holographic theory by (perhaps minimally) breaking symmetries in steps and to understand their meaning in the dual gravity language. 

\section*{Acknowledgements} We would like to thank Yingfei Gu, Pengfei Zhang, Thomas Schuster, Bryce Kobrin, Cheng Peng, and David Kolchmeyer for stimulating and helpful discussions. PG is supported by the US Department of Defense (DOD) grant  KK2014.

\appendix

\section{Computation of $h_{\mu}$\label{sec:Computation-of}}

Expanding the two exponentials $e^{\pm\mu V}$ in (\ref{eq:79}),
we have
\begin{align}
h_{\mu}= & -i\sum_{I,J}\tilde{K}_{IJ}\avg{0|F(-\b/2)\G_{I}^{l}\G_{I}^{r}\psi_{j}^{l}(\tau_{1})\G_{J}^{l}\G_{J}^{r}\psi_{j}^{r}(\tau_{2})F(-\b/2)|0}\nonumber \\
= & -i\sum_{I,J}\tilde{K}_{IJ}i^{|I|}\avg{0|F_{|I\cap K|=\Z_{+}}(-\b/2)F_{|I\cap K|=\Z_{-}}(\b/2)\psi_{j}^{l}(\tau_{1})\G_{J}^{l}\G_{J}^{r}\psi_{j}^{r}(\tau_{2})F(-\b/2)|0}\nonumber \\
= & \sum_{I,J}\tilde{K}_{IJ}i^{|I|}\avg{0|F(-\tau_{1})\psi_{j}^{r}F_{|I\cap K|=\Z_{+}}(\tau_{1}-\b/2)F_{|I\cap K|=\Z_{-}}(\tau_{1}+\b/2)\G_{J}^{l}\G_{J}^{r}\psi_{j}^{r}(\tau_{2})F(-\b/2)|0}\nonumber \\
= & \sum_{I,J}\tilde{K}_{IJ}i^{|I|+|J|}(-)^{|J|}\Tr_{r}\left(\G_{J}^{r}F(-\tau_{1})\psi_{j}^{r}F_{|I\cap K|=\Z_{+}}(\tau_{1}-\b/2)F_{|I\cap K|=\Z_{-}}(\tau_{1}+\b/2)\G_{J}^{r}F(\tau_{2})\psi_{j}^{r}F(-\tau_{2}-\b/2)\right)\nonumber \\
= & \sum_{I,J}\tilde{K}_{IJ}i^{|I|+|J|}(-)^{|j\cap J|}\Tr\left(F_{|I\cap K|=\Z_{+}}(\tau_{1}-\b/2)F_{|I\cap K|=\Z_{-}}(\tau_{1}+\b/2)F_{|J\cap K|=\Z_{+}}(\tau_{2})F_{|J\cap K|=\Z_{-}}(-\tau_{2})\right.\nonumber \\
 & \left.\psi_{j}^{r}F_{|J\cap K|=\Z_{+}}(-\tau_{2}-\tau_{1}-\b/2)F_{|J\cap K|=\Z_{-}}(\tau_{2}-\tau_{1}+\b/2)\psi_{j}^{r}\right)\nonumber \\
= & \f 12\sum_{I,J}\tilde{K}_{IJ}i^{|I|+|J|}(-)^{|j\cap J|}\Tr\left(F_{|I\cap K|=\Z_{+}}(\tau_{1}-\b/2)F_{|I\cap K|=\Z_{-}}(\tau_{1}+\b/2)F_{j\in K,|J\cap K|=\Z_{+}}(2\tau_{2}+\tau_{1}+\b/2)\right.\nonumber \\
 & \left.F_{j\notin K,|J\cap K|=\Z_{+}}(-\tau_{1}-\b/2)F_{j\in K,|J\cap K|=\Z_{-}}(-2\tau_{2}+\tau_{1}-\b/2)F_{j\notin K,|J\cap K|=\Z_{-}}(-\tau_{1}+\b/2)\right)\label{eq:65}
\end{align}
where $\tilde{K}_{IJ}=\f 1Z(\cosh\f{\mu}2)^{2N-|I|-|J|}(i\sinh\f{\mu}2)^{|I|+|J|}(-)^{|J|}$
and we can define
\begin{equation}
K_{IJ}=i^{|I|+|J|}\tilde{K}_{IJ}=\f 1Z(-)^{|I|}(\cosh\f{\mu}2)^{2N-|I|-|J|}(\sinh\f{\mu}2)^{|I|+|J|}
\end{equation}
All $F$ terms in (\ref{eq:65}) can be factorized as a product of
eight terms, which are summarized as Table \ref{tab:Eight-cases-for}.
For each $F_{\text{condition}}(x)$, taking ensemble average leads
to
\begin{equation}
\overline{F_{\text{condition}}(x)}=\exp\left(\f{c_{\text{condition}}x^{2}\s^{2}}{2^{q+1}}\right)
\end{equation}

Let us again consider $q=4$ and $j=1$. In order to count the terms
correctly, we first separate the indices into 9 groups, each group
includes a pattern of $X_{i}$ overlapping with $I$ and $J$. For
each $X_{i}$, if it overlaps $a$ indices with $I$ and $b$ indices
with $J$, we call it $(a,b)$ type, where $a,b=0,1,2$. We count
each type as $i_{ab}$. For example, if $N=10$ and $I=\{1,3,4,5,6\}$
and $J=\{1,2,3,5,7,8\}$, we have $i_{12}=i_{02}=1,i_{21}=2$ and
all other $i_{ab}=0$. It is easy to find the following relations
\begin{equation}
\sum_{ab}i_{ab}=N/2,\quad\sum_{ab}ai_{ab}=|I|,\quad\sum_{ab}bi_{ab}=|J|
\end{equation}
To compute (\ref{eq:65}), besides the cases in Table \ref{tab:Eight-cases-for},
we need to consider more conditions, which depends on whether $X_{1}$
has overlap with $I$ and $J$. This is because under the condition
$1\in K$ or $1\notin K$, the counting for $c_{\text{condition}}$
will be different. Before we proceed to the counting, let us first
note a cancellation due to the factor $(-)^{|j\cap J|}$. If $|X_{1}\cap J|=1$,
there are two cases $1\in J,2\notin J$ and $1\notin J,2\in J$. If
we consider two $J$ of these two cases with all other indices to
be identical, $c_{\text{condition}}$ is the same for all cases in
Table \ref{tab:Eight-cases-for}. However, they will have the opposite sign due to $(-)^{|j\cap J|}$ and thus cancel each other exactly.
Therefore, we only have the following 6 cases to consider: $|X_{1}\cap I|=\g,|X_{1}\cap J|=\d$
with $\g=0,1,2$ and $\d=0,2$, which are labelled by $(\g,\d)$ in
Table \ref{tab:c-cond}. The countings for all nontrivial \#'s in
Table \ref{tab:Eight-cases-for} are as follows.

\begin{table}
\begin{centering}
\begin{tabular}{|c|c|c|}
\hline 
\# & conditions & $x$\tabularnewline
\hline 
1 & $j\notin K,|I\cap K|=\Z_{+},|J\cap K|=\Z_{+}$ & $-\b$\tabularnewline
\hline 
2 & $j\notin K,|I\cap K|=\Z_{+},|J\cap K|=\Z_{-}$ & 0\tabularnewline
\hline 
3 & $j\notin K,|I\cap K|=\Z_{-},|J\cap K|=\Z_{+}$ & 0\tabularnewline
\hline 
4 & $j\notin K,|I\cap K|=\Z_{-},|J\cap K|=\Z_{-}$ & $\b$\tabularnewline
\hline 
5 & $j\in K,|I\cap K|=\Z_{+},|J\cap K|=\Z_{+}$ & $2(\tau_{1}+\tau_{2})$\tabularnewline
\hline 
6 & $j\in K,|I\cap K|=\Z_{+},|J\cap K|=\Z_{-}$ & $2(\tau_{1}-\tau_{2})-\b$\tabularnewline
\hline 
7 & $j\in K,|I\cap K|=\Z_{-},|J\cap K|=\Z_{+}$ & $2(\tau_{1}+\tau_{2})+\b$\tabularnewline
\hline 
8 & $j\in K,|I\cap K|=\Z_{-},|J\cap K|=\Z_{-}$ & $2(\tau_{1}-\tau_{2})$\tabularnewline
\hline 
\end{tabular}
\par\end{centering}
\caption{Eight cases for $F_{\text{condition}}(x)$. \label{tab:Eight-cases-for}}
\end{table}
\begin{table}
\begin{centering}
\begin{tabular}{|c|c||c||c||c|c||c|}
\hline 
 & \multicolumn{6}{c|}{$c_{\text{condition}}$(for different $(\g,\d)$)}\tabularnewline
\hline 
\# & \multicolumn{4}{c|}{$(0,0)$,$(2,0)$,$(0,2)$,$(2,2)$} & \multicolumn{2}{c|}{$(1,0)$,$(1,2)$}\tabularnewline
\hline 
1+4 & \multicolumn{4}{c|}{$\f 12(u-1)(u-2)+\f 12v(v-1)$} & \multicolumn{2}{c|}{$\f 12u(u-1)+\f 12(v-1)(v-2)$}\tabularnewline
\hline 
5 & \multicolumn{4}{c|}{$u-i_{11}-1$} & \multicolumn{2}{c|}{$i_{10}+i_{12}-1$}\tabularnewline
\hline 
6 & \multicolumn{4}{c|}{$i_{01}+i_{21}$} & \multicolumn{2}{c|}{$i_{11}$}\tabularnewline
\hline 
7 & \multicolumn{4}{c|}{$i_{10}+i_{12}$} & \multicolumn{2}{c|}{$u-i_{11}$}\tabularnewline
\hline 
8 & \multicolumn{4}{c|}{$i_{11}$} & \multicolumn{2}{c|}{$i_{01}+i_{21}$}\tabularnewline
\hline 
\end{tabular}
\par\end{centering}
\caption{Counting for $c_{\text{condition}}$. \label{tab:c-cond}}
\end{table}
\paragraph{\#1} For all $1\notin K$, we have two indices of $X_{k_{1}}$
and $X_{k_{2}}$ to choose from $2$ to $N/2$. We have 9 cases: $(|I\cap K|,|J\cap K|)=(0,0),(0,2),(2,0),(2,2),(4,0),(0,4),(4,2),(2,4),(4,4)$,
which for $(\g,\d)=(0,0)$ case contributes to $c_{\text{condition}}$
(the total choice of $k_{1}$ and $k_{2}$) respectively as $C_{i_{00}-1}^{2},C_{i_{01}}^{2}+i_{02}(i_{00}-1),C_{i_{10}}^{2}+i_{20}(i_{00}-1),i_{20}i_{02}+C_{i_{11}}^{2}+(i_{00}-1)i_{22}+i_{10}i_{12}+i_{01}i_{21},C_{i_{20}}^{2},C_{i_{02}}^{2},i_{20}i_{22}+C_{i_{21}}^{2},i_{02}i_{22}+C_{i_{12}}^{2},C_{i_{22}}^{2}$.
Note that $i_{00}-1$ appears here because $k_{1,2}\neq1$ in the
$(\g,\d)=(0,0)$ case. Summing over all cases leads to
\begin{align}
c_{\text{condition}}= & \frac{1}{2}\left[i_{00}^{2}+\left(2i_{02}+2i_{20}+2i_{22}-1\right)i_{00}+i_{01}^{2}+i_{10}^{2}+\left(i_{02}+i_{20}+i_{22}\right){}^{2}-i_{02}+\left(i_{11}-1\right)i_{11}\right.\nonumber \\
 & \left.+\left(i_{12}-1\right)i_{12}+i_{10}\left(2i_{12}-1\right)-i_{20}+\left(i_{21}-1\right)i_{21}+i_{01}\left(2i_{21}-1\right)-i_{22}\right]_{i_{00}\ra i_{00}-1}\label{eq:69}
\end{align}
For other $(\g,\d)$, one just needs to replace the $i_{00}\ra i_{00}-1$
with $i_{\g\d}\ra i_{\g\d}-1$ in (\ref{eq:69}). One can check that
$c_{\text{condition}}$ are identical for $(\g,\d)=(1,0),(1,2)$,
and also identical for $(\g,\d)=(0,0),(0,2),(2,0),(2,2)$ but with
a different value.
\paragraph{\#4} We have 4 cases: $(|I\cap K|,|J\cap K|)=(1,1),(1,3),(3,1),(3,3)$,
which for $(\g,\d)=(0,0)$ case contributes to $c_{\text{condition}}$
respectively as $i_{10}i_{01}+(i_{00}-1)i_{11},i_{11}i_{02}+i_{12}i_{01},i_{11}i_{20}+i_{21}i_{10},i_{11}i_{22}+i_{12}i_{21}$.
Summing over all cases leads to
\begin{align}
c_{\text{condition}}= & \left(i_{10}+i_{12}\right)\left(i_{01}+i_{21}\right)+i_{11}\left(i_{00}+i_{02}+i_{20}+i_{22}\right)|_{i_{00}\ra i_{00}-1}\label{eq:70}
\end{align}
For other $(\g,\d)$, one just needs to replace the $i_{00}\ra i_{00}-1$
with $i_{\g\d}\ra i_{\g\d}-1$ in (\ref{eq:70}). Again, we see that
$c_{\text{condition}}$ are identical for $(\g,\d)=(1,0),(1,2)$,
and also identical for $(\g,\d)=(0,0),(0,2),(2,0),(2,2)$ but with
a different value.
\paragraph{\#1+4} Since $x^{2}=\b^{2}$ for both \#1 and \#4, we can add
their $c_{\text{condition}}$ together and have
\begin{equation}
c_{\text{condition}}=\begin{cases}
\f 12(u-1)(u-2)+\f 12v(v-1), & (\g,\d)=(0,0),(0,2),(2,0),(2,2)\\
\f 12u(u-1)+\f 12(v-1)(v-2), & (\g,\d)=(1,0),(1,2)
\end{cases}
\end{equation}
where 
\begin{equation}
u=i_{00}+i_{02}+i_{20}+i_{22}+i_{11},\quad v=i_{10}+i_{12}+i_{01}+i_{21}
\end{equation}
\paragraph{\#5} For all $1\in K$ cases, we set $k_{1}=1$ and we only need
to choose $k_{2}$ from 2 to $N/2$. For $(\g,\d)=(0,0)$, we have
four cases: $(|I\cap k_{2}|,|J\cap k_{2}|)=(0,0),(0,2),(2,0),(2,2)$,
which contributes to $c_{\text{condition}}$ respectively as $i_{00}-1,i_{02},i_{20},i_{22}$.
Similarly, for $(\g,\d)=(2,0),(0,2),(2,2)$ cases, just need to replace
$i_{00}-1$ with $i_{00}$ and $i_{\g\d}$ with $i_{\g\d}-1$. In
all these cases, the total $c_{\text{condition}}$ are the same 
\begin{equation}
c_{\text{condition}}=i_{00}+i_{02}+i_{20}+i_{22}-1
\end{equation}
For $(\g,\d)=(1,0)$, the counting is different and we have $(|I\cap k_{2}|,|J\cap k_{2}|)=(1,0),(1,2)$,
which contributes to $c_{\text{condition}}$ respectively as $i_{10}-1,i_{12}$;
For $(\g,\d)=(1,2)$, the counting leads to the contribution to $c_{\text{condition}}$
respectively as $i_{10},i_{12}-1$. In both cases, the total $c_{\text{condition}}$
are the same 
\begin{equation}
c_{\text{condition}}=i_{10}+i_{12}-1
\end{equation}
\paragraph{\#6} For $(\g,\d)=(0,0),(2,0),(0,2),(2,2)$, we have four cases:
$(|I\cap k_{2}|,|J\cap k_{2}|)=(0,1),(2,1)$, which contributes to
$c_{\text{condition}}$ respectively as $i_{01},i_{21}$, which together
gives
\begin{equation}
c_{\text{condition}}=i_{01}+i_{21}
\end{equation}
For $(\g,\d)=(1,0),(1,2)$, we have $(|I\cap k_{2}|,|J\cap k_{2}|)=(1,1)$,
whose contribution to $c_{\text{condition}}$ is
\begin{equation}
c_{\text{condition}}=i_{11}
\end{equation}
\paragraph{\#7} For $(\g,\d)=(0,0),(2,0),(0,2),(2,2)$, we have four cases:
$(|I\cap k_{2}|,|J\cap k_{2}|)=(1,0),(1,2)$, which contributes to
$c_{\text{condition}}$ respectively as $i_{10},i_{12}$, which together
gives
\begin{equation}
c_{\text{condition}}=i_{10}+i_{12}
\end{equation}
For $(\g,\d)=(1,0),(1,2)$, we have $(|I\cap k_{2}|,|J\cap k_{2}|)=(0,0),(2,0),(0,2),(2,2)$,
which contributes to $c_{\text{condition}}$ respectively as $i_{00},i_{20},i_{02},i_{22}$,
which together gives
\begin{equation}
c_{\text{condition}}=i_{00}+i_{02}+i_{20}+i_{22}
\end{equation}
\paragraph{\#8} For $(\g,\d)=(0,0),(0,2),(2,0),(2,2)$, we have $(|I\cap k_{2}|,|J\cap k_{2}|)=(1,1)$,
whose contribution to $c_{\text{condition}}$ is
\begin{equation}
c_{\text{condition}}=i_{11}
\end{equation}
For $(\g,\d)=(1,0),(1,2)$, we have $(|I\cap k_{2}|,|J\cap k_{2}|)=(0,1),(2,1)$,
whose contribution to $c_{\text{condition}}$ is
\begin{equation}
c_{\text{condition}}=i_{01}+i_{21}
\end{equation}
Above results are summarized in Table \ref{tab:c-cond}.

The next step is to count the number $d_{\g,\d}$ of configurations
of $I$ and $J$ for given $i_{ab}$ and $(\g,\d)$. This is staightforward
and the result is
\begin{equation}
d_{\g,\d}=\f{i_{\g\d}(N/2-1)!}{\prod_{ab}i_{ab}!}\prod_{ab}\left(\f 4{a!b!(2-a)!(2-b)!}\right)^{i_{ab}}\equiv i_{\g\d}d(i_{ab})
\end{equation}
In (\ref{eq:65}) $(\g,\d)=(0,0),(2,0),(1,0)$ gives $(-)^{|j\cap J|}=1$
and $(\g,\d)=(0,2),(2,2),(1,2)$ gives $(-)^{|j\cap J|}=-1$. Therefore,
the net contribution from $(\g,\d)=(0,0),(0,2),(2,0),(2,2)$ to (\ref{eq:65})
is
\begin{align}
h_{\mu}^{1}= & \f 12\sum_{i_{ab}}K_{IJ}(i_{00}+i_{20}-i_{02}-i_{22})d(i_{ab})e^{\f{\left[\f 12(u-1)(u-2)+\f 12v(v-1)\right]\b^{2}+4(u-i_{11}-1)(\tau_{1}+\tau_{2})^{2}}{4(N/2-1)}}\nonumber \\
 & \times e^{\f{(i_{01}+i_{21})(2(\tau_{1}-\tau_{2})-\b)^{2}+(i_{10}+i_{12})(2(\tau_{1}+\tau_{2})+\b)^{2}+4i_{11}(\tau_{1}-\tau_{2})^{2}}{4(N/2-1)}})\label{eq:82}
\end{align}
and the net contribution from $(\g,\d)=(1,0),(1,2)$ to (\ref{eq:65})
is
\begin{align}
h_{\mu}^{2}= & \f 12\sum_{i_{ab}}K_{IJ}(i_{10}-i_{12})d(i_{ab})e^{\f{\left[\f 12(v-1)(v-2)+\f 12u(u-1)\right]\b^{2}+4(i_{10}+i_{12}-1)(\tau_{1}+\tau_{2})^{2}}{4(N/2-1)}}\nonumber \\
 & \times e^{\f{i_{11}(2(\tau_{1}-\tau_{2})-\b)^{2}+(u-i_{11})(2(\tau_{1}+\tau_{2})+\b)^{2}+4(i_{01}+i_{21})(\tau_{1}-\tau_{2})^{2}}{4(N/2-1)}}\label{eq:83}
\end{align}
where the sum over $i_{ab}$ is restricted to $\sum_{ab}i_{ab}=N/2$
and $K_{IJ}$ in terms of $i_{ab}$ is
\begin{equation}
K_{IJ}=\f 1{\overline{Z}}(-)^{\sum_{ab}ai_{ab}}(\cosh\f{\mu}2)^{2N-\sum_{ab}(a+b)i_{ab}}(\sinh\f{\mu}2)^{\sum_{ab}(a+b)i_{ab}}
\end{equation}
For both (\ref{eq:82}) and (\ref{eq:83}), the exponents are fixed
if we fix $u,w=i_{11},y=i_{01}+i_{21}$ and $z=i_{10}+i_{12}$. Therefore,
we can sum over other degrees of freedom first without changing the
exponents. The remaining numbers obey
\begin{equation}
u+y+z=u+v=N/2
\end{equation}
For (\ref{eq:82}), it turns out that
\begin{align}
h_{\mu}^{1}= & (\cosh\f{\mu}2)^{2N}\sum_{u,v,w}\sum_{z+y=v}\f{(-)^{z+w}2^{2w+v-1}(\tanh\f{\mu}2)^{2w+v}(1+\tanh^{2}\f{\mu}2)^{2u-2w+v}\G(N/2)}{\overline{Z}\cosh\mu\G(u-w)\G(1+y)\G(1+z)\G(1+w)}\nonumber \\
 & \times e^{\f{y(2\tau_{12}-\b)^{2}+z(2(\tau_{1}+\tau_{2})+\b)^{2}}{4(N/2-1)}}e^{\f{\left[\f 12(u-1)(u-2)+\f 12v(v-1)\right]\b^{2}+4(u-w-1)(\tau_{1}+\tau_{2})^{2}+4w(\tau_{1}-\tau_{2})^{2}}{4(N/2-1)}}\nonumber \\
= & (\cosh\f{\mu}2)^{2N}\sum_{u,v}\sum_{w=0}^{u}\f{(-)^{w+v}2^{2w+2v-1}(\tanh\f{\mu}2)^{2w+v}(1+\tanh^{2}\f{\mu}2)^{2u-2w+v}\G(N/2)}{\overline{Z}\cosh\mu\G(u-w)\G(1+v)\G(1+w)}e^{-\f{w\tau_{1}\tau_{2}}{N/2-1}}\nonumber \\
 & \times e^{\f{\left[(u-1)(u-2)+v(v-1)\right]\b^{2}/8+(u-1)(\tau_{1}+\tau_{2})^{2}+v(\tau_{1}^{2}+(\tau_{2}+\b/2)^{2})}{N/2-1}}\left[\sinh\f{\tau_{1}(2\tau_{2}+\b)}{N/2-1}\right]^{v}\nonumber \\
= & (\cosh\f{\mu}2)^{2N}\sum_{u+v=N/2}\f{(-)^{v}2^{2v-1}(\tanh\f{\mu}2)^{v}(1+\tanh^{2}\f{\mu}2)^{2u+v}\G(N/2)}{\overline{Z}\cosh\mu\G(u)\G(1+v)}\left(1-e^{-\f{4\tau_{1}\tau_{2}}{N/2-1}}\tanh^{2}\mu\right)^{u-1}\nonumber \\
 & \times e^{\f{\left[(u-1)(u-2)+v(v-1)\right]\b^{2}/8+(u-1)(\tau_{1}+\tau_{2})^{2}+v(\tau_{1}^{2}+(\tau_{2}+\b/2)^{2})}{N/2-1}}\left[\sinh\f{\tau_{1}(2\tau_{2}+\b)}{N/2-1}\right]^{v}\nonumber \\
= & (\cosh\mu)^{N-1}\sum_{v=0}^{N/2-1}\f{(-)^{v}2^{v-1}(\tanh\mu)^{v}\G(N/2)}{\G(N/2-v)\G(1+v)}\left(1-e^{-\f{4\tau_{1}\tau_{2}}{N/2-1}}\tanh^{2}\mu\right)^{N/2-v-1}\nonumber \\
 & \times \left[\sinh\f{\tau_{1}(2\tau_{2}+\b)}{N/2-1}\right]^{v} e^{\f{v^{2}\b^{2}/4-v(\tau_{2}(2\tau_{1}-\b)+(N-4)\b^{2}/8)}{N/2-1}}e^{-\b^{2}/4+(\tau_{1}+\tau_{2})^{2}}\label{eq:86}
\end{align}
where in the last step we drop off $v=N/2$ term because $1/\G(0)=0$.
Smilarly, for (\ref{eq:83}), we have 
\begin{align}
h_{\mu}^{2}= & (\cosh\f{\mu}2)^{2N}\sum_{u,v,w}\sum_{z+y=v}\f{(-)^{z+w}2^{2w+v-1}(\tanh\f{\mu}2)^{2w+v}(1+\tanh^{2}\f{\mu}2)^{2u+v-2w}\G(N/2)}{\overline{Z}\cosh\mu\G(1+u-w)\G(1+y)\G(1+w)\G(z)}\nonumber \\
 & \times e^{\f{y\tau_{12}^{2}+z(\tau_{1}+\tau_{2})^{2}}{N/2-1}} e^{\f{\left[\f 12(v-1)(v-2)+\f 12u(u-1)\right]\b^{2}-8w\tau_{1}(2\tau_{2}+\b)+u(2(\tau_{1}+\tau_{2})+\b)^{2}-4(\tau_{1}+\tau_{2})^{2}}{4(N/2-1)}}\nonumber \\
= & (\cosh\f{\mu}2)^{2N}\sum_{u,v}\sum_{w=0}^{u}\f{(-)^{w+v}2^{2w+2v-2}(\tanh\f{\mu}2)^{2w+v}(1+\tanh^{2}\f{\mu}2)^{2u+v-2w}\G(N/2)}{\overline{Z}\cosh\mu\G(1+u-w)\G(1+w)\G(v)}\nonumber \\
 & \times e^{-\f{2w\tau_{1}(2\tau_{2}+\b)}{N/2-1}} e^{\f{\left[(v-1)(v-2)+u(u-1)\right]\b^{2}/8+u(\tau_{1}+\tau_{2}+\b/2)^{2}+(v-1)(\tau_{1}^{2}+\tau_{2}^{2})}{N/2-1}} \left[\sinh\f{2\tau_{1}\tau_{2}}{N/2-1}\right]^{v-1}\nonumber \\
= & (\cosh\f{\mu}2)^{2N}\sum_{u+v=N/2}\f{(-)^{v}2^{2v-2}(\tanh\f{\mu}2)^{v}(1+\tanh^{2}\f{\mu}2)^{2u+v}\G(N/2)}{\overline{Z}\cosh\mu\G(1+u)\G(v)}\left(1-e^{-\f{2\tau_{1}(2\tau_{2}+\b)}{N/2-1}}\tanh^{2}\mu\right)^{u}\nonumber \\
 & \times e^{\f{\left[(v-1)(v-2)+u(u-1)\right]\b^{2}/8+u(\tau_{1}+\tau_{2}+\b/2)^{2}+(v-1)(\tau_{1}^{2}+\tau_{2}^{2})}{N/2-1}}\left[\sinh\f{2\tau_{1}\tau_{2}}{N/2-1}\right]^{v-1}\nonumber \\
= & (\cosh\mu)^{N-1}\sum_{v=0}^{N/2-1}\f{(-)^{v+1}2^{v-1}(\tanh\mu)^{v+1}\G(N/2)}{\G(N/2-v)\G(v+1)}\left(1-e^{-\f{2\tau_{1}(2\tau_{2}+\b)}{N/2-1}}\tanh^{2}\mu\right)^{N/2-v-1}\nonumber \\
 & \times \left[\sinh\f{2\tau_{1}\tau_{2}}{N/2-1}\right]^{v} e^{\f{v^{2}\b^{2}/4-v(2\tau_{1}\tau_{2}+(\tau_{1}+\tau_{2})\b+N\b^{2}/8)}{N/2-1}}e^{(\tau_{1}+\tau_{2})(\tau_{1}+\tau_{2}+\b)}\label{eq:87}
\end{align}
where in the last line we used $1/\G(0)=0$ and shift $v\ra v+1$. 

As a consistency check, at $\mu=0$, we should get back to two point
function $G(\b/2-\tau_{1}-\tau_{2})$. From (\ref{eq:86}), this leaves
one term with $v=0$, which gives
\begin{equation}
h_{\mu}^{1}=\f 12e^{-\b^{2}/4+(\tau_{1}+\tau_{2})^{2}}\label{eq:88}
\end{equation}
From (\ref{eq:87}), we see no term survives under $\mu\ra0$. Comparing
(\ref{eq:88}) with (\ref{eq:17}) (for $q=4$ and $\tau=\b/2-\tau_{1}-\tau_{2}$),
we see they exactly match. 

To evaluate $h_{\mu}^{1}$ and $h_{\mu}^{2}$, we can rewrite the
sum over $v$ in terms of Gaussian integral. Using the same trick
(\ref{eq:trick}), we can write the normalized (\ref{eq:86}) as 
\begin{align}
h_{\mu}^{1}= & \f{(\cosh\mu)^{N-1}}{2\sqrt{a\pi}}\left(1-e^{-\f{4\tau_{1}\tau_{2}}{N/2-1}}\tanh^{2}\mu\right)^{N/2-1}e^{-\b^{2}/4+(\tau_{1}+\tau_{2})^{2}}\nonumber \\
 & \times\int dxe^{-\f 1ax^{2}}\sum_{v=0}^{N/2-1}\f{\G(N/2)}{\G(N/2-v)\G(1+v)}\left(-2\tanh\mu\f{\sinh\f{\tau_{1}(2\tau_{2}+\b)}{N/2-1}}{1-e^{-\f{4\tau_{1}\tau_{2}}{N/2-1}}\tanh^{2}\mu}e^{2x-\f{(\tau_{2}(2\tau_{1}-\b)+(N-4)\b^{2}/8)}{N/2-1}}\right)^{v}\nonumber \\
= & \f{e^{-\b^{2}/4+(\tau_{1}+\tau_{2})^{2}}\cosh\mu}{2\sqrt{a\pi}}\int dxe^{-\f 1ax^{2}}\left(\cosh^{2}\mu-e^{-b_{1}}\sinh^{2}\mu-\sinh2\mu\sinh\f{\tau_{1}(2\tau_{2}+\b)}{N/2-1}e^{2x-b_{2}}\right)^{N/2-1}
\end{align}
where
\begin{align}
a & =\f{\b^{2}}{4(N/2-1)},\quad b_{1}=\f{4\tau_{1}\tau_{2}}{N/2-1}\quad b_{2}=\f{\tau_{2}(2\tau_{1}-\b)+(N-4)\b^{2}/8}{N/2-1}
\end{align}
Similarly, we can write the normalized (\ref{eq:87}) as 
\begin{align}
h_{\mu}^{2}= & -\f{(\cosh\mu)^{N-1}\tanh\mu}{2\sqrt{a\pi}}\left(1-e^{-\f{2\tau_{1}(2\tau_{2}+\b)}{N/2-1}}\tanh^{2}\mu\right)^{N/2-1}e^{(\tau_{1}+\tau_{2})(\tau_{1}+\tau_{2}+\b)}\nonumber \\
 & \times\int dxe^{-\f 1ax^{2}}\sum_{v=0}^{N/2-1}\f{\G(N/2)}{\G(N/2-v)\G(1+v)}\left(-2\tanh\mu\f{\sinh\f{2\tau_{1}\tau_{2}}{N/2-1}e^{2x-\f{(2\tau_{1}\tau_{2}+(\tau_{1}+\tau_{2})\b+N\b^{2}/8)}{N/2-1}}}{1-e^{-\f{2\tau_{1}(2\tau_{2}+\b)}{N/2-1}}\tanh^{2}\mu}\right)^{v}\nonumber \\
= & -\f{e^{(\tau_{1}+\tau_{2})(\tau_{1}+\tau_{2}+\b)}\sinh\mu}{2\sqrt{a\pi}}\int dxe^{-\f 1ax^{2}}\left(\cosh^{2}\mu-e^{-c_{1}}\sinh^{2}\mu-\sinh2\mu\sinh\f{2\tau_{1}\tau_{2}}{N/2-1}e^{2x-c_{2}}\right)^{N/2-1}
\end{align}
where
\begin{align}
c_{1} & =\f{2\tau_{1}(2\tau_{2}+\b)}{N/2-1},\quad c_{2}=\f{2\tau_{1}\tau_{2}+(\tau_{1}+\tau_{2})\b+N\b^{2}/8}{N/2-1}
\end{align}

\bibliographystyle{JHEP.bst}
\bibliography{main.bib}

\end{document}